%% file: os4-all.tex
\def\spacingNumerator{5}
\def\spacingDenominator{4}
\input jfomacros

\chapno=\cHintroIV
\font \bigteni = cmmi10 scaled\magstep2
\font \bigsevenrm = cmr7 scaled\magstep2

%=====================================================================
%========================== TITLE PAGE ===============================
%=====================================================================

{\nopagenumbers
\multiply\baselineskip by \spacingDenominator\divide \baselineskip by\spacingNumerator

\null\vskip3truecm

%  
%   (1) Put title in next two lines
%
\centerline{\tafontt Single Scale Analysis of Many Fermion Systems}

\vskip0.1in
\centerline{\tbfontt Part 4: Sector Counting}

\vskip0.75in
\centerline{Joel Feldman{\parindent=.15in\footnote{$^{*}$}{Research supported 
in part by the
 Natural Sciences and Engineering Research Council of Canada and the Forschungsinstitut f\"ur Mathematik, ETH Z\"urich}}}
\centerline{Department of Mathematics}
\centerline{University of British Columbia}
\centerline{Vancouver, B.C. }
\centerline{CANADA\ \   V6T 1Z2}
\centerline{feldman@math.ubc.ca}
\centerline{http:/\hskip-3pt/www.math.ubc.ca/\squiggle
feldman/}
\vskip0.3in
\centerline{Horst Kn\"orrer, Eugene Trubowitz}
\centerline{Mathematik}
\centerline{ETH-Zentrum}
\centerline{CH-8092 Z\"urich}
\centerline{SWITZERLAND}
\centerline{knoerrer@math.ethz.ch, trub@math.ethz.ch}
\centerline{http:/\hskip-3pt/www.math.ethz.ch/\squiggle
knoerrer/}

\vskip0.75in
\noindent
%
%   (3) Put abstract below here
{\bf Abstract.\ \ \ } 
For a two dimensional, weakly coupled system of fermions at temperature 
zero, one principal ingredient used to control the composition of the
associated renormalization group maps is the careful counting of the 
number of quartets of sectors that are consistent with conservation of
momentum. A similar counting argument is made to show that 
particle-particle ladders are irrelevant in the case of an asymmetric 
Fermi curve.

\vfill
\eject

%=====================================================================
%======================= TABLE OF CONTENTS ===========================
%=====================================================================

\titleb{Table of Contents}
\halign{\hfill#\ &\hfill#\ &#\hfill&\ p\ \hfil#&\ p\ \hfil#\cr
\noalign{\vskip0.05in}
\S XVIII&\omit Introduction to Part 4                  \span&\:\pgOSXVIII\cr
\noalign{\vskip0.05in}
\S XIX&\omit Comparison of Norms                     \span&\:\pgOSXIX\cr
\noalign{\vskip0.05in}
\S XX&\omit Sums of Momenta and $\epsilon$ -- separated Sets \span&\:\pgOSXX\cr
&&$\epsilon$ -- separated sets                        &\omit&\:\pgOSXXa\cr
&&Sums of Momenta                                     &\omit&\:\pgOSXXb\cr
&&Pairs of Momenta                                    &\omit&\:\pgOSXXc\cr
&&Sectors that are Compatible with Conservation of Momentum &\omit&\:\pgOSXXd\cr
\noalign{\vskip0.05in}
\S XXI&\omit Sectors Compatible with Conservation of Momentum\span&\:\pgOSXXI\cr
&&Comparison of the $1$--norm and the $3$--norm for Four--legged Kernels
                                                      &\omit&\:\pgOSXXIa\cr
&&Auxiliary Norms                                     &\omit&\:\pgOSXXIb\cr
&&Change of Sectorization                             &\omit&\:\pgOSXXIc\cr
\noalign{\vskip0.05in}
\S XXII&\omit Sector Counting for Particle--Particle Ladders\span&\:\pgOSXXII\cr
\noalign{\vskip0.05in}
{\bf Appendices}\span\cr
\noalign{\vskip0.05in}
\S E&\omit Sectors for  $k_0$ Independent Functions   \span&\:\pgOSE\cr
\noalign{\vskip0.05in}
 &\omit References                                    \span&\:\pgOSIVref \cr
\noalign{\vskip0.05in}
 &\omit Notation                                      \span&\:\pgOSIVnot \cr
}
\vfill\eject
\multiply\baselineskip by \spacingNumerator\divide \baselineskip by\spacingDenominator}
\pageno=1

%=====================================================================
%=======================  INTRODUCTION  ==============================
%=====================================================================

\chap{Introduction to Part 4}\PG\pgOSXVIII

In the application of the results of Parts 1 through 3 to many fermion systems
([FKTf1--f3]) the effective potential and all the quantities derived from it
will conserve particle number. Particle number conservation implies that sectorized functions
$\varphi \big((\,\cdot\,,s_1),\cdots,(\,\cdot\,,s_n) \big) \in \cF_0(n;\Si)$,
where $\Si$ is a sectorization, vanish unless the configuration 
$s_1,\cdots,s_n$ of sectors is consistent with conservation of momentum (for a more precise statement see Definition \:\defOSadmissablesectors\ and Remark \:\remOSadmissablesectors). We shall count the number of configurations $s_1,\cdots,s_n$ of sectors consistent with conservation of momentum that satisfy certain constraints. The results are used to compare different norms for four point functions (Proposition \:\propOSthreetoonenorm), and to compare norms associated to different sectorizations at different scales
(Proposition \:\propOSresectorI). The latter is crucial for a multi scale analysis of many fermion systems ([FKTf1--f3]). Notation
tables are provided at the end of the paper.

We retain the assumptions that the dispersion relation $e(\k)$  is 
$r+d+1$ times differentiable, with $r\ge 2$ and $d=2$, and that its 
gradient does not vanish on the Fermi curve 
$
F = \set{\k \in\bbbr^d}{e(\k)=0}
$. 
All the above results hold under additional geometric assumptions on 
the geometry of the Fermi curve $F$. First of all, we assume throughout the rest of the paper that the Fermi curve $F$ is {\bf strictly convex},
with curvature bounded away from zero. If the dispersion 
relation $e(\k)$ is that of a background
electric field alone then $e(\k)=e(-\k)$ and the Fermi curve $F$ is symmetric about the origin.
That is, $\k\in F$ if and only if $-\k$ in $F$.

\definition{\STM\defOSantp}{
\Item i) 
Since $F$ is strictly convex, for each point $\k\in F$ there is a unique
point $a(\k)\in F$ different from $\k$ such that the tangent lines to $F$ at $\k$ and $a(\k)$ are parallel.
$a(\k)$ is called the antipode of $\k$.
\Item ii)
We say that $F$ is symmetric about a point $\p\in\bbbr^2$ if
$\,F=\{\,2\p-\k\,\big|\,\k\in F\,\}$.
}

\example{\STM\exOSantp}{
If $F$ is symmetric about a point $\p$ then $\,a(\k) = 2\p-\k\,$ for all $\k\in F$.
}

Symmetry of the Fermi curve about a point allows for the formation of Cooper pairs and the phase transition to a superconducting state. In [FKTf1--3] 
we show that this is the only instability in
a broad class of short range many fermion models. We now make a
precise asymmetry assumption on the geometry of the Fermi surface.

\definition{\STM\defModI}{ 
Choose an orientation for $F$.
\Item i)
Let $\k\in F$, $\tanV$ the oriented unit tangent vector to $F$ at $\k$
and $\normV$ the inward pointing unit normal vector to $F$ at $\k$. Then
there is a function $\varphi_\k(s)$, defined on a neighbourhood of $0$
in $\bbbr$, such that $\, s\mapsto \k+ s\tanV+\varphi_\k(s)\normV\ $ is an
oriented parametrization of $F$ near $\k$.
\Item ii)
We say that $F$ is strongly asymmetric if there is $n_0\in\bbbn$, with
$n_0\le r$, such
that for each $\k\in F$ there exists an $n\le n_0$ such that
$$
\varphi_\k^{(n)}(0)\ne \varphi_{a(\k)}^{(n)}(0)
$$
}

\remark{\STM\remModII}{

\noindent
i) By construction, $\varphi_\k(0)=\dot\varphi_\k(0)=0$ and
$\ddot\varphi_\k(0)$ is the curvature of $F$ at $\k$.
\Item ii)
If $F$ is symmetric under inversion in some point $\p\in\bbbr^2$,
then $\varphi_\k=\varphi_{a(\k)}$ for all $\k\in F$.
\Item iii)
In [FKTa] we show that independent electrons in a suitably chosen periodic
electromagnetic background field have a dispersion relation whose associated
Fermi curve, for suitably chosen chemical potential, is smooth, 
strictly convex, strongly asymmetric and has nonzero curvature everywhere.
\Item iv) 
In [FKTf1--3] we show that a many fermion system with a strongly 
asymmetric Fermi surface and weak, short range interaction is a Fermi liquid. 
}

Throughout the rest of the paper we assume, unless otherwise stated, that the Fermi surface is strictly convex and either symmetric about a point or strictly asymmetric in the sense of Definition \defModI. In Section \CHppladsect, 
we derive a sector counting result that holds only for strongly 
asymmetric Fermi curves and use it to get an estimate on 
particle--particle bubbles that is better than the logarithmic divergence that,
in the case of a symmetric Fermi surface, is responsible for the Cooper instability.

We emphasize that for the sector counting arguments of Section \CHsumsmom, 
the fact that the model is in two space dimensions is crucial. 
Propositions \:\propSecX\ and \:\proSecXI\ would not hold in a 
three dimensional situation.  See [FKTf1, \S \CHintroOverview, subsection 8].

\vfill\eject
%=====================================================================
%================= COMPARISON of NORMS ===============================
%=====================================================================

\chap{Comparison of Norms}\PG\pgOSXIX

Theorem \thOSrengroupestimate\ indicates that  ladders give the dominant
contributions to $w_{0,4}$. The $\v \,\cdot\,\v_{3,\Si}$ norm of ladders will be
estimated in \S\CHppladsect\ and [FKTl]. To control the $N(w;\,\cdots\,)$
norms of $w$, we develop a bound on the $\v\,\cdot\,\v_{1,\Si}$ norm
of a ladder in terms of its $\v\,\cdot\,\v_{3,\Si}$ norm.

\proposition{\STM\propOSthreetoonenorm}{ Let $\Si$ be a sectorization 
of length $\sfrac{1}{M^{2j/3}}\le \fl \le \sfrac{1}{M^{j/2}}$ 
at scale $j\ge 4$. Furthermore let $\varphi\in\cF_0(4,\Si)$ and
$f\in \check\cF_{4;\Si}$ be particle number conserving functions. Then 
$$
\v\varphi\v_{1,\Si} \le \abcst \sfrac{1}{\fl}\,\v\varphi\v_{3,\Si}
\qquad {\rm and} \qquad
\v f\tv_{1,\Si} \le \abcst \sfrac{1}{\fl}\,\v f\tv_{3,\Si}
$$
with a constant $\abcst$ that is independent of $M,j, \Si$
}
This Proposition is proven after Lemma \:\lemOSonetothree.
In the renormalization group analysis, we go from scale to scale. After
integrating out scale $j$, we shall have an effective potential $\cW$ with a
representative $w$, sectorized at scale $j$; and we will have an estimate on the
norm of $w$. To apply Theorem \thOSrengroupestimate\ at scale $j+1$ we then need a
representative for $\cW$ that is sectorized at scale $j+1$ and estimates on it. 
This change of sectorizations is implemented by

\definition{\STM\defOSresector}{ Let $j,i\ge 2$. Let $\Si$ and $\Si'$ be  
sectorizations of length $\fl$ at scale  $j$ and length $\fl'$ at scale $i$,
respectively. If $i\ne j$, define, for functions $\varphi$ on $\cB^m\times\big(\cB\times\Si'\big)^n$ and $f$ on $\check\cB^m\times\big(\cB\times\Si'\big)^n$,
$$\eqalign{
\varphi_\Si({\sst\eta_1,\cdots,\eta_m;\,(\xi_1,s_1),\cdots,(\xi_n,s_n)} )
&=\smsum_{s'_1,\cdots,s'_{n} \in \Si'} \int {\sst d\xi'_1\cdots d\xi'_{n}}\,
\varphi({\sst\eta_1,\cdots,\eta_m;\,(\xi'_1,s'_1),\cdots,(\xi'_n,s'_n)} )
\smprod_{\ell=1}^n\hat\chi_{s_\ell}(\xi'_\ell,\xi_\ell) \cr
f_\Si({\sst\check\eta_1,\cdots,\check\eta_m;\,(\xi_1,s_1),\cdots,(\xi_n,s_n)} )
&=\smsum_{s'_1,\cdots,s'_{n} \in \Si'} \int {\sst d\xi'_1\cdots d\xi'_{n}}\,
f({\sst\check\eta_1,\cdots,\check\eta_m;\,(\xi'_1,s'_1),\cdots,(\xi'_n,s'_n)} )
\smprod_{\ell=1}^n\hat\chi_{s_\ell}(\xi'_\ell,\xi_\ell) \cr
}$$
where $\chi_s,\ s\in\Si$ is the partition of unity of Lemma 
\:\lemOSsectpartunit\ and (\:\eqnOSpartunit). 
If $\varphi$ is translation invariant and antisymmetric
under permutation of its $\et$ arguments, then $\varphi_\Si\in\cF_m(n;\Si)$. 
For $i= j$ and $\Si'=\Si$, define $\varphi_{\Si}=\varphi$ and $f_\Si=f$.
}

\goodbreak
\remark{\STM\remOSresector}{ 
\Item i)
 If $u\in\cF_0(2;\Si')$ is an antisymmetric, spin
independent and particle number conserving function  then
$$
\check u_\Si(k)=\check u(k)\,\big(\tilde\nu^{(\ge j)}(k)\big)^2
$$
\Item ii)
For a function $\varphi$ on $\cB^m\times\big(\cB\times\Si'\big)^n$ one has
$\ \big(\varphi_\Si\big)^\sim = \big(\varphi^\sim\big)_\Si$.
\Item iii)
Let $j,i_1,i_2\ge 2$ with $i_2>i_1$. Let $\Si$, $\Si_1$
and $\Si_2$ be  sectorizations at scales $j$, $i_1$ and $i_2$
respectively. Then, for each function $\varphi$ on $\cB^m\times\big(\cB\times\Si\big)^n$ and each function $f$ on $\check\cB^m\times\big(\cB\times\Si\big)^n$
$$
\big(\varphi_{\Si_1}\big)_{\Si_2}=\varphi_{\Si_2}
\qquad {\rm and} \qquad
\big(f_{\Si_1}\big)_{\Si_2}=f_{\Si_2}
$$
}

\proposition{\STM\propOSresectorI}{ 
Let $j>i\ge 2$, $\sfrac{1}{M^{j-3/2}}\le\fl\le\sfrac{1}{M^{(j-1)/2}}$ and 
$\sfrac{1}{M^{i-3/2}}\le\fl'\le\sfrac{1}{M^{(i-1)/2}}$ with $4\fl<\fl'$. 
Let $\Si$ and $\Si'$ be sectorizations of length  $\fl$ at scale $j$ and 
length $\fl'$ at scale $i$, respectively. Let $\varphi\in\cF_m(n;\Si')$ and
$f\in\check\cF_m(n;\Si' )$ be particle number conserving functions.
\Item i) 
If $m\ne 0$
$$
\V\varphi_\Si\V_{1,\Si}\le\abcst^n\,\cb_{j-1}\,\big[\sfrac{\fl'}{\fl}\big]^n\,
\v\varphi\v_{1,\Si'}
$$

\Item ii) 
If $f$ is antisymmetric in its $(\xi,s)$ arguments, then for all $p$
$$
\V f_\Si\tV_{p,\Si}
\le\abcst^n\,\cb_{j-1}\,\big[\sfrac{\fl'}{\fl}\big]^{n+m-p-1}
\v f \tv _{p,\Si'}
$$
\phantom{ii)} Moreover, if $\ \fl \ge \sfrac{1}{M^{2/3(j-1)}}$, $\fl'\le \sfrac{1}{6}\sqrt{\fl}$ and 
$n\ge 3$
$$
\V f_\Si \tV _{1,\Si}
\le\abcst^n\,\cb_{j-1}\,\big[\sfrac{\fl'}{\fl}\big]^{n+m-3}
\Big( \v f\tv_{1,\Si'} + \sfrac{1}{\fl'}\, \v f\tv_{3,\Si'} \Big)
$$

\Item iii) 
If $f$ is antisymmetric in its $(\xi,s)$ arguments, then for all $p$
$$
\V f_{\Si'}\tV_{p,\Si'}
\le\abcst^n\,\cb_{i-1}\,\big[\sfrac{\fl'}{\fl}\big]^{p-m}
\v f \tv _{p,\Si}
$$

\noindent
Here $\abcst$ is  a constant that is independent of $M,j, \Si$

}
\noindent
This Proposition is proved after Lemma \:\lemOSseccountI.

\remark{\STM\remOStoresectorI}{
Since for $m=0$ the norms $\v \varphi \v_{p,\Si}$ and 
$\v \varphi\tv_{p,\Si}$ agree, Proposition \propOSresectorI.ii implies that,
in the case that $\ \fl \ge \sfrac{1}{M^{2/3(j-1)}}$ and
$4\fl<\fl'<\sfrac{1}{6}\sqrt{\fl}$,
for antisymmetric  $\varphi\in\cF_0(n;\Si')$  
$$\meqalign{
\V\varphi_\Si\V_{1,\Si}
&\le\abcst\,\cb_{j-1} \v\varphi\v_{1,\Si'}
&\qquad {\rm if}\ n=2  \cr
\V\varphi_\Si\V_{3,\Si}
&\le\abcst\,\cb_{j-1} \v\varphi\v_{3,\Si'}
&\qquad {\rm if}\ n=4  \cr
\V\varphi_\Si\V_{1,\Si}
&\le\abcst^n\,\cb_{j-1}\,\big[\sfrac{\fl'}{\fl}\big]^{n-3}
\Big(  \v\varphi\v_{1,\Si'}+ \sfrac{1}{\fl'}\, \v\varphi\v_{3,\Si'}\Big)
&\qquad {\rm if}\ n\ge 4  \cr
}$$
}

The resectorization of functions on $\fX_\Si^n\ =\ \big({\check\cB\dunion(\cB\times\Si)}\big)^n$
is defined just as in Definition \defOSresector. To be precise, recall from
Remark \remOSbigdisjointunion\ and Definition \defOSdisjointOrd.iii that
$$
\fX_\Si^n\ =\ 
\bigcup_{i_1,\cdots,i_n \in \{0,1\}}\kern-2.8em\cdot\kern2.8em \fX_{i_1}(\Si) \times\cdots\times \fX_{i_n}(\Si)
$$
where $\fX_0(\Si)=\check\cB$ and $\fX_1(\Si)=\cB\times\Si$. Furthermore, 
for each $\vec\imath=(i_1,\cdots,i_n)\in\{0,1\}^n$, the map $\ord$ gives a bijection
between functions on $\fX_{i_1}(\Si) \times\cdots\times \fX_{i_n}(\Si)$ and
functions on $\check\cB^{m(\vec\imath)}\times (\cB\times\Si)^{n-m(\vec\imath)}$,
where $m(\vec\imath)=n-i_1-\cdots-i_n$.
\definition{\STM\defOSresectorII}{ Let $j,i\ge 2$. Let $\Si$ and $\Si'$ be  
sectorizations of length $\fl$ at scale  $j$ and length $\fl'$ at scale $i$,
respectively.
\Item i)
Let $\vec\imath=(i_1,\cdots,i_n)\in\{0,1\}^n$ and $f$ a function 
 on $\fX_{i_1}(\Si') \times\cdots\times \fX_{i_n}(\Si')$. Then
$f_\Si$ is the function on $\fX_{i_1}(\Si) \times\cdots\times \fX_{i_n}(\Si)$
determined by $\ord (f_\Si)=(\ord f)_\Si$.
\Item ii) If $f$ is a function on $\fX_{\Si'}^n$, its resectorization
$f_\Si$ is the function on $\fX_\Si^n$ determined by
$$ 
f_\Si\big|_{\vec\imath}=\big(f\big|_{\vec\imath}\big)_\Si\qquad
\hbox{for all}\quad \vec\imath\in\{0,1\}^n
$$
}
From Proposition \propOSresectorI, we have
\corollary{\STM\corOSresectorI}{
Let $j>i\ge 2$, $\sfrac{1}{M^{j-3/2}}\le\fl\le\sfrac{1}{M^{(j-1)/2}}$ and 
$\sfrac{1}{M^{i-3/2}}\le\fl'\le\sfrac{1}{M^{(i-1)/2}}$ with $4\fl<\fl'$. 
Let $\Si$ and $\Si'$ be 
sectorizations of length  $\fl$ at scale $j$ and length $\fl'$ 
at scale $i$, respectively.
 Let $f\in\check\cF_{n;\Si'}$ be an antisymmetric particle number conserving function. Then for all $p$
$$
\V f_\Si\tV_{p,\Si}
\le\abcst^n\,\cb_{j-1}\,\big[\sfrac{\fl'}{\fl}\big]^{n-p-1}
\v f \tv _{p,\Si'}
$$
Moreover, if $\ \fl \ge \sfrac{1}{M^{2/3(j-1)}}$, $\fl'\le \sfrac{1}{6}\sqrt{\fl}$ and 
$n\ge 4$
$$
\V f_\Si \tV _{1,\Si}
\le\abcst^n\,\cb_{j-1}\,\big[\sfrac{\fl'}{\fl}\big]^{n-3}
\Big( \v f\tv_{1,\Si'} + \sfrac{1}{\fl'}\, \v f\tv_{3,\Si'} \Big)
$$
}

\vskip .5cm

In the renormalization group analysis of [FKTf1--f3], the numbers $\rho_{0;n}$ 
used as weights in the norms $N_j$  of Definition \defOSscalednorms\ do 
not depend on
the scale $j$. As pointed out in Remark \remOSscalednorms, boundedness 
in $j$ of the norms
$N_j$ implies that the coefficient of $t^\0$ in $\v w_{0,2}\v_{1,\Si}$ 
has positive power counting (that is, tends to zero as a power of  
$\sfrac{1}{M^j}$) and the coefficient of $t^\0$ in $\v w_{0,4}\v_{3,\Si}$ has
neutral power counting. The other contributions $w_{m,n}$ behave well with respect to resectorization.

\corollary{\STM\corOSirrelevantresect}{
Fix $ \sfrac{1}{2} < \aleph < \sfrac{2}{3}$ and let 
$j\ge \sfrac{3}{2-3\aleph}$. Let $\Si_{j+1}$ and $\Si_j$ be  
sectorizations of length $ \fl_{j+1} = \sfrac{1}{M^{\aleph(j+1)}}$ at scale
$j+1$ and $\fl_j =\sfrac{1}{M^{\aleph j}}$ at scale $j$,
respectively. Let $\vec\rho=\big(\rho_{m;n}\big)$
be a system of positive real numbers obeying (\eqnOSrhomn) and set
$$
\rho'_{m;n}=\cases{\rho_{m;n} & if $m=0$\cr
\noalign{\vskip.05in}
                   \root{4}\of{\sfrac{\fl_jM^j}{\fl_{j+1} M^{j+1}}}\ \rho_{m;n}
                   =\sfrac{1}{M^{(1-\aleph)/4}}\ \rho_{m;n} & if $m>0$}
$$
Let 
$$\eqalign{
w(\phi,\psi) = \sum_{m,n\atop{m+n{\rm\ even}}}\ \sum_{s_1,\cdots,s_n\in\Si_{j+1}}\ 
\int {\sst d\eta_1\cdots d\eta_m\,d\xi_1\cdots d\xi_n}\ & 
w_{m,n}({\sst \eta_1,\cdots, \eta_m\,(\xi_1,s_1),\cdots ,(\xi_n,s_n)})\cr
& \hskip0.5cm \phi({\sst \eta_1})\cdots \phi({\sst \eta_m})\
\psi({\sst (\xi_1,s_1)})\cdots \psi({\sst (\xi_n,s_n)\,})\cr
}$$
with $w_{m,n}\in\cF_m(n;\Si_j)$,
be an even $\Si_j$--sectorized particle number conserving Grassmann function
with $w_{0,2}=0$ and $w_{m,0}=0$ for all $m$.  If $M$ is big enough, then
$$
N_{j+1}(w_{\Si_{j+1}};64\al;X,\Si_{j+1},\vec\rho\,)
\le \abcst\,\fe_{j+1}(X)\, 
N_j\big(w;\sfrac{\al}{2};X,\Si_j,\vec\rho^{\,\prime}\big)
$$
with the  constant $\abcst$  independent of $M,\ j,\ \Si_j$ and $\Si_{j+1}$.
If, in addition $w_{0,4}=0$, then
$$
N_{j+1}(w_{\Si_{j+1}};64\al;X,\Si_{j+1},\vec\rho\,)
\le \sfrac{1}{M^{(1-\aleph)/8}}\,\fe_{j+1}(X)\,
N_j\big(w;\sfrac{\al}{2};X,\Si_j,\vec\rho^{\,\prime}\big)
$$
}
\prf We apply Proposition \propOSresectorI\ 
with $j$ replaced by $j+1$, $i=j$, $\fl=\fl_{j+1}$ and $\fl'=\fl_j$.
Observe that the hypotheses of part (ii) are fulfilled. 
In this proof, use $\v\ \cdot\ \v_{\Si,\vec\rho}$ to designate the norm
of Definition \defOSscalednorms\ using the indicated $\vec\rho$.

If $m,n\ge 1$, by Proposition \propOSresectorI.i,
$$\eqalign{
&\sfrac{M^{2(j+1)}}{\fl_{j+1}}\,\fe_{j+1}(X)\,\big(64\al\big)^{n}\,
\big(\sfrac{\fl_{j+1}\,\IB}{M^{j+1}}\big)^{n/2} 
\,\v (w_{m,n})_{\Si_{j+1}}\v_{\Si_{j+1},\vec\rho} \cr
%%%
& \hskip 1cm = \,\fe_{j+1}(X)\,\big(64\al\big)^{n}\,
\big(\sfrac{\fl_{j+1}\,\IB}{M^{j+1}}\big)^{n/2}\,
\rho_{m;n}
\,\V (w_{m,n})_{\Si_{j+1}}\V_{1,\Si_{j+1}} \cr
%%%
&\hskip 1cm \le
\abcst^n\,\cb_j\,\fe_{j+1}(X)\,\big(\sfrac{\fl_j}{\fl_{j+1}}\big)^n\ 
\Big( \sfrac{2^{14}}{M}\,\sfrac{\fl_{j+1}}{\fl_j} \Big)^{n/2}
\sfrac{\rho_{m;n}}{\rho'_{m;n}}\ 
\big(\sfrac{\al}{2}\big)^{n}
\big(\sfrac{\fl_j\,\IB}{M^j}\big)^{n/2}
\rho'_{m;n}\,\v w_{m,n}\v_{1,\Si_j} \cr
%%%
&\hskip 1cm \le
\abcst^n\,\fe_{j+1}(X)\,
\Big( \sfrac{1}{M^{1-\aleph}} \Big)^{(2n-1)/4}\ 
\sfrac{M^{2j}}{\fl_j}\,\cb_j
\big(\sfrac{\al}{2}\big)^{n}
\big(\sfrac{\fl_j\,\IB}{M^j}\big)^{n/2}
\v w_{m,n}\v_{\Si_j,\vec\rho'} \cr
%%%
&\hskip 1cm \le
\sfrac{1}{M^{(1-\aleph)/8}}\,\fe_{j+1}(X)\ 
\sfrac{M^{2j}}{\fl_j}\,\fe_{j}(X)\,
\big(\sfrac{\al}{2}\big)^{n}
\big(\sfrac{\fl_j\,\IB}{M^j}\big)^{n/2}
\v w_{m,n}\v_{\Si_j,\vec\rho'} \cr
}$$
if $M$ is large enough.
If $m=0$ and $n\ge 4$, by Proposition \propOSresectorI.ii
and Remark \remOSsecdiffdecaynorm,
$$\eqalign{
&\sfrac{M^{2(j+1)}}{\fl_{j+1}}\,\fe_{j+1}(X)\,\big(64\al\big)^{n}\,
\big(\sfrac{\fl_{j+1}\,\IB}{M^{j+1}}\big)^{n/2} 
\,\v (w_{m,n})_{\Si_{j+1}}\v_{\Si_{j+1},\vec\rho} \cr
%%%
& \hskip 1cm = \sfrac{M^{2(j+1)}}{\fl_{j+1}}\,\fe_{j+1}(X)\,\big(64\al\big)^{n}\,
\big(\sfrac{\fl_{j+1}\,\IB}{M^{j+1}}\big)^{n/2}\,
\rho_{0;n} 
\cr& \hskip 3cm  
\Big[  \V (w_{0,n})_{\Si_{j+1}}\V_{1,\Si_{j+1}} 
     + \sfrac{1}{\fl_{j+1}}\,\V (w_{0,n})_{\Si_{j+1}}\V_{3,\Si_{j+1}}
     + \sfrac{1}{\fl_{j+1}^2}\,\V (w_{0,n})_{\Si_{j+1}}\V_{5,\Si_{j+1}} \Big]  \cr 
%%%
&\hskip 1cm \le
\abcst^n\,\cb_j\,\fe_{j+1}(X)\,\sfrac{M^{2j}}{\fl_j}\,\big(\sfrac{\al}{2}\big)^n\,
\big(\sfrac{\fl_j\,\IB}{M^j}\big)^{n/2}\, 
\sfrac{1}{M^{(n-4)/2}} \big(\sfrac{\fl_{j+1}}{\fl_j}\big)^{(n-2)/2}\,  
\rho_{0;n} \, \cr
& \hskip 3cm  \Big[\, 
   \big(\sfrac{\fl_j}{\fl_{j+1}}\big)^{n-3}\, \v w_{0,n}\v_{1,\Si_j} 
   +\big(\sfrac{\fl_j}{\fl_{j+1}}\big)^{n-3}\,\sfrac{1}{\fl_j}\,
      \v w_{0,n}\v_{3,\Si_j}
   +\big(\sfrac{\fl_j}{\fl_{j+1}}\big)^{n-4}\,\sfrac{1}{{\fl_j}^2}\,
      \v w_{0,n}\v_{5,\Si_j} \Big] \cr 
%%%
&\hskip 1cm \le
\abcst^n\,\fe_{j+1}(X)\,\sfrac{M^{2j}}{\fl_j}\,\cb_j\,
\big(\sfrac{\al}{2}\big)^n\,
\big(\sfrac{\fl_j\,\IB}{M^j}\big)^{n/2}\, 
\sfrac{1}{M^{(n-4)/2}} \big(\sfrac{\fl_j}{\fl_{j+1}}\big)^{(n-4)/2}\, \rho'_{0;n}\,  \cr
& \hskip 3cm  \Big[\, \v w_{0,n}\v_{1,\Si_j} 
   +\sfrac{1}{\fl_j}\,\v w_{0,n}\v_{3,\Si_j}
   +\sfrac{1}{{\fl_j}^2}\,\v w_{0,n}\v_{5,\Si_j} \Big] \cr 
&\hskip 1cm =
\abcst^n\,\fe_{j+1}(X)\,\Big( \sfrac{1}{M^{1-\aleph}}  \Big)^{(n-4)/2}
\sfrac{M^{2j}}{\fl_j}\,\cb_j\,\big(\sfrac{\al}{2}\big)^n\,
\big(\sfrac{\fl_j\,\IB}{M^j}\big)^{n/2}\, 
 \v w_{0,n}\v_{\Si_j,\vec\rho'} \cr
&\hskip 1cm \le
\,\Big( \sfrac{1}{M^{(1-\aleph)/8}} +\abcst\,\de_{n,4} \Big)\fe_{j+1}(X)\ 
\sfrac{M^{2j}}{\fl_j}\,\fe_{j}(X)\,\big(\sfrac{\al}{2}\big)^n\,
\big(\sfrac{\fl_j\,\IB}{M^j}\big)^{n/2}\, 
 \v w_{0,n}\v_{\Si_j,\vec\rho'} 
}$$
\endproof

The analog of Corollary \corOSirrelevantresect\ for the $N_j^\sim$ norms
is

\corollary{\STM\corOStildeirrelevantresect}{
Fix $ \sfrac{1}{2} < \aleph < \sfrac{2}{3}$ and let 
$j\ge \sfrac{3}{2-3\aleph}$.  Let $\Si_{j+1}$ and $\Si_j$ be  
sectorizations of length $ \fl_{j+1} = \sfrac{1}{M^{\aleph(j+1)}}$ at scale
$j+1$ and $\fl_j =\sfrac{1}{M^{\aleph j}}$ at scale $j$,
respectively. Let $\vec\rho=\big(\rho_{m;n}\big)$
be a system of positive real numbers obeying (\eqnOStilderhomn).
Let 
$$\eqalign{
w(\phi,\psi) = \sum_n 
\int_{\fX_\Si^n} {\sst dx_1\cdots dx_n}\ 
f_{n}({\sst x_1,\cdots, x_n})\ 
\Psi({\sst x_1})\cdots \Psi({\sst x_n})\cr
}$$
with $f_n \in \check \cF_{n;\Si}$ antisymmetric,
be an even $\Si_j$--sectorized particle number conserving Grassmann function
with $f_2=0$.  If $M$ is big enough, then
$$
N^\sim_{j+1}(w_{\Si_{j+1}};64\al;X,\Si_{j+1},\vec\rho)
\le \abcst\,\fe_{j+1}(X)\, N^\sim_j\big(w;\sfrac{\al}{2};X,\Si_j,\vec\rho\big)
$$
with the  constant $\abcst$  independent of $M,\ j,\ \Si_j$ and $\Si_{j+1}$.
If, in addition $f_4=0$, then
$$
N^\sim_{j+1}(w_{\Si_{j+1}};64\al;X,\Si_{j+1},\vec\rho)
\le \sfrac{1}{M^{(1-\aleph)/8}}\,\fe_{j+1}(X)\,
N^\sim_j\big(w;\sfrac{\al}{2};X,\Si_j,\vec\rho\big)
$$
}
\prf 
If $n\ge 4$, by Proposition \propOSresectorI.ii 
with $j$ replaced by $j+1$, $i=j$, $\fl=\fl_{j+1}$ and $\fl'=\fl_j$,
$$\eqalign{
&\sfrac{M^{2(j+1)}}{\fl_{j+1}}\,\fe_{j+1}(X)\,\big(64\al\big)^{n}\,
\big(\sfrac{\fl_{j+1}\,\IB}{M^{j+1}}\big)^{n/2} 
\,\v (f_n)_{\Si_{j+1}}\tv_{\Si_{j+1}} \cr
%%%
& \hskip 0.3cm \le \sfrac{M^{2(j+1)}}{\fl_{j+1}}\,\fe_{j+1}(X)\,\big(64\al\big)^{n}\,
\big(\sfrac{\fl_{j+1}\,\IB}{M^{j+1}}\big)^{n/2} 
\Big\{\V (f_n)_{\Si_{j+1}}\tV_{1,\Si_{j+1},\vec\rho} + 
\smsum_{p=2}^6 \sfrac{1}{\fl^{[(p-1)/2]}_{j+1}}
\V (f_n)_{\Si_{j+1}}\tV_{p,\Si_{j+1},\vec\rho}  \Big\}  \cr 
%%%
&\hskip 0.3cm \le
\abcst^n\,\cb_j\,\fe_{j+1}(X)\,\sfrac{M^{2j}}{\fl_j}\,\big(\sfrac{\al}{2}\big)^n\,
\big(\sfrac{\fl_j\,\IB}{M^j}\big)^{n/2}\, 
\sfrac{1}{M^{(n-4)/2}} \big(\sfrac{\fl_{j+1}}{\fl_j}\big)^{(n-2)/2}\cr
& \hskip 2.3cm 
\Big\{ 
\big(\sfrac{\fl_j}{\fl_{j+1}}\big)^{n-3}
       \big(\V f_n\tV_{1,\Si_{j},\vec\rho}+\sfrac{1}{\fl_j}
       \V f_n\tV_{3,\Si_{j},\vec\rho}\big)
+\smsum_{p=2}^6 \sfrac{1}{\fl^{[(p-1)/2]}_{j+1}}\big(\sfrac{\fl_j}{\fl_{j+1}}\big)^{n-p-1}
\V f_n\tV_{p,\Si_{j},\vec\rho}  \Big\}\cr
%%%
&\hskip 0.3cm \le
\abcst^n\,\fe_{j+1}(X)\,\sfrac{M^{2j}}{\fl_j}\,\cb_j\,
\big(\sfrac{\al}{2}\big)^n\,
\big(\sfrac{\fl_j\,\IB}{M^j}\big)^{n/2}\, 
\sfrac{1}{M^{(n-4)/2}} \big(\sfrac{\fl_j}{\fl_{j+1}}\big)^{(n-4)/2}  \cr
& \hskip 2.3cm 
\Big\{ \V f_n\tV_{1,\Si_{j},\vec\rho}+\sfrac{1}{\fl_j}
       \V f_n\tV_{3,\Si_{j},\vec\rho}
+\smsum_{p=2}^6 \big(\sfrac{\fl_{j+1}}{\fl_j}\big)^{p-2-[(p-1)/2]}
\sfrac{1}{\fl^{[(p-1)/2]}_{j}}\V f_n\tV_{p,\Si_{j},\vec\rho}  \Big\}\cr
%%%
&\hskip 0.3cm \le
\abcst^n\,\fe_{j+1}(X)\,\Big( \sfrac{1}{M^{1-\aleph}}  \Big)^{(n-4)/2}
\sfrac{M^{2j}}{\fl_j}\,\fe_j(X)\,\big(\sfrac{\al}{2}\big)^n\,
\big(\sfrac{\fl_j\,\IB}{M^j}\big)^{n/2}\, 
 \v f_n\tv_{\Si_j} \cr
%%%
&\hskip 0.3cm \le
\,\Big( \sfrac{1}{M^{(1-\aleph)/8}} +\abcst\,\de_{n,4} \Big)\fe_{j+1}(X)\ 
\sfrac{M^{2j}}{\fl_j}\,\fe_{j}(X)\,\big(\sfrac{\al}{2}\big)^n\,
\big(\sfrac{\fl_j\,\IB}{M^j}\big)^{n/2}\, 
 \v f_n\tv_{\Si_j} 
}$$
\endproof

The positive power counting of $\v w_{0,2}\v_{1,\Si}$ is achieved by
renormalization. That is, we choose the counterterm in such a way that, at each
scale, the restriction of the Fourier transform of $w_{0,2}$ to the Fermi surface
is small. The following Proposition ensures that then $\v w_{0,2}\v_{1,\Si}$
is also small. 

\definition{\STM\defOSvanishkzero}{
The function $u\in \cF_0(2;\Si)$ is said to vanish at $k_0=0$ if
$$
\check u\big(((0,\k),\si,a,s), ((0,\k),\si',a',s')\big)=0
$$
for all $a,a'\in\{0,1\}$, $\si,\si'\in\{\uparrow,\downarrow\}$ 
and $s,s'\in\Si$.
}

\proposition{\STM\propOSIntUp}{ 
There is a constant $\abcst$, independent of $M$,
such that the following holds: 
Let $j\ge i\ge 2$ and $\Si$ and $\Si'$ be sectorizations 
at scale $j$ and $i$, respectively. If $i=j$ assume that $\Si=\Si'$.
Let $u\in \cF_0(2;\Si')$ be a function that vanishes at $k_0=0$. Then
\Item i)
$$
\v u_\Si\,\v_{1,\Si}
\le \abcst\,\sfrac{1}{M^{j-1}}\,\cb_{j-1}\,\V\cD^{(1,0,0)}_{1,2}u\,\V_{1,\Si'} 
+\smsum_{\de \in \bbbn_0\times\bbbn_0^d\atop \de_0\ne 0}\infty t^\de
$$
\Item ii)
$$
\V \cD^{(1,0,0)}_{1,2}u_\Si\,\V_{1,\Si}
\le \abcst\,\cb_{j-1}\,\V\cD^{(1,0,0)}_{1,2}u\,\V_{1,\Si'} 
$$
}

\prf i)
Fix $s_1,s_2 \in \Si$. If $i<j$, by Lemma \lemOSelloneinfty, Lemma \lemOSprepintup.i, 
Lemma \:\lemOSmorepartunity\ and (\eqnOSprodcontrbound)
$$\eqalign{
\big\| u_\Si({\sst(\xi_1,s_1),(\xi_2,s_2)}) \big\|_{1,\infty} 
& \le \abcst \max_{s_1',s_2'\in \Si'}
\Big\| \int {\sst d\eta_1\,d\eta_2}\,u({\sst(\eta_1,s'_1),(\eta_2,s'_2)})\
\hat\chi_{s_1}({\sst \eta_1,\xi_1})\,\hat\chi_{s_2}(\sst{\eta_2,\xi_2})
\,\Big\|_{1,\infty} \cr 
& \le \abcst\, \big\|\hat\chi_{s_2}\big\|_{1,\infty}
\max_{s_1',s_2'\in \Si'}
\Big\| \int {\sst d\eta_1}\,u({\sst(\eta_1,s'_1),(\,\cdot\,,s'_2)})\
\hat\chi_{s_1}({\sst \eta_1,\,\cdot\,})\,\Big\|_{1,\infty}  \cr
& \le \abcst\, \cb_{j-1}\,
\big\|\sfrac{\partial \chi_{s_1}'}{\partial x_0}\big\|_{L^1}\,
\max_{s_1',s_2'\in \Si'}
\big\| \cD^{(1,0,0)}_{1,2}u({\sst(\,\cdot\,,s'_1),(\,\cdot\,,s'_2)}\,
\big\|_{1,\infty} 
+\smsum_{\de \in \bbbn_0\times\bbbn_0^d\atop \de_0\ne 0}
\hskip-10pt \infty t^\de\cr
& \le \abcst\, \sfrac{1}{M^{j-1}}\,\cb_{j-1}^2\,
\V \cD^{(1,0,0)}_{1,2}u\,\V_{1,\Si'}  
+\smsum_{\de \in \bbbn_0\times\bbbn_0^d\atop \de_0\ne 0}\infty t^\de\cr
 & \le \abcst\,
\sfrac{1}{M^{j-1}}\,\cb_{j-1}\,\V\cD^{(1,0,0)}_{1,2}u\,\V_{1,\Si'}  
+\smsum_{\de \in \bbbn_0\times\bbbn_0^d\atop \de_0\ne 0}\infty t^\de\cr  
}$$
Similarly, if $i=j$ and $\Si=\Si'$, then setting
$$
\chi^{(e)}_s(k) 
= \varphi\big(\half M^{2j-2}(k_0^2+e(\k)^2)\big)\,\Th_s(\k'(k))
$$
(which just differs by a $\half$ from the definition of $\chi_s(k)$ in 
(\eqnOSpartunit)), we have, using the support property of Definition 
\defOSsectrepr.ii,
$$\eqalign{
\big\| u_\Si({\sst(\xi_1,s_1),(\xi_2,s_2)}) \big\|_{1,\infty}
& =  \big\| u({\sst(\xi_1,s_1),(\xi_2,s_2)}) \big\|_{1,\infty} \cr
& = \Big\|\smsum_{s'_1,s'_2\in\Si} \int {\sst d\eta_1\,d\eta_2}\,u({\sst(\eta_1,s_1),(\eta_2,s_2)})\
\hat{\chi}^{(e)}_{s'_1}({\sst\eta_1,\xi_1})\,
\hat{\chi}^{(e)}_{s'_2}(\sst{\eta_2,\xi_2})\,\Big\|_{1,\infty} \cr 
& \le \abcst \max_{s_1',s_2'\in \Si} \Big\|\int {\sst d\eta_1\,d\eta_2}\,u({\sst(\eta_1,s_1),(\eta_2,s_2)})\
\hat{\chi}^{(e)}_{s'_1}({\sst\eta_1,\xi_1})\,
\hat{\chi}^{(e)}_{s'_2}(\sst{\eta_2,\xi_2})\,\Big\|_{1,\infty} \cr 
 & \le \abcst\,
\sfrac{1}{M^{j-1}}\,\cb_{j-1}\,\V \cD^{(1,0,0)}_{1,2}u\,\V_{1,\Si}  
+\smsum_{\de \in \bbbn_0\times\bbbn_0^d\atop \de_0\ne 0}\infty t^\de\cr 
}$$
Since for every $s_1\in\Si$ there are at most three sectors $s_2$ with
$\tilde s_1 \cap \tilde s_2 \ne \emptyset$, in both cases
$$
\v u_\Si\,\v_{1,\Si}
\le \abcst\,\sfrac{1}{M^{j-1}}\,\cb_{j-1}\,\V\cD^{(1,0,0)}_{1,2}u\,\V_{1,\Si'}
+\smsum_{\de \in \bbbn_0\times\bbbn_0^d\atop \de_0\ne 0}\infty t^\de 
$$
\vskip.1in
\Item ii) 
If $i=j$ and $\Si=\Si'$ the statement is trivial. So assume that $i<j$.
Fix $s_1,s_2 \in \Si$. By Lemma  \lemOSprepintup.ii 
(twice), 
Lemma \:\lemOSmorepartunity\  and (\eqnOSprodcontrbound)
$$\eqalign{
\big\| &\cD^{(1,0,0)}_{1,2}u_\Si({\sst(\xi_1,s_1),(\xi_2,s_2)}) \big\|_{1,\infty} \cr 
& \le \abcst \max_{s_1',s_2'\in \Si'}
\Big\| \cD^{(1,0,0)}_{1,2}
\int {\sst d\eta_1\,d\eta_2}\,u({\sst(\eta_1,s'_1),(\eta_2,s'_2)})\
\hat\chi_{s_1}({\sst \eta_1,\xi_1})\,\hat\chi_{s_2}(\sst{\eta_2,\xi_2})
\,\Big\|_{1,\infty} \cr 
& \le \abcst\, \Big( \|\hat\chi_{s_2}\|_{1,\infty} 
  + \big\|x_0\, \sfrac{\partial\chi'_{s_2}}{\partial x_0}(x) \big\|_{L^1} \Big)
\max_{s_1',s_2'\in \Si'}
\Big\| \cD^{(1,0,0)}_{1,2}
\int {\sst d\eta_1}\,u({\sst(\eta_1,s'_1),(\,\cdot\,,s'_2)})\
\hat\chi_{s_1}({\sst \eta_1,\,\cdot\,})\,\Big\|_{1,\infty} \cr 
&\hskip4.3in
+\smsum_{\de \in \bbbn_0\times\bbbn_0^d\atop \de_0>r_0}\infty t^\de\cr
& \le \abcst\,\cb_{j-1} \Big( \|\hat\chi_{s_1}\|_{1,\infty} 
  + \big\|x_0\, \sfrac{\partial\chi'_{s_1}}{\partial x_0}(x) \big\|_{L^1} \Big)
\max_{s_1',s_2'\in \Si'}   \big\|
\cD^{(1,0,0)}_{1,2} u({\sst(\,\cdot\,,s'_1),(\,\cdot\,,s'_2)})\,\big\|_{1,\infty} 
\cr 
& \le \abcst\,\cb_{j-1} \V \cD^{(1,0,0)}_{1,2} u \V_{1,\Si'} 
\cr
}$$
\endproof

\corollary{\STM\corOSIntUp}{ There is a constant $\abcst$, independent of $M$,
such that the following holds: 
Let $\Si$ be a sectorization of scale $j\ge 2$ and $u\in \cF_0(2;\Si)$ be a
function that vanishes at $k_0=0$. Let $X\in\fN_{d+1}$. If
$$
\V  \cD^{(1,0,0)}_{1,2} u  \V_{1,\Si}\le \cb_{j-1} X+\sum_{\de_0=r_0}\infty\,t^\de
$$
then
$$
\V  u  \V_{1,\Si}\le \abcst\,\sfrac{M}{M^j}\cb_{j-1}X
$$
}
\prf By Proposition \propOSIntUp.i and (\eqnOSprodcontrbound)
$$
\v  u\,\v_{1,\Si}
\le \abcst\,\sfrac{1}{M^{j-1}} \cb_{j-1} \V  \cD^{(1,0,0)}_{1,2} u  \V_{1,\Si}
+\smsum_{\de_0\ne 0}\infty\,t^\de
\le \abcst\,\sfrac{1}{M^{j-1}} \cb_{j-1}X+\smsum_{\de_0\ne 0}\infty\,t^\de
$$
Also
$$\eqalign{
\V  u  \V_{1,\Si}
&\le t_0\sfrac{\partial\hfill}{\partial t_0}\V  u  \V_{1,\Si}
+\smsum_{\de_0=0}\infty\,t^\de\cr
&= t_0\V  \cD^{(1,0,0)}_{1,2} u  \V_{1,\Si}
+\smsum_{\de_0=0}\infty\,t^\de\cr
&\le t_0\cb_{j-1}X+\sum_{\de_0=r_0+1}\infty\,t^\de
+\smsum_{\de_0=0}\infty\,t^\de\cr
&\le \sfrac{1}{M^{j-1}}\cb_{j-1}X
+\smsum_{\de_0=0}\infty\,t^\de\cr
}$$
since $t_0\cb_{j-1}\le \sfrac{1}{M^{j-1}}\cb_{j-1}$. The Corollary now follows by
taking the  minimum of the two estimates on $\v  u \, \v_{1,\Si}$.
\endproof

\corollary{\STM\corOSresectorvanishkzero}{ Let $j>i\ge 2$ and $\Si$ and $\Si'$ 
be  sectorizations at scale $j$ and $i$, respectively. Let $u\in \cF_0(2;\Si')$
vanish at $k_0=0$.
Assume that $\v u\,\v_{1,\Si'}\le\la\cb_i$ for some $\la>0$. Then
$$
\v u_\Si\v_{1,\Si}\le\abcst\,\la\, M\sfrac{M^i}{M^j}\cb_{j-1}
$$
}
\prf 
By hypothesis
$$
\V  \cD^{(1,0,0)}_{1,2} u  \V_{1,\Si'} 
\le \abcst\,\la M^i \cb_i  +\smsum_{\de_0=r_0} \infty \,t^\de
$$
Therefore, by Proposition \propOSIntUp.i and (\eqnOSprodcontrbound)
$$
\v  u_\Si\,\v_{1,\Si}
\le \abcst\,\sfrac{1}{M^{j-1}} \cb_{j-1} \V  \cD^{(1,0,0)}_{1,2} u  \V_{1,\Si'}
+\smsum_{\de_0\ne 0}\infty\,t^\de
\le \abcst\,\la\,\sfrac{M^i}{M^{j-1}}\cb_{j-1}+\smsum_{\de_0\ne 0}\infty\,t^\de
$$
Also, by Proposition \propOSIntUp.ii
$$\eqalign{
\V  u_\Si  \V_{1,\Si}
&\le t_0\V  \cD^{(1,0,0)}_{1,2} u_\Si  \V_{1,\Si}
+\smsum_{\de_0=0}\infty\,t^\de\cr
&\le \abcst\, ( t_0 \,\cb_{j-1})\,\V\cD^{(1,0,0)}_{1,2}u\,\V_{1,\Si'} 
+\smsum_{\de_0=0}\infty\,t^\de\cr
&\le \abcst\, \big(\sfrac{1}{M^{j-1}}\cb_{j-1}\big)\,\la M^i \cb_i 
+\smsum_{\de_0=r_0+1} \infty \,t^\de+\smsum_{\de_0=0}\infty\,t^\de\cr 
&\le \abcst\,\la \sfrac{ M^i}{M^{j-1}}\cb_{j-1}
+\smsum_{\de_0=0}\infty\,t^\de\cr  
}$$
Again, the Corollary follows by
taking the  minimum of the two estimates on $\v u_\Si \,\v_{1,\Si}$.
\endproof

When we start the multi scale analysis in [FKTf2], the effective potential after
integrating out the first scales does not have a natural sectorized representative (see also Theorem \thmOSfirststep). Therefore we need analogs of Definition \defOSresector\ and Proposition \propOSresectorI\ that pass from unsectorized functions to sectorized functions (see also Example \exOSsectrepr).

\definition{\STM\defOScreateSectoriz}{ 
Let  $\Si$ be a sectorization of scale $j\ge 2$. For a function $f$ on $\cB^m\times \cB^n$ define the function $f_\Si$
on $\cB^m \times \big( \cB\times \Si \big)^n$ by
$$
f_\Si({\sst\eta_1,\cdots,\eta_m;\,(\xi_1,s_1),\cdots,(\xi_n,s_n) } )
= \int\smprod_{i=1}^n\, ({\sst d\xi'_i\ \hat\chi_{s_i}(\xi'_i,\xi_i)})
\ f({\sst\eta_1,\cdots,\eta_m;\,\xi'_1,\cdots,\xi'_n} )
$$
where $\chi_s$ is the partition of unity of Lemma \lemOSsectpartunit.
}

\proposition{\STM\propOScreateSectoriz}{  
Let  $\Si$ be a sectorization of scale $j\ge 0$ and $f\in \cF_m(n)$,
$f'\in \check\cF_m(n)$  
particle number conserving functions that are antisymmetric in their $\xi$--variables.
\Item i) 
If $m=0$ and $f$ is translation invariant, then for all $p<n$
$$
\V f_\Si\V_{p,\Si}\le \abcst^n\sfrac{1}{{\fl^{}}^{n-p-1}} \cb_{j-1}\,
\| f\|_{1,\infty}
$$
\Item ii) 
If $m\ne 0$
$$
\V f_\Si\V_{1,\Si}\le \big[\sfrac{\abcst}{\fl}\big]^n \,\cb_{j-1}\,
\| f\|_{1,\infty}
$$

\Item iii) 
If $0<m\le p\le m+n$
$$
\V f'_\Si\tV_{p,\Si}\le \big[\sfrac{\abcst}{\fl}\big]^{m+n-p} \,\cb_{j-1}\,
\| f\tnorm
$$
}

\noindent
The proof of part (i) of this Proposition is analogous to that of Proposition \propOSresectorI, and part (ii) was already proven in Example \exOScommsectnorms.
The proof of part (iii) is similar to that of part (ii).

\vfill\eject
%=====================================================================
%=========== SUMS of MOMENTA and EPS SEPAATED SETS ===================
%=====================================================================

\chap{Sums of Momenta and $\textfont1=\bigteni\epsilon$ -- separated Sets}
\PG\pgOSXX

In the next section we shall exploit conservation of momentum to prove
Proposition \propOSthreetoonenorm, relating the $1$-- and $3$--norms of a
four--legged kernel, and Proposition \propOSresectorI, concerning the behaviour of norms under
change of sectorization. Conservation of momentum is equivalent to translation
invariance in position space. 

The following Definition is motivated by Definition \defOSsymmetries.N 
of [FKTo2], of conservation of particle number, and 
Definition \defOSsectcheckcF.i, of the spaces $\check\cF_m(n;\Si)$.

\definition{\STM\defOSadmissablesectors}{
 A configuration 
$(\check\eta_1,\cdots,\check\eta_m;\,s_1,\cdots, s_n)\in\check\cB^m\times\Si^n$,
where $\Si$ is a sectorization of some scale $j$,
is consistent with conservation of momentum for the sequence
$(a_1,\cdots,a_n)$ of creation--annihilation indices if there are $k_1,\cdots,k_n
\in \bbbr \times
\bbbr^2$, with $k_i$ in the  extended sector $\tilde s_i$ for each $i=1,\cdots,n$,
such that
$$
\smsum_{i=1}^m (-1)^{b_i} p_i + \smsum_{i=1}^n (-1)^{a_i}k_i=0
$$
where $\check\eta_i=(p_i,\si_i,b_i)$.

\noindent
We say that the configuration $(\check\eta_1,\cdots,\check\eta_m;\,s_1,\cdots,
s_n)$ is consistent with conservation of momentum if it is consistent with
conservation of momentum for some sequence
$(a_1,\cdots,a_n)\in \{0,1\}^n$ of creation--annihilation indices such that
$$
\#\set{i}{a_i=0} +\#\set{\ell}{b_\ell=0} =\#\set{i}{a_i=1}+\#\set{\ell}{b_\ell=1}
$$
}

\remark{\STM\remOSadmissablesectors}{
Let $\Si$ be a sectorization of scale $j$.
\Item i) 
If $f \in \check\cF_m(n;\Si)$ preserves particle number then
$$
f({\sst \check\eta_1,\cdots,\check\eta_m;(\,\cdot\,,s_1),\cdots,(\,\cdot\,,s_n)
}) =0
$$
unless the configuration $(\check\eta_1,\cdots,\check\eta_m;\,s_1,\cdots,s_n)$ 
is consistent with conservation of momentum.
\Item ii)
If a configuration $(\check\eta_1,\cdots,\check\eta_m;\,s_1,\cdots,s_n)$ 
is consistent with conservation of momentum for the sequence $(a_1,\cdots,a_n)$ of
creation--annihilation indices then there are 
$\k_1,\cdots,\k_n \in \bbbr^2$ such that  
$$ 
\smsum_{i=1}^m (-1)^{b_i} \p_i + \smsum_{i=1}^n (-1)^{a_i}\k_i=\0
$$ 
and
$$
\pi_F((0,\k_i))\in s \qquad,\qquad |e(\k_i)| \le \sfrac{\sqrt{2}}{M^{j-1}}
$$
for $i=1,\cdots,n$. 
}

The comparison of the $1$-- and $3$-- norms of a four--legged kernel
(Proposition \propOSthreetoonenorm) uses an estimate on the maximal number of
triples $(s_2,s_3,s_4)$ of sectors that complete a given sector $s_1$ to a
quadruple $(s_1,s_2,s_3,s_4)$ that is consistent with conservation of momentum
(Proposition \:\propSecX). Similarly, the estimate on the behaviour of norms
under change of sectorization (Proposition \propOSresectorI) is based on estimates
of the number of $(2n)$--tuples  $(s_1,\cdots,s_{2n})$ of sectors that
are consistent with conservation of momentum and such that each $s_i$ intersects
a given bigger sector from another sectorization (Proposition \:\proSecXI).

We reduce these counting problems to problems of estimating volumes of sets
in momentum space that are characterized by the geometric constraints 
that the sectors are required to satisfy. To pass from volume estimates
to sector counting we use the concept of $\ep$--separated sets (see also
[G, p.22]).

\titlec{ $\epsilon$ -- separated sets}\PG\pgOSXXa

Let $M$ be a Riemannian manifold of dimension $n$. For any two points $x,y\in M$
we denote by $d(x,y)$ the distance between $x$ and $y$ in $M$. For $x\in M$ and $r>0$ let$$
B_r(x) = \set{y\in M}{d(x,y)<r}
$$
be the open ball of radius $r$ around $x$. We set, for $\ep>0$,
$$
V_{M,\ep} = \inf_{x\in M,\,0<r\le\ep} \sfrac{1}{r^n} \,{\rm vol}\,B_{r/2}(x) 
$$
where {\rm vol} denotes the volume with respect to the Riemannian metric on $M$. For a subset $A\subset M$ and $\de >0$ we call
$$
A_\de = \set{x\in M}{ \inf_{y\in A} d(x,y) \le \de}
$$
the (closed) $\de$--neighbourhood of $A$.
If $X$ is a tangent vector to $M$ at the point $x$ we denote by $\|X\|$ the length of $X$ with respect to the Riemannian metric on $M$. 

If $f$ is a differentiable map from $M$ to another Riemannian manifold $N$ we
denote by $Df(x)$ the derivative of $f$ at the point $x\in M$.
The point $x$ is said to be a critical point of $f$ if $Df(x)$ has rank strictly
less than the dimension of $N$.

\definition{\STM\defSecI}{ Let $\epsilon>0$. A subset $\,\Gamma\,$ of $M$ is called 
$\epsilon$--{\it separated} if for any two different 
$\,\gamma,\ \gamma' \in \Gamma\,$
$$
 d(\gamma, \gamma') \ \ge \ \epsilon
$$
}

\example{}{
Let $\Sigma$ be a sectorization of length $\frak l$ and
let $\Ga$ be the set of centers of the intervals $s\cap F,\ s\in\Si$.
If  $\epsilon < 7\,\fl/8 $, then $\Ga$ is an $\epsilon$--separated 
subset of the Fermi curve $F$ and more generally $\Ga^n$ 
is an $\epsilon$--separated subset of $F^n$.
}

We wish to count, for example, for a given sector $s_4$,
 the number of triples of sectors $(s_1,s_2,s_3)$ such that there exist
$\k_i\in s_i$ obeying $\k_1+\k_2-\k_3-\k_4=0$. If $(s_1,s_2,s_3)$ are such
sectors, then the map
$$\eqalign{
f:F\times F\times F&\longrightarrow \bbbr^2\cr
(\k_1,\k_2,\k_3)&\longmapsto \k_1+\k_2-\k_3
}$$
maps $F^3\cap (s_1\times s_2\times s_3)$ to a neighbourhood of $s_4$.
We start with an abstract lemma counting the number of points of an 
$\ep$--separated set $\Ga$ in the preimage of a specified set $A$ under 
a specified map $f$.

\lemma{\STM\lemSecIII }{
Let $M$ be a Riemannian manifold of dimension $n$ and $f:M\mapsto \bbbr^d$ a differentiable map. For $x\in M$ denote by $Df(x)$ the derivative of $f$ at the point $x$. Let ${\vec n}_1,\cdots,{\vec n}_d\,$ be an orthonormal basis of $\bbbr^d$.
Set, for $\,i=1,\cdots, d\,$
$$
C_i\ = \ \sup_{x\in M} \sup \set{\big| {\vec n}_i \cdot Df(x)v \big|}
{v\ {\rm is\ a\ unit\ tangent\ vector\ to\ } M {\rm \ at\ }x}
$$
Furthermore, for any subset $A$ of $\bbbr^d$ and any $\epsilon>0$ set
$$
A'(\epsilon)\ =\ \set{y\in \bbbr^d}
{\exists (t_1,\cdots,t_d)\in(-\epsilon,\epsilon)^d \ {\rm such\ that\ }
y+\smsum_{i=1}^d t_i C_i {\vec n}_i \in A}
$$
Then for all $A\subset \bbbr^d$, $\ep_0>0$, $0<\ep<\ep_0$ and all  $\epsilon$--separated subsets $\Gamma$ of $M$
$$
\#\big( f^{-1}(A) \cap \Gamma \big)\ \le\ \sfrac{1}{\epsilon^n\,V_{M,\ep_0}}\,
 {\rm vol} \Big( f^{-1}\big(A'(\epsilon) \big)\Big)
$$
}
\prf
If  $\gamma \in f^{-1}(A) \cap \Gamma$ and $x\in M$ with $d(x,\gamma) < \epsilon/2$, then, by the assumption on the derivative of $f$ for $\,i=1,\cdots, d\,$
$$
\big| {\vec n}_i \cdot \big( f(x) -f(\gamma) \big) \big| < C_i \epsilon
$$
so that $\,f(x)\in A'(\epsilon)\,$.
Obviously the sets 
$\,B_{\epsilon/2}(\gamma),\ \gamma \in  f^{-1}(A) \cap \Gamma\,$ 
are pairwise disjoint.
Consequently, by the definition of $V_{M,\ep_0}$
$$\eqalign{
V_{M,\ep_0}\, \epsilon^n \, \#\big( f^{-1}(A) \cap \Gamma \big)\
&\le \smsum_{\gamma \in  f^{-1}(A) \cap \Gamma}\, 
{\rm vol}\, \big( B_{\epsilon/2}(\gamma)\big) \cr
&= {\rm vol}\, \big( \bigcup_{\gamma \in  f^{-1}(A) \cap \Gamma}\, 
 B_{\epsilon/2}(\gamma) \big) \cr
&\le  {\rm vol} \Big( f^{-1}\big(A'(\epsilon)\big) \Big)
}$$
\endproof

\vskip 4cm

\titlec{ Sums of Momenta}\PG\pgOSXXb

For the proofs of Propositions \:\propSecX\ and \:\proSecXI\ we shall apply 
the discussion of the previous
subsection with $\Ga$ being the set of centers of sectors of a given sectorization.
The proofs of these Propositions then lead to the problem of estimating the
number of points $(\k_1,\cdots,\k_n,\k_{n+1},\cdots,\k_{2n-1})\in\Ga^{2n-1}$
such that $\k_1+\cdots +\k_n - \k_{n+1}-\cdots-\k_{2n-1}$ is close to
$\q$. Thus we are lead to studying the maps
$$\eqalign{
F^{2n-1} &\longrightarrow \bbbr^2\cr
(\k_1,\cdots,\k_n,\k_{n+1},\cdots,\k_{2n-1}) &\longmapsto 
\k_1+\cdots +\k_n - \k_{n+1}-\cdots-\k_{2n-1}\cr
}$$
and the intersection of preimages of sets in $\bbbr^2$ with $\Ga^{2n-1}$.
Outside a neighbourhood of the set of critical points of this map this can usually
be done using Lemma \lemSecIII.
The critical points of the map are exactly those points 
$(\k_1,\cdots,\k_{2n-1})\in F^{2n-1}$ for which the tangent lines of $F$ at 
$\k_1,\cdots,\k_{2n-1}$ are all parallel. This is the case if and only all the points
$\k_2,\cdots,\k_{2n-1}$  coincide either with $\k_1$ or its antipode $a(\k_1)$.

For $\k\in F$ and $0<\Lambda \le \fl$, we call 
$$
 s_{\Lambda,\fl}(\k) = \set{\q \in \bbbr^2}
{ |e(\q)| \le \Lambda, \ d_F(\q',\k)\le \fl/2 }
$$
the two dimensional sector of length $\l$ and width $\Lambda$ around $\k$. 
Here $\q\mapsto \q'$ is the projection to the Fermi curve introduced in
 \S\CHintroIII\ and used in Definition \defOSsectors.i of [FKTo3] and $d_F$ is
the intrinsic metric on $F$.

Near critical points of the map discussed above we shall use

\proposition{\STM\propSecVI}{ Let $\,\k,\,\k_1,\cdots,\k_{2n-3}\in F\,$ and
$\omega>0$ be such that for $i=1,\cdots,2n-3\ $ one has  $\,\|\k_i-\k\| <\omega\,$ or $\,\|\k_i-a(\k)\|<\omega\,$. Let 
$\,\epsilon_1,\cdots \epsilon_{2n-3} \in \{\pm1\}\,$ and set
$$
\q = \ \epsilon_1 \k_1+ \cdots + \epsilon_{2n-3} \k_{2n-3}
$$
Furthermore, let $\,0<\Lambda\le \fl\le \omega$ and let ${\vec{\rm n}}$ resp. 
${\vec{\rm t}}$ be unit normal resp. tangent vector of $F$ at $\k$.
Then
$$
\set{\epsilon_1 x_1+ \cdots + \epsilon_{2n-3} x_{2n-3}}
{x_i \in s_{\Lambda,\fl}(\k_i)}
$$
is contained in the rectangle
$$
\set{\q+t_1 {\vec{\rm n}} + t_2 {\vec{\rm t}}}
{ |t_1| \le n \,\abcst\, (\Lambda + \fl \omega),\ |t_2| \le 4 n \fl }
$$
The constant $\abcst$ depends only on the geometry of $F$.
}
\prf Without loss of generality we may assume that $\,{\vec{\rm n}} = (0,1)\,$
and $\,{\vec{\rm t}} = (1,0)\,$. The angle between $F$ and the $k_1$ direction
at a point $\q\in F\,$ is bounded by 
$\,\abcst\,\|\q-\k\|\,$ and
$\,\abcst\,\|\q-a(\k)\|\,$. Therefore 
$s_{\Lambda,\fl}(\k_i)$ is contained in a rectangle that is centered at $\k_i$ and
has two edges parallel to the $k_1$ axis of length $2\fl$ and  two edges  parallel
to the $k_2$ axis of length 
$\,\abcst\, \big(\Lambda + \fl\min\big\{ \|\k_i-\k\|,\|\k_i-a(\k)\|\big\}\big) 
\le \,\abcst\,  (\Lambda + \fl\, \omega)$. 
The claim now follows. 
\endproof

\proposition{\STM\propOSSecV}{ Let $\,\k,\,\k_1,\cdots,\k_{2n-1}\in F\,$,  
$I\subset \{1,\cdots,n\},\ J\subset \{n+1,\cdots,2n-1\}\,$ and $\omega$ a
positive real number smaller than the diameter of $F$ such that   
$\,\|\k_i-\k\| <\omega\,$  for $ i\in I\cup J$ and $\,\|\k_j-a(\k)\|<\omega\,$ 
for $ j \notin I\cup J $.
Furthermore, let $\p\in\bbbr^2$ and $\,0<\Lambda\le \fl\le \omega$. 
Assume that there are points
$x_i \in s_{\Lambda,\fl}(\k_i),\ i=1,\cdots,2n-1\,$  such that
$$
\|x_1+\cdots+x_n \,-\, x_{n+1}-\cdots -x_{2n-1}-\p\|\le 2\fl
$$
Then
\Item{i)}
$$
\|\p-(|I|-|J|)\k + (|I|-|J|-1)a(\k)\| \le \abcst\, n\,\omega
$$
\Item{ii)}
If $\p\in F$ then 
$$
\|\p-\k \|  \le \abcst\, n\,\omega \qquad {\rm or}\qquad 
\|\p-a(\k) \|  \le \abcst\, n\,\omega
$$
The constants $\abcst$ depend only on the geometry of $F$.
}

\prf 
By Proposition \propSecVI,
$$\eqalign{
&\big\|x_1+\cdots+x_n \,-\, x_{n+1}-\cdots -x_{2n-1}
-\big(\k_1+\cdots +\k_n - \k_{n+1}-\cdots-\k_{2n-1}\big)\big\|\cr
&\hskip2.5in\le (n+1)\abcst\,(\La+\fl\om)+4(n+1)\fl
\le\abcst\,n\om
}$$
Since
$$
\big\|\k_1+\cdots +\k_n - \k_{n+1}-\cdots-\k_{2n-1}
-|I|\k-(n-|I|)a(\k)+|J|\k+(n-1-|J|)a(\k)\big\|
\le\abcst\,n\om
$$
part (i) follows. To prove part (ii) assume
that $\p \in F$. Set $r=|I|, s=|J|$. By possibly interchanging
$\k$ with its antipode, we may assume that $r\ge s$.
If $r=s$ or $r=s+1$ part (ii) it follows directly from part (i) that 
$\|\p-\k \|  \le \abcst\, n\,\omega$. So we assume that $r-s\ge 2$.

Let $x_i \in s_{\Lambda,\fl}(\k_i),\ i=1,\cdots,2n-1\,$ such that
$$
y=x_1+\cdots+x_n \,-\, x_{n+1}-\cdots -x_{2n-1}
$$
has distance at most $2\fl$ from $\p$.
Let $ \vec n$ be the outward pointing unit normal vector of $F$ at $\k$. 
Then $\,(\k-a(\k))\cdot \vec n \ge \abcst_1\,$ and
$$
|(x_i-\k)\cdot \vec n| \le \|x_i-\k_i\|+\|\k_i-\k\|
\le\abcst\,(\La+\fl)+\om
\le \abcst\, \omega
$$
for $i\in I\cup J$. Similarly for $j\notin I\cup J$
$$
|(x_j-a(\k))\cdot \vec n| \le \abcst\, \omega
$$
Consequently
$$
\big| ( x_1+\cdots+x_n \,-\, x_{n+1}-\cdots -x_{2n-1} -\k) \cdot \vec n - 
(r-s-1)\,(\k-a(\k))\cdot \vec n \big| 
\le (2n-1)\, \abcst\, \omega 
$$
and therefore
$$ ( y -\k) \cdot \vec n 
 \ge A \big( (r-s-1) -  \abcst\, n \omega \big)
$$
with strictly positive constants $A,\abcst$. 

The tangent line to $F$ at $\k$ is a ``supporting hyperplane" for the convex hull of $F$. Therefore
$$
F\cap \set{x\in \bbbr^2}{(x-\k)\cdot \vec n >0} =\emptyset
$$
So
$$\eqalign{
0 \ge (\p-\k)\cdot \vec n 
&\  = (\p-y) \cdot \vec n + (y-\k) \cdot \vec n \cr
&\ \ge (\p-y) \cdot \vec n +  A \big( (r-s-1) - n\, \abcst\,\omega
\big)\cr }$$
As $\,|(\p-y)\cdot \vec n| \le 2\fl\,$ and hence $(\p-y)\cdot\vec n\ge-2\fl\ge-2\om\ge
-n\,\abcst\,\om$,
this shows that 
$$
n\ge \abcst' \sfrac{r-s-1}{\omega} \ge \abcst' \sfrac{1}{\omega}
$$
since $r-s\ge 2$. Thus $n\,\omega$ is larger than some strictly positive
constant  and the estimate of part (ii) holds.
\endproof

\vskip 1cm

\titlec{ Pairs of Momenta}\PG\pgOSXXc

\lemma{\STM\lemSecVII}{
Let $\ep_1,\ \ep_2\in\{\pm 1\}$.
There is a subset $X$ of $F\times F$ and a constant $C$ such that
\Item{i)} For every $\p\in\bbbr^2$
$$
\#\set{(\k_1,\k_2)\in X}{\ep_1\k_1+\ep_2\k_2=\p}\le C
$$
\Item{ii)} $(F\times F)\setminus X$ has measure zero.
}
\prf We may assume that $\ep_1= +1$. If $\ep_2=-1$, then we claim that,
for every $\p\ne 0$
$$
\#\set{(\k_1,\k_2)\in F\times F}{\k_1-\k_2=\p}\le 2
\EQN\eqnSecII$$
In this case, the Lemma with $C=2$ and $X=\set{(\k_1,\k_2)}{\k_1,\k_2\in
F,\ \k_1\ne\k_2}$ follows directly from (\eqnSecII).
To prove (\eqnSecII), choose a nonzero vector $\vec v$ perpendicular to $\p$. Assume 
that there are distinct pairs $(\k_1,\k_2),\ (\k_3,\k_4), (\k_5,\k_6)$
such that $\k_1-\k_2=\k_3-\k_4=\k_5-\k_6=\p$. Without loss of generality
we assume that $\k_1\cdot\vec v<\k_3\cdot\vec v<\k_5\cdot\vec v$. By
convexity, the parallelogram, P, with vertices $\k_1,\k_2,\k_5,\k_6$ 
is contained in the convex hull of $F$.
\vadjust{\null\hfill \figplace{pgram}{0 in}{0 in} \hfill\null}
The segment joining $\k_3$ and $\k_4$ must cross this parallelogram. Therefore
$\k_3$ and $\k_4$ lie on the edges of $P$. This contradicts the strict
convexity of $F$. 

Formula (\eqnSecII) may be phrased in more geometrical terms as follows. Let $s$
be a secant of $F$ (that is, a straight line segment joining two different
points $\k_1,\k_2\in F$). Then there is at most one other secant $s'$ for
$F$ that is parallel to $s$ and has the same length. In the case that there
is no second such secant, we set $s'=s$. Clearly there are $r,R>0$ such that
for all  $\k\in F$
\item{i)} $B_r(\k)\cap B_R(a(\k))=\emptyset$
\item{ii)} For any secant $s\subset B_r(\k)$ one has $s'\subset B_R(a(\k))$

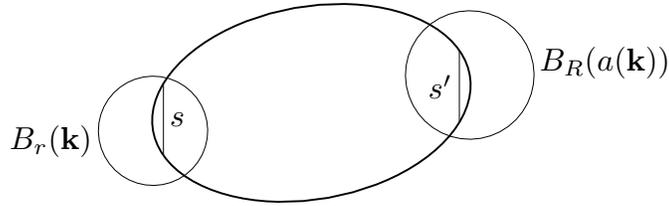
\centerline{\figplace{secants}{0 in}{0 in}}

Now consider $\ep_2=+1$. If $F$ is invariant
under inversion in some point $\p_0\in\bbbr^2$, then, for $\p\ne2\p_0$,
$$
\#\set{(\k_1,\k_2)\in F\times F}{\k_1+\k_2=\p}
=\#\set{(\k_1,\k'_2)\in F\times F}{\k_1+(2\p_0-\k'_2)=\p}\le 2
$$
by the case $\ep_2=-1$.

Now we discuss the case that $F$ is strongly asymmetric in the sense of
Definition \defModI. Since $F\times F$ is compact, it suffices to show that
for each point $(\k_1,\k_2)\in F\times F$ there exists a neighbourhood
$U$ of $(\k_1,\k_2)$ in $F\times F$, a subset $U'$ of $U$ whose complement
$U\setminus U'$ has measure zero and a number $m$ such that, for all $\p\in\bbbr^2$,
$$
\#\set{(\q_1,\q_2)\in U'}{\q_1+\q_2=\p}\le m
$$
If $\k_2\ne\k_1,a(\k_1)$, then the map $(\q_1,\q_2)\mapsto\q_1+\q_2$ has
rank 2 at $(\k_1,\k_2)$. By the inverse function Theorem, there is a 
neighbourhood $U$ of $(\k_1,\k_2)$ such that for all $\p\in\bbbr^2$ 
$\#\set{(\q_1,\q_2)\in U}{\q_1+\q_2=\p}\le 1$. 

Next assume that $\k_1=\k_2=\k$. Let 
$U_r=\set{(\q_1,\q_2)\in F^2}{\|\q_1-\k\|,\|\q_2-\k\|<r}$,
where $r$ is defined in the discussion of secants above. We claim that
for $(\q_1,\q_2), (\q'_1,\q'_2)\in U_r$
$$
\q_1+\q_2=\q'_1+\q'_2\quad\iff\quad (\q_1,\q_2)=(\q'_1,\q'_2)\hbox{ or }
(\q_1,\q_2)=(\q'_2,\q'_1)
$$
Assume that $\q_1+\q_2=\q'_1+\q'_2$ and 
$(\q_1,\q_2)\ne (\q'_1,\q'_2),(\q'_2,\q'_1)$. Then 
$\q_1-\q'_1=\q'_2-\q_2\ne 0$, so that the sector $s$ of $F$ joining $\q_1$
to $\q'_1$ is parallel to and of the same length as, but disjoint
from  the sector $\tilde s$ joining $\q'_2$ to $\q_2$. Therefore, $\tilde
s=s'$ where, as above, $s'$ is the unique second secant parallel to and 
of the same length as $s$. But this is impossible, as $\tilde s\subset
B_r(\k)$, $s'\subset B_R(a(\k))$ and $B_r(\k)\cap B_R(a(\k))=\emptyset$.

Finally, assume that $\k_1=\k$ and $\k_2=a(\k)$ for some $\k\in F$. We
may assume, without loss of generality, that the oriented unit tangent
vector to $F$ at $\k$ is $(1,0)$  and that the inward pointing unit normal
vector to $F$ at $\k$ is $(0,1)$. Then in the notation of Definition \defModI,
$$
t\longmapsto \k+\big(t,\varphi_\k(t)\big)
$$ 
is a parametrization of $F$ near $\k$ and
$$
t\longmapsto a(\k)-\big(t,\varphi_{a(\k)}(t)\big)
$$ 
is a parametrization of $F$ near $a(\k)$. Then
$$
(t_1,t_2)\longmapsto\big(\k+\big(t_1,\varphi_\k(t_1)\big),
a(\k)-\big(t_2,\varphi_{a(\k)}(t_2)\big)\big)
$$
is a parametrization of $F\times F$ near $(\k_1,\k_2)$. With respect to these
coordinates, the map $(\q_1,\q_2)\longmapsto \q_1+\q_2-\big(\k+a(\k)\big)$
from $F\times F$ to $\bbbr^2$ is
$$
\tilde f:(t_1,t_2)\longmapsto\big(t_1-t_2,
\varphi_{\k}(t_1)-\varphi_{a(\k)}(t_2)\big)
$$
Since $F$ is strongly asymmetric, there is $2\le n\le n_0$ such that 
$$
\varphi^{(n)}_{\k}(0)\ne\varphi^{(n)}_{a(\k)}(0)
$$
We show that, if $\ep$ is small enough, then, for any $\p=(p_1,p_2)\in\bbbr^2$, 
the equation
$$
\tilde f(t_1,t_2)=\p
\EQN\eqnSecIII$$
has at most $n$ solutions in $(-\ep,\ep)^2$. Fix $\p$ and
set $g(t)=\varphi_{\k}(p_1+t)-\varphi_{a(\k)}(t)-p_2$. Then $(t_1,t_2)$
is a solution of (\eqnSecIII) if and only if $(t_1,t_2)=(p_1+t,t)$
with $t$ a zero of $g(t)$. Hence it suffices to prove that $g(t)$ has at most
$n$ zeros. But, since 
$\varphi^{(n)}_{\k}(0)-\varphi^{(n)}_{a(\k)}(0)\ne 0$,
the $n^{\rm th}$ derivative 
$g^{(n)}(t)=\varphi^{(n)}_{\k}(p_1+t)-\varphi^{(n)}_{a(\k)}(t)$
never vanishes for $|p_1+t|,|t|<\ep$, for $\ep$ sufficiently small.
Consequently $g$ can have at most $n$ zeros on this set.
\endproof

\lemma{\STM\lemSecVIII}{ 
Again, assume either that $F$ is  symmetric about a point or that $F$ is 
strongly asymmetric.
Let $\p\in F$,  $0<\omega_1< \half\omega_2 $ and set 
$$\eqalign{
M=\{ (\k_1,\k_2)\in F\times F \,\big|\,
&\min [d(\k_1,\k_2),d(a(\k_1),\k_2)] \ge \omega_1 \cr
&{\rm and \ } \min [d(\k_i,\p),d(a(\k_i),\p)] \le \omega_2 \ {\rm for\ } i=1,2\} \cr
}$$
Let $\,\epsilon_1,\epsilon_2 \in \{\pm1\}\,$ and let $f$ be the map from 
$F\times F$ to $\bbbr^2$ given by 
$$
f(\k_1,\k_2) = \epsilon_1 \k_1 + \epsilon_2 \k_2
$$   
There are constants that depend only on the geometry of $F$ such that
\Item{ i)} for all measurable subsets $A$ of $\bbbr^2$
$$
{\rm vol\,}\big( f^{-1}(A)\cap M \big) \
\le \sfrac{\abcst}{ \omega_1}\,{\rm vol\,}(A)
$$
\Item{ ii)} 
$
V_{M,\om_1} \ge \abcst
$
\Item{ iii)} Let $\vec n$ a unit normal vector to $F$ at $\p$.
Then for all $(\k_1,\k_2) \in M$
$$
\sup_{\vec v\in T_{(\k_1,\k_2)}M\atop \|\vec v\|\le 1}\big| 
{\vec n} \cdot Df(\k_1,\k_2)\vec v \big| \le\abcst\,\omega_2
$$
\Item{ iv)} Let $ 0<\epsilon <\omega_1/4$ and let $\Gamma$ be an $\epsilon$--separated subset of $F$. Furthermore let $R$ be a rectangle in $\bbbr^2$ having one pair of sides parallel to $\vec n$ of length $A>0$
and  a second pair of sides perpendicular to $\vec n$ of length 
$B>0$. Then
$$
\# f^{-1}(R) \cap M \cap \Gamma^2
\le \sfrac{\abcst}{\omega_1 \epsilon^2}(A+\epsilon \omega_2)(B+\epsilon)
$$
}

\prf 
\Item{ i)} For $\k_1,\k_2\in F$ let $\theta(\k_1,\k_2)$ be the angle between the normal vectors to $F$ at $\k_1$ and at $\k_2$. Then the Jacobian determinant of $f$ at $(\k_1,\k_2)$ is
$$
| \sin \theta(\k_1,\k_2) | \ge 
\abcst\, \min \big( d(\k_1,\k_2), d(\k_1,a(\k_2))\big)
\ge \abcst\, \omega_1
$$
The claim follows from the rule for the change of variables in integrals 
and Lemma \lemSecVII.
\Item{ ii)} is trivial.
\Item{ iii)} For $\q\in F$, let $\vartheta(\q)$ be the angle between $\vec n$ and the normal vector to $F$ at $\q$. Then
$$\eqalign{
&\sup_{\vec v\in T_{(\k_1,\k_2)}M\atop \|\vec v\|\le 1}\big| 
{\vec n} \cdot Df(\k_1,\k_2)\vec v \big| 
\le 2\max\big( |\sin \vartheta(\k_1)| , |\sin \vartheta(\k_2)|\big)  \cr
&\hskip1in \le \abcst \,\max \big( \min\big( \| \p-\k_1 \| , \| \p-a(\k_1) \|\big) ,
\min\big( \| \p-\k_2 \| ,\| \p-a(\k_2) \|\big)\big)  \cr
&\hskip1in \le \abcst\,\omega_2 \cr
}$$ 
\Item{ iv)}
Let $\vec t$ be the tangent vector to $F$ at $\p$. Obviously
$$
\sup_{\vec v\in T_{(\k_1,\k_2)}M\atop \|\vec v\|\le 1}\big| 
{\vec t} \cdot Df(\k_1,\k_2)\vec v \big| \le 2
$$
for all $(\k_1,\k_2)\in M$. 
So by parts ii and iii of this Lemma and Lemma \lemSecIII
$$
\# f^{-1}(R) \cap M \cap \Gamma^2  
\le \sfrac{\abcst}{\epsilon^2} {\rm vol}\big(  f^{-1}(R') \big)
$$
where $R'$ is a rectangle of side lengths $A+\abcst\,\epsilon \omega_2$ and 
$B+4\epsilon$. By part i,
$$
{\rm vol}\big(  f^{-1}(R') \big) \le 
\sfrac{1}{\omega_1}(A+\abcst\,\epsilon \omega_2)(B+4\epsilon)
$$
\endproof

\titlec{ Sectors that are Compatible with Conservation of Momentum}\PG\pgOSXXd

Let $0\le \Lambda \le \fl, \ \Lambda \ge \fl^2$, let $\p\in F$, $\q\in\bbbr^2$
and let $\Gamma$ be a discrete subset of $F$. We define for $n\ge 2$ 
$$\eqalign{
Mom_{2n-1}(\Gamma,\p) &=\{(\k_1,\cdots,\k_{2n-1}) \in \Gamma^{2n-1}\,\big|\,
\exists  \, x_i \in s_{\Lambda,\fl}(\k_i),\ i=1,\cdots,2n-1\cr
&{\rm such\ that\ }\  
x_1+\cdots+x_n \,-\, x_{n+1}-\cdots -x_{2n-1} \in s_{\Lambda,\fl}(\p) \}\cr
Mom^\sim_{2n-1}(\Gamma,\q) &=\{(\k_1,\cdots,\k_{2n-1}) \in \Gamma^{2n-1}\,\big|\,
\exists  \, x_i \in s_{\Lambda,\fl}(\k_i),\ i=1,\cdots,2n-1\cr
&{\rm such\ that\ }\  
x_1+\cdots+x_n \,-\, x_{n+1}-\cdots -x_{2n-1} =\q \}
}$$

\lemma{\STM\lemSecIX}{ 
Let $\p\in F$, $\q\in\bbbr^2$ and $\Gamma$ an $\fl$--separated subset of
$F$. 
\Item {i)} 
If $\omega \ge \Lambda/\fl$ then
$$\eqalign{
\# \big\{ (\k_1,\k_2,\k_3) \in Mom_3(\Gamma,\p) \ \big|\ 
\omega \le \max_{1\le \mu\ne\nu\le 3} 
\min [d(\k_\mu,\k_\nu),d(a(\k_\mu),\k_\nu) ] 
\le 2\omega \big\}                       
&\le \abcst\, \sfrac{\omega}{\fl} \cr
\# \big\{ (\k_1,\k_2,\k_3) \in Mom^\sim_3(\Gamma,\q) \ \big|\ 
\omega \le \max_{1\le \mu\ne\nu\le 3} 
\min [d(\k_\mu,\k_\nu),d(a(\k_\mu),\k_\nu) ] 
\le 2\omega \big\}                       
&\le \abcst\, \sfrac{\omega}{\fl} \cr
}$$ 
\Item {ii)} 
If $4\fl \le \omega \le \Lambda/\fl$ then
$$\eqalign{
\# \big\{ (\k_1,\k_2,\k_3) \in Mom_3(\Gamma,\p) \ \big|\ 
\omega \le \max_{1\le \mu\ne\nu\le 3} 
\min [d(\k_\mu,\k_\nu),d(a(\k_\mu),\k_\nu) ] 
&\le 2\omega \big\}                       
\le \abcst\, \sfrac{\Lambda}{\fl^2} \cr
\# \big\{ (\k_1,\k_2,\k_3) \in Mom^\sim_3(\Gamma,\q) \ \big|\ 
\omega \le \max_{1\le \mu\ne\nu\le 3} 
\min [d(\k_\mu,\k_\nu),d(a(\k_\mu),\k_\nu) ] 
&\le 2\omega \big\}                       
\le \abcst\, \sfrac{\Lambda}{\fl^2} \cr
}$$ 
The constants $\abcst$ above depend only on the geometry of $F$.
}

\prf
If $(\k_1,\k_2,\k_3) \in Mom_3(\Gamma,\p) $ and 
$\max_{1\le \mu\ne\nu\le 3} 
\min [d(\k_\mu,\k_\nu),d(a(\k_\mu),\k_\nu) ] 
\le 2\omega$, then by Proposition \propOSSecV.ii, with $n=2$, $\om$ replaced by $2\om$ and $\k=\k_i$ or $a(\k_i)$, $i=1,2,3$
$$
 \min [d(\k_i,\p),d(a(\k_i),\p) ]\le 4\,\abcst_0 \,\omega
\ \ {\rm for\ } 1\le i\le 3
\EQN\OSthreemomI
$$
where $\abcst_0$ is the constant of Proposition \propOSSecV. Therefore we set for $1\le \mu\ne\nu\le 3$ 
$$\eqalign{
{\cal S}_{\mu,\nu} &= \big\{ (\k_1,\k_2,\k_3) \in Mom_3(\Gamma,\p) \ \big|\ 
 \omega \le \min [d(\k_\mu,\k_\nu),d(a(\k_\mu),\k_\nu) ] \le 2\omega \cr 
&\hskip1.7in\max_{1\le \al\ne\be\le 3} 
       \min [d(\k_\al,\k_\be),d(a(\k_\al),\k_\be) ]\le 2\omega \cr 
&\hskip1.2in{\rm and}\  \min [d(\k_i,\p),d(a(\k_i),\p) ]\le 4\abcst_0 \,\omega
\ \ {\rm for\ } 1\le i\le 3 \big\} \cr
}$$
The discussion above shows that
$$\big\{ (\k_1,\k_2,\k_3) \in Mom_3(\Gamma,\p) \ \big|\ 
\omega \le \max_{1\le \mu\ne\nu\le 3} 
\min [d(\k_\mu,\k_\nu),d(a(\k_\mu),\k_\nu) ] 
\le 2\omega \big\}
\subset \hskip -7pt \bigcup_{1\le \mu\ne\nu\le 3} {\cal S}_{\mu,\nu}
\EQN\OSthreemomII
$$

Similarly, if $(\k_1,\k_2,\k_3) \in Mom^\sim_3(\Gamma,\q) $ and 
$\min [d(\k_\mu,\k_\nu),d(a(\k_\mu),\k_\nu) ] \le 2\omega$ 
for all $1\le \mu\ne\nu\le 3$, then by Proposition \propOSSecV.i,
we have for $i=1,2,3$
$$\meqalign{
 d(\k_i,\p)&\le 4\,\abcst_0 \,\omega && \quad&{\rm or}  && 
 d(a(\k_i),\p)&\le 4\,\abcst_0\,\omega && \quad&{\rm or} \cr 
 d(2\k_i-a(\k_i),\p)&\le 4\,\abcst_0 \,\omega && &{\rm or}  && \quad
 d(2a(\k_i)-\k_i,\p)&\le 4\,\abcst_0\,\omega\cr 
}\EQN\OSthreemomIII $$
Setting for $1\le \mu\ne\nu\le 3$ 
$$\eqalign{
{\cal S}^\sim_{\mu,\nu} = \big\{ (\k_1,\k_2,\k_3) \in Mom^\sim_3(\Gamma,\q) \ 
\big|\ 
\hskip.25in\omega \le &\min [d(\k_\mu,\k_\nu),d(a(\k_\mu),\k_\nu) ]\le 2\omega \cr 
\max_{1\le \al\ne\be\le 3} 
       &\min [d(\k_\al,\k_\be),d(a(\k_\al),\k_\be) ]\le 2\omega \cr 
&\hskip1.5in{\rm and\ (\OSthreemomIII)\ holds} \big\} \cr
}$$
we get
$$\big\{ (\k_1,\k_2,\k_3) \in Mom^\sim_3(\Gamma,\q) \ \big|\ 
\omega \le \max_{1\le \mu\ne\nu\le 3} 
\min [d(\k_\mu,\k_\nu),d(a(\k_\mu),\k_\nu) ] 
\le 2\omega \big\}
\subset \hskip -7pt \bigcup_{1\le \mu\ne\nu\le 3} {\cal S}^\sim_{\mu,\nu}
\EQN\OSthreemomIV
$$

We show that for $1\le \mu\ne\nu\le 3$ one has
 $\,\# {\cal S}_{\mu,\nu},\,\# {\cal S}^\sim_{\mu,\nu}  \le
\abcst\,\sfrac{\omega}{\fl}\,$  in case (i), and that
 $\,\# {\cal S}_{\mu,\nu},\,\# {\cal S}^\sim_{\mu,\nu}   \le
\abcst\,\sfrac{\Lambda}{\fl^2}\,$ in case (ii). We only discuss the case
$\mu=1,\nu=2$, the other cases are similar. 

Set ${\cal S}={\cal S}_{1,2}$ or ${\cal S}={\cal S}^\sim_{1,2}$. By construction,
if $(\k_1,\k_2,\k_3)\in {\cal S}$, 
$$
\min[\|\k_1-\k_3\|,\|\k_1-a(\k_3)\|],\ \min[\|\k_2-\k_3\|,\|\k_2-a(\k_3)\|] 
\le \abcst\, \omega
\EQN\eqnOSSecsumI$$
and  (\OSthreemomI) respectively  (\OSthreemomIII) hold for $i=3$.
Since the maps $\k\mapsto \k$, $ \k\mapsto a(\k)$, $\k\mapsto 2\k-a(\k)$ and 
$\k\mapsto 2a(\k) -\k$ are embeddings of $F$, 
there are at most $\,\abcst\, \sfrac{\omega}{\fl}$ choices of $\k_3 \in \Gamma$
for which (\OSthreemomI) or (\OSthreemomIII) are satisfied. Fix such a $\k_3$.
Let ${\vec{\rm n}}$ be a unit normal vector to $F$ at $\k_3$ and
${\vec{\rm t}}$ be a unit tangent vector to $F$ at $\k_3$. If $(\k_1,\k_2,\k_3)\in
{\cal S}$, by (\eqnOSSecsumI), the sectors $\,s_{\Lambda,\fl}(\k_i),\
i=1,2,3\,$ are each contained in a rectangle two of whose edges are parallel to
${\vec{\rm t}}$ and have length at most $\,\abcst\,\fl\,$, and two of whose edges are parallel to
${\vec{\rm n}}$ and have length at most
$$
\abcst\, (\Lambda + \fl \, \om ) 
\le \cases{ \abcst\, \fl\,\omega & in case (i)\cr
             \abcst\, \Lambda    & in case (ii)\cr}
$$
The same holds for $\,s_{\Lambda,\fl}(\p)$ when ${\cal S}={\cal S}_{1,2}$.
In particular, if ${\cal S}={\cal S}_{1,2}$, the set
$$
\set{x_3 + y}  { x_3 \in s_{\Lambda,\fl}(\k_3)\ , \ y \in s_{\Lambda,\fl}(\p)}
$$
is contained in a rectangle $R$ whose one pair of edges is parallel to 
${\vec{\rm t}}$ and have length at most $\,\abcst\, \fl\,$, and whose other pair
of edges is parallel to ${\vec{\rm n}}$ and has length at most 
$\,\abcst\, \fl\,\omega\,$ in case (i) and length at most 
$\,\abcst\, \Lambda\,$ in case (ii). If ${\cal S}={\cal S}^\sim_{1,2}$, the set
$$
\set{x_3 + \q}  { x_3 \in s_{\Lambda,\fl}(\k_3)}
$$
is contained in such a rectangle $R$.

Let
$$
{\cal M}(\k_3)\ =\ \{(\k_1,\k_2) \in \,\Gamma^2\, \big|\,
(\k_1,\k_2,\k_3)\in {\cal S}\}
$$
By definition, if $(\k_1,\k_2) \in {\cal M}(\k_3)$, there are 
$x_1\in s_{\Lambda,\fl}(\k_1)$, $x_2\in s_{\Lambda,\fl}(\k_2)$ such that 
$x_1+x_2 \in R$. The shape of $s_{\Lambda,\fl}(\k_1)$, $s_{\Lambda,\fl}(\k_2)$ and
$R$ determined above implies that the map
$\,f: (\k_1,\k_2) \mapsto \k_1+\k_2\,$
maps ${\cal M}(\k_3)$
to a rectangle $R'$ that contains $R$ and has one pair of edges parallel to 
${\vec{\rm t}}$ and of length at most $\,\abcst'\, \fl\,$, and  a second
 pair of edges parallel to ${\vec{\rm n}}$ and of length at most 
$\,\abcst'\, \fl\,\omega\,$ in case (i) and $\abcst'\,\La$ in case (ii). Observe
that 
$$\eqalign{
{\cal M}(\k_3) \subset \{(\k_1,\k_2) \in \,\Gamma^2 \,\big|\,&
\omega \le \min [ d(\k_1,\k_2),d(a(\k_1),\k_2) ]\cr 
&{\rm and\ } 
\min [ d(\k_i,\k_3),d(a(\k_i),\k_3)] \le \abcst\, \omega \ {\rm for}\ i=1,2\}
}$$
It follows from part iv) of Lemma \lemSecVIII, with $\p=\k_3$,
$A=\abcst'\,\fl\om$ or $\abcst'\,\La$, $B=\abcst'\,\fl$, $\om_1=\om$,
$\om_2=\abcst\,\om$ and $\epsilon=\fl$ that
$$
\#{\cal M}(\k_3) \le  
\cases {
\sfrac{\abcst}{\omega \fl^2} (\fl \, \omega) (\fl) = \abcst & in case (i) \cr
\sfrac{\abcst}{\omega \fl^2} \Lambda \,\fl = \abcst \sfrac{\Lambda}{\omega \fl}
& in case (ii) \cr
}
$$
Together with the observation made above, that there are at most
$\,\abcst\, \sfrac{\omega}{\fl}$ choices of $\k_3 \in \Gamma$
for which there exist $(\k_1,\k_2)\in \Gamma^2$ with 
$(\k_1,\k_2,\k_3)\in {\cal S}$, this completes the proof of the Lemma.
\endproof

\proposition{\STM\propSecX}{
For all $\fl$--separated subsets $\Gamma$ of $F$ and all $\p\in F$, $\q\in\bbbr^2$
$$\eqalign{
\# Mom_3(\Gamma,\p)  &\le \, \sfrac{\abcst}{\fl} 
\big( 1+ \sfrac{\Lambda}{\fl} \log \sfrac{\Lambda}{\fl^2}  \big) \cr
\# Mom^\sim_3(\Gamma,\q)  &\le \, \sfrac{\abcst}{\fl} 
\big( 1+ \sfrac{\Lambda}{\fl} \log \sfrac{\Lambda}{\fl^2}  \big) \cr
}$$ 
with a constant $\abcst$ that depends only on the geometry of $F$.
}

\prf
We give the proof for $Mom_3(\Gamma,\p)$, the proof for $Mom^\sim_3(\Gamma,\q)$ is
similar. Applying part (i) of  Lemma \lemSecIX\ successively to
$$
\omega = \Lambda/\fl,\ 2\Lambda/\fl,\ 4\Lambda/\fl, \cdots ,\abcst
$$
one sees that
$$\eqalign{
&\# \big\{ (\k_1,\k_2,\k_3) \in Mom_3(\Gamma,\p)\, \big|\  
 \max_{1\le \mu\ne\nu\le 3} 
\min [d(\k_\mu,\k_\nu),d(a(\k_\mu),\k_\nu)] \ge \Lambda/\fl\big\}   \cr        
&\hskip1in\le\sum_{j=1}^{\ln_2(\abcst{\fl\over\La})}\abcst
2^j\sfrac{\La}{\fl^2}
\le\abcst\sfrac{\fl}{\La}\sfrac{\La}{\fl^2}\le \, \sfrac{\abcst}{\fl} \cr
}$$ 
Similarly, if $4\fl\le\sfrac{\La}{\fl}$ and one applies  part (ii) of 
Lemma \lemSecIX\  successively to
$$
\omega = 4\fl,\, 8\fl\,, 16\fl,\,\cdots, \,
2^{1 +[\log_2{\La\over\fl^2}] }\fl
$$
one sees that
$$\eqalign{
\# \big\{ (\k_1,\k_2,\k_3) \in Mom_3(\Gamma,\p)\, \big|\  
4\fl \le \max_{1\le \mu\ne\nu\le 3} 
\min [&d(\k_\mu,\k_\nu),d(a(\k_\mu),\k_\nu)] \le \Lambda/\fl
\big\} \cr                        
&\le \, \sfrac{\abcst\, \Lambda}{\fl^2} \,
\big(1+\log_2\sfrac{\Lambda}{\fl^2}\big)\cr
}$$ 
Finally, it is obvious that
$$
\# \big\{ (\k_1,\k_2,\k_3) \in Mom_3(\Gamma,\p)\, \big|\  
 \max_{1\le \mu\ne\nu\le 3} 
\min [d(\k_\mu,\k_\nu),d(a(\k_\mu),\k_\nu)] \le 4\fl
\big\}                        
\le \, \sfrac{\abcst}{\fl}
$$ 
\endproof

\proposition{\STM\proSecXI}{ 
Let $n\ge 2$, $\delta \ge \fl$ and let $I_1,\cdots,I_{2n-1}$ be intervals of
length $\delta$ in $F$. Assume that
$$
\sfrac{1}{3}\omega  = \max _{1\le i\ne j\le 2n-1} \min
\Big( {\rm dist\,}(I_i,I_j),\, {\rm dist\,}(I_i,a(I_j)) \Big)\ 
>\ \max \big( \delta,\,4\fl \big)
$$ 
Then for all $\fl$--separated subsets $\Gamma$ of $F$, all $\p\in F$ and all
$\q\in \bbbr^2$
$$\eqalign{
\# Mom_{2n-1}(\Gamma,\p) \cap (I_1 \times \cdots \times I_{2n-1})   
&\le \abcst\,n^2
\,\big( \sfrac{\delta}{\fl} +1 \big)^{2n-3}\,
\big(1+\sfrac{\Lambda}{\fl \omega}\big) \cr
\# Mom^\sim_{2n-1}(\Gamma,\q) \cap (I_1 \times \cdots \times I_{2n-1})   
&\le \abcst\,n^2
\,\big( \sfrac{\delta}{\fl} +1 \big)^{2n-3}\,
\big(1+\sfrac{\Lambda}{\fl \omega}\big) \cr
}$$ 
with a constant $\abcst$ that depends only on the geometry of $F$.
}

\prf The proof is similar to the proof of Proposition \propSecX.
Set
$$
\epsilon_i\ =\ \cases{ +1 &  for $1 \le i \le n$ \cr
                       -1 &  for $ \ n+ 1 \le i \le 2n-1$ \cr}
$$
Choose a point $\,\k \in I_1\,$. Then for all 
$\,x\in \bigcup \limits_{i=1}^{2n-1} I_i\,$
$$
d(x,\k) \le \omega \ \ \  \ {\rm or} \ \ \ \ \ d(x,a(\k)) \le \omega
$$
Choose $1\le i_0<j_0 \le 2n-1$ such that
$$
\min\big({\rm dist\,}(I_{i_0},I_{j_0}),\,
{\rm dist\,}(I_{i_0},a(I_{j_0}))\big)
\ =\ \sfrac{1}{3}\omega
$$
Since $\Ga$ is $\fl$--separated and each $I_j$ is of length $\de$,
$$
\# \smprod_{i=1\atop i\ne i_0,j_0}^{2n-1} \Gamma \cap I_i   
\le \Big( \sfrac{\delta}{\fl}+1  \Big)^{2n-3}
\EQN\eqnSecI$$
Fix $\k_i\in  \Ga\cap I_i \,$ for $i=1,\cdots,2n-1,\,i\ne i_0,j_0$ .
Let ${\vec{\rm n}}$ be a unit normal vector to $F$ at $\k$ and ${\vec{\rm t}}$ be 
a unit tangent vector to $F$ at $\k$.

By Proposition  \propSecVI\ 
$$
-\set{\smsum_{i=1 \atop i\ne i_0,j_0}^{2n-1} \epsilon_i x_i\,  -\q}  
{ x_i \in s_{\Lambda,\fl}(\k_i)}
$$
is contained in a rectangle $R^\sim$ having one pair of edges parallel to 
${\vec{\rm t}}$ and of length at most $\,\abcst\, n\,\fl\,$ and  a second
pair of edges  parallel to ${\vec{\rm n}}$ and of length at most
$\abcst\,n \big[ \Lambda + \fl \, \omega\big]$. As each $s_{\Lambda,\fl}(\k_i)$ 
is contained in a rectangle having one pair of edges parallel to 
${\vec{\rm t}}$ and of length at most $\,\abcst\,\,\fl\,$ and  a second
pair of edges parallel to ${\vec{\rm n}}$ and of length at most
$\abcst\ \big[ \Lambda + \fl \, \omega\big]$, the map 
$\,f: (\k_{i_0},\k_{j_0}) \mapsto \epsilon_{i_0}\k_{i_0}+\epsilon_{j_0}\k_{j_0}\,$
maps the set
$$\eqalign{
{\cal M}^\sim\ =\ \{(\k_{i_0},\k_{j_0}) \in &\,\Gamma^2\cap (I_{i_0}\times
I_{j_0})\, \big|\,
\exists  \, x_i \in s_{\Lambda,\fl}(\k_i),\ i=1,\cdots,2n-1\cr
&{\rm such\ that\ }\  
x_1+\cdots+x_n \,-\, x_{n+1}-\cdots -x_{2n-1} = \q \}                 \cr
}$$
to a rectangle $R'$ having one pair of edges parallel to 
${\vec{\rm t}}$ and of length at most $B=\,n\,\abcst'\, \fl\,$, and  
a second pair of edges parallel to ${\vec{\rm n}}$ and of length at most
$$
A=\abcst'\,n \big[ \Lambda + \fl \, \omega \big]
$$
By part iv of Lemma \lemSecVIII, with $\p$ replaced by $\k$, 
$\,\omega_1 = \sfrac{1}{3}\omega ,\  \omega_2 = \omega\,$ and $\ep=\fl$
$$
\# {\cal M}^\sim \le  \sfrac{\abcst}{\fl^2\omega}(n\Lambda+n\fl \omega)(n\fl)
\le \abcst\, n^2 \big(1+\sfrac{\Lambda}{\fl \omega}\big)
$$
This, together with (\eqnSecI), proves 
$$
\# Mom^\sim_{2n-1}(\Gamma,\q) \cap (I_1 \times \cdots \times I_{2n-1})          
             \le \abcst\,
\,\big( \sfrac{\delta}{\fl} +1 \big)^{2n-3}\,
n^2 \big(1+\sfrac{\Lambda}{\fl \omega}\big)
$$ 

By Proposition \propOSSecV.ii
$$
\| \p-\k \| \le \abcst\,n\, \omega \qquad\hbox{or}
\qquad \| \p-a(\k) \| \le \abcst\, n\,\omega
$$
Therefore $\,s_{\Lambda,\fl}(\p)\,$ is contained 
in a rectangle, two of whose edges are parallel to
${\vec{\rm t}}$ and have length at most 
$\,\abcst\,\fl\,$ and two of whose edges are parallel to ${\vec{\rm n}}$ and have
length at most
$$
\abcst\, (\Lambda + \fl\,\|\p-\k \| ) 
\le \abcst\, (\Lambda + n \,\fl\, \omega)  
$$
This, and Proposition \propSecVI\ imply that
$$
-\set{\smsum_{i=1 \atop i\ne i_0,j_0}^{2n-1} \epsilon_i x_i\,  -y}  
{ x_i \in s_{\Lambda,\fl}(\k_i)\  , \ y \in s_{\Lambda,\fl}(\p)}
$$
is contained in a rectangle $R$ having one pair of edges parallel to 
${\vec{\rm t}}$ and of length at most $\,\abcst\, n\,\fl\,$ and  a second
other pair of edges parallel to ${\vec{\rm n}}$ and of length at most
$$
\abcst\, \big[ \,n\Lambda \,+\, n\, \fl \, \omega \,+
\,n\,\fl\,  \omega \ \big]  
\le \abcst\,n \big[ \Lambda + \fl \, \omega\big]
$$
As above, this implies that
$$
\# Mom_{2n-1}(\Gamma,\p) \cap (I_1 \times \cdots \times I_{2n-1})          
             \le \abcst\,
\,\big( \sfrac{\delta}{\fl} +1 \big)^{2n-3}\,
n^2 \big(1+\sfrac{\Lambda}{\fl \omega}\big)
$$ 
\endproof

\vfill\eject
%=====================================================================
%========= SECTORS COMPATIBLE with MOM CONSERVATION ==================
%=====================================================================

\chap{Sectors Compatible with Conservation of Momentum}

\titlec{ Comparison of the $1$--norm and the $3$--norm for Four--legged Kernels}

\PG\pgOSXXI
\PG\pgOSXXIa

\lemma{\STM\lemOSonetothree}{ There is a constant $\abcst$ independent of $M$ such that the following holds.
Let $\Si$ be a sectorization of length $\fl$ at scale $j$ with 
$\,\sfrac{2}{M^{j-1}}\le\fl \le \sfrac{1}{M^{(j-1)/2}}\,$. Furthermore let $\varphi \in \cF_0(4;\Si)$ and
$f\in\check\cF_1(3;\Si)$ be particle number conserving functions. Then
$$\eqalign{
\v \varphi \v_{1,\Si}
&\le   \sfrac{\abcst}{\fl}
\big( 1+\sfrac{1}{\fl M^{j-1}} \log \sfrac{1}{\fl^2 M^{j-1}}  \big)
 \v \varphi \v_{3,\Si}  \cr
\v f \tv_{1,\Si}
&\le  \sfrac{\abcst}{\fl}
\big( 1+\sfrac{1}{\fl M^{j-1}} \log \sfrac{1}{\fl^2 M^{j-1}} \big) 
\v f \tv_{3,\Si}  \cr
}$$
}

\prf 
By Definition \defOSsectnorm\ of [FKTo3] and Remark \remOSadmissablesectors.i
$$
\v \varphi\v_{1,\Si}
=\max_{1\le i_1 \le 4}\ \ \max_{s_{i_1}\in\Si}\ \ 
\sum_{s_i\in\Si\ {\rm for}\ i\ne i_1
\atop {s_1,s_2,s_3,s_4 {\rm \ compatible \ with}
\atop{ {{\rm conservation \ of \ momentum} }}}}
\ \big\| \varphi\big((\,\cdot\,,s_1),\cdots,(\,\cdot\,,s_4)\big)\big\|_{1,\infty} 
$$
Let $1\le i_1 \le 4$ and $\,s_{i_1}\in\Si\,$. Choose $i_2,i_3,i_4$ such that
$\,\{1,2,3,4\} = \{i_1,i_2,i_3,i_4\}\,$.
By Remark \remOSadmissablesectors.ii\ and Proposition \propSecX, with
$\La=\sfrac{\sqrt{2}}{M^{j-1}}$, there are at most 
$\,\sfrac{\abcst}{\fl}\big( 1+\sfrac{M}{\fl M^j} \log \sfrac{M}{\fl^2 M^j} 
\big)\,$ triples $\, (s_{i_2},s_{i_3},s_{i_4})\,$ such that 
$\,(s_1,s_2,s_3,s_4)\,$ is compatible with conservation of momentum. Consequently
$$\eqalign{
\v \varphi\v_{1,\Si} 
& \le \sfrac{\abcst}{\fl}\big( 1+\sfrac{M}{\fl M^j} \log \sfrac{M}{\fl^2 M^j} \big)
\max_{s_1,s_2,s_3,s_4 \in \Si}
\ \big\| \varphi\big((\,\cdot\,,s_1),\cdots,(\,\cdot\,,s_4)\big)\big\|_{1,\infty} \cr
& \le \sfrac{\abcst}{\fl}\big( 1+
\sfrac{1}{\fl M^{j-1}} \log \sfrac{1}{\fl^2 M^{j-1}} \big)
\max_{1\le i_1<i_2<i_3\le 4\atop s_{i_1},s_{i_2},s_{i_3} \in \Si}
\sum_{s_i \in \Si\ \atop{\rm for\ } i\ne i_1,i_2,i_3}
\ \big\| \varphi\big((\,\cdot\,,s_1),\cdots,(\,\cdot\,,s_4)\big)\big\|_{1,\infty}
 \cr
& = \sfrac{\abcst}{\fl}
\big( 1+\sfrac{1}{\fl M^{j-1}} \log \sfrac{1}{\fl^2 M^{j-1}}  \big)
\ \v \varphi\v_{3,\Si}
}$$
The argument for $\v f \tv_{1,\Si}$ is analogous.
\endproof

\proof{ of Proposition \propOSthreetoonenorm}
Under the hypotheses of this Proposition, the term
$\ \sfrac{1}{\fl M^{j-1}} \log \sfrac{1}{\fl^2 M^{j-1}} \ $ is bounded by an
$M$--independent constant and the first inequality,
$\ 
\v\varphi\v_{1,\Si} \le \abcst \sfrac{1}{\fl}\,\v\varphi\v_{3,\Si}
\ $,
follows. If $f\in\check\cF_{4;\Si}$ and $\vec i\in\{0,1\}^4$, then
$\v f\big|_{\vec i}\tv_{1,\Si}=0$ unless $m(\vec i)\le 1$. Therefore
the second inequality, $\ 
\v f\tv_{1,\Si} \le \abcst \sfrac{1}{\fl}\,\v f\tv_{3,\Si}
\ $, also follows from Lemma \lemOSonetothree.
\endproof

%%%%%%%%%%%%

\titlec{Auxiliary Norms}\PG\pgOSXXIb

Let $\Si$ be a sectorization of scale $j$ and length $\fl\ge \sfrac{1}{M^{j-1}}$.
For $\om>0$ we define auxiliary norms on functions
$\varphi$ in $\cF_0(n;\Si)$ and $f\in \cF_1(n;\Si)$ that are antisymmetric in
their $(\xi,s)$ arguments by
$$\eqalign{
\v \varphi\v_{1,\Si,\om} 
&=\max_{s_1\in\Si} 
\sum_{s_2,\cdots,s_n\in\Si
\atop { {\rm dist}(s_k,s_\ell) \ge \omega \ {\rm and} 
\atop { {\rm dist}(s_k,a(s_\ell)) \ge \omega   
\atop {\rm for\ some\ } 2\le k\ne \ell \le n  }  }}
\big\|\varphi\big((\,\cdot\,,s_1),\cdots,(\,\cdot\,s_n)\big)\big\|_{1,\infty} \cr
\v f\tv_{1,\Si,\om} 
&=\sum_{\de\in \bbbn_0\times\bbbn_0^2} \ 
\sup_{\check\eta \in\cB} 
\sum_{s_1,\cdots,s_n\in\Si
\atop { {\rm dist}(s_k,s_\ell) \ge \omega \ {\rm and} 
\atop { {\rm dist}(s_k,a(s_\ell)) \ge \omega   
\atop {\rm for\ some\ } 1\le k\ne \ell \le n  }  }}
\max_{\rD\, {\rm dd-operator} \atop{\rm with\ } \de(\rD) =\de}\  \TN\rD f\,({\sst\check\eta;(\xi_1,s_1),\cdots,(\xi_n,s_n)})
\TN_{1,\infty}\ \sfrac{t^\de}{\de!}
}$$
The norm
$\tn\,\cdot\,\tn_{1,\infty}$ of Example \exOSSymmNorm\ refers to the variables 
${\sst \xi_1,\cdots,\xi_n}$. Furthermore, maxima, like $\max_{s_1\in\Si}$,
that  act on a formal power series $\sum_\de a_\de t^\de$ are
to be applied separately to each coefficient $a_\de$.

\lemma{\STM\lemSecXI}{
Let $\om \ge \max \big\{ \fl, \sfrac{1}{M^{(j-1)/2}}\big\}$, $n\ge 3$ and  
let $\varphi \in\cF_0(n;\Si)$  and $f\in \cF_1(n;\Si)$ be particle number conserving functions that
are antisymmetric in their $(\xi,s)$ arguments. Then
$$\eqalign{
\v \varphi \v_{1,\Si} &\le 
\v \varphi\v_{1,\Si,\om} +\abcst\, n \, \sfrac{\omega^2}{\fl^2}\,
\v \varphi\v_{3,\Si}  \cr
\v f \tv_{1,\Si} &\le 
\v f \tv_{1,\Si,\om} +\abcst\, n^2 \, \sfrac{\omega^2}{\fl^2}\,
\v f  \tv_{3,\Si}  \cr
}$$
}

\prf  
By definition of $\v \varphi \v_{1,\Si} $ and the antisymmetry of $\varphi$
$$
\v \varphi \v_{1,\Si} \le  \v \varphi\v_{1,\Si,\om}
+ \max_{s_1\in\Si}
\sum_{s_2,\cdots,s_n\in\Si
\atop { {\rm dist}(s_k,s_\ell) \le \omega \ {\rm or} 
\atop { {\rm dist}(s_k,a(s_\ell)) \le \omega   
\atop {\rm for\ all\ } 2\le k\ne \ell \le n  }  }}
\big\| \varphi\big((\,\cdot\,,s_1),\cdots,(\,\cdot\,s_n)\big)\big\|_{1,\infty} 
$$
Fix $s_1\in\Si$. If $s_2,\cdots,s_n\in\Si$ are such that for all 
$2\le k\ne \ell \le n$ one has ${\rm dist}(s_k,s_\ell) \le \omega$ or 
${\rm dist}(s_k,a(s_\ell)) \le \omega$  and such that $s_1,\cdots,s_n$ 
are compatible with conservation of momentum for
some choice of annihilation/creation indices $(b_1,\cdots,b_n)$, then,
by Proposition \propOSSecV.ii with $\La=\sfrac{\sqrt{2}}{M^{j-1}}$,
$\p$ the center of $s_1$ and $\k$ the center of $s_2$,
$$
{\rm dist}(s_1,s_k) \le \abcst\, n \,\omega  
\ \ \ \ \ \  {\rm or} \ \ \ \ \ \ \ 
{\rm dist}(s_1,a(s_k)) \le \abcst\, n \,\omega
$$
for $2\le k \le n$. Set 
$$\eqalign{
{\rm Sect} = \big\{ (s_2,s_3)\in \Si^2 \,\big| \,
{\rm dist}(s_1,s_2) \le \abcst\,n\, \omega  
 \ \ &{\rm or}\ \  {\rm dist}(s_1,a(s_2)) \le \abcst\,n\, \omega \cr
{\rm and\ } {\rm dist}(s_2,s_3) \le \abcst\, \omega  
 \ \ &{\rm or}\ \  {\rm dist}(s_2,a(s_3)) \le \abcst\, \omega \big\} \cr
}$$
Clearly $\,\big|{\rm Sect}\big| \le   \abcst\, n\,\sfrac{\omega^2}{\fl^2}\ $.
Consequently
$$\eqalign{
\sum_{s_2,\cdots,s_n\in\Si
\atop { {\rm dist}(s_k,s_\ell) \le \omega \ {\rm or} 
\atop { {\rm dist}(s_k,a(s_\ell)) \le \omega   
\atop {\rm for\ all\ } 2\le k\ne \ell \le n  }  }}
\big\| \varphi\big((\,\cdot\,,s_1),\cdots,(\,\cdot\,s_n)\big)\big\|_{1,\infty}  
& \le \smsum_{s_2,s_3\in {\rm Sect}}\ 
\smsum_{s_4,\cdots,s_n\in\Si}\ 
\big\| \varphi\big((\,\cdot\,,s_1),\cdots,(\,\cdot\,s_n)\big)\big\|_{1,\infty}  \cr
&\le 
\abcst\, n \,\sfrac{\omega^2}{\fl^2}\ \v \varphi\v_{3,\Si}
}$$

Similarly,
$$
\v f \tv_{1,\Si} \le  \v f\tv_{1,\Si,\om}
+ 
\sum_{\de\in \bbbn_0\times\bbbn_0^2} 
\max_{\check\eta \in \check \cB}
\sum_{s_1,\cdots,s_n\in\Si
\atop { {\rm dist}(s_k,s_\ell) \le \omega \ {\rm or} 
\atop { {\rm dist}(s_k,a(s_\ell)) \le \omega   
\atop {\rm for\ all\ } 1\le k\ne \ell \le n  }  }}
\max_{\rD\, {\rm dd-operator} \atop{\rm with\ } \de(\rD) =\de}\  \TN\rD f\,({\sst\check\eta;(\xi_1,s_1),\cdots,(\xi_n,s_n)})
\TN_{1,\infty}\ \sfrac{t^\de}{\de!}
$$
Fix $\check\eta =(p_0,\p,\si,b) \in \check \cB$. If $s_1,\cdots,s_n\in\Si$ are such that for all 
$1\le k\ne \ell \le n$ one has ${\rm dist}(s_k,s_\ell) \le \omega$ or 
${\rm dist}(s_k,a(s_\ell)) \le \omega$  and such that the configuration 
$(\check\eta;\, s_1,\cdots,s_n)$ 
is compatible with conservation of momentum, then,
by Proposition \propOSSecV.i with $\La=\sfrac{\sqrt{2}}{M^{j-1}}$ and $\k$ the center of $s_1$, there is an integer $r$ with $|r|\le n$ such  that
$$
\|\p-r\k +(r-1)a(\k) \| \le \abcst \,n\,\omega
\EQN\eqnOSauxnorms$$
The maps $\,F\longrightarrow \bbbr^2,\ \k \mapsto r\k -(r-1)a(\k) $ are embeddings. Therefore there are at most $ \abcst \,n\omega/\fl$ sectors $s_1$ 
containing a $\k$ such that (\eqnOSauxnorms) holds. Set
$$\eqalign{
{\rm Sect} = \big\{ (s_1,s_2)\in \Si^2 \,\big| \,s_1 
&{\rm\ contains\ a\ point\ } \k 
{\rm\ for\ which\ (\eqnOSauxnorms)\ holds\ with\ some\ } |r|\le n \cr
&{\rm and\ } {\rm dist}(s_1,s_2) \le \abcst\, \omega  
 \ \ {\rm or}\ \  {\rm dist}(s_1,a(s_2)) \le \abcst\, \omega \big\} \cr
}$$
Again $\,\big|{\rm Sect}\big| \le   \abcst\, n^2\,\sfrac{\omega^2}{\fl^2}\ $.
Consequently
$$\eqalign{
&\sum_{s_1,\cdots,s_n\in\Si
\atop { {\rm dist}(s_k,s_\ell) \le \omega \ {\rm or} 
\atop { {\rm dist}(s_k,a(s_\ell)) \le \omega   
\atop {\rm for\ all\ } 1\le k\ne \ell \le n  }  }}
\smsum_{\de\in \bbbn_0\times\bbbn_0^2} \sfrac{1}{\de!}
\max_{\rD\, {\rm dd-operator} \atop{\rm with\ } \de(\rD) =\de}\  \TN\rD f\,({\sst\check\eta;(\xi_1,s_1),\cdots,(\xi_n,s_n)})
\TN_{1,\infty}\ t^\de\cr  
& \hskip.5in\le \smsum_{s_1,s_2\in {\rm Sect}}\ 
\smsum_{s_3,\cdots,s_n\in\Si}\ 
\smsum_{\de\in \bbbn_0\times\bbbn_0^2} \sfrac{1}{\de!}
\max_{\rD\, {\rm dd-operator} \atop{\rm with\ } \de(\rD) =\de}\  \TN\rD f\,({\sst\check\eta;(\xi_1,s_1),\cdots,(\xi_n,s_n)})
\TN_{1,\infty}\ t^\de  \cr
& \hskip.5in\le 
\abcst\, n^2 \,\sfrac{\omega^2}{\fl^2}\ \v f\tv_{3,\Si}
}$$
\endproof

\vskip .5cm

\titlec{Change of Sectorization}\PG\pgOSXXIc

To prepare for the proof of Proposition \propOSresectorI, we note

\lemma{\STM\lemOSprepresectorI}{ Let $j>i\ge 2$,
$\sfrac{1}{M^{j-3/2}}\le\fl\le\sfrac{1}{M^{(j-1)/2}}$ and 
$\sfrac{1}{M^{i-3/2}}\le\fl'\le\sfrac{1}{M^{(i-1)/2}}$. Let $\Si$ and $\Si'$ be 
sectorizations of length  $\fl$
at scale $j$ and length $\fl'$ 
at scale $i$, respectively. Suppose that $\fl<\fl'$. 
Let $\varphi\in\cF_m(n;\Si')$ and $f\in\check\cF_m(n;\Si')$ be particle 
number conserving.
\Item i)
For $s_1,\cdots,s_n\in \Si$
$$\eqalign{
\big\| \varphi_\Si({\sst\eta_1,\cdots,\eta_m;\,(\xi_1,s_1),\cdots,(\xi_n,s_n)})
\big\|_{1,\infty} 
& \le  \abcst^n\, \cb_{j-1}
\sum_{s'_1,\cdots,s'_{n} \in \Si' \atop \tilde s_i' \cap \tilde s_i \ne \emptyset}
\big\| \varphi({\sst\eta_1,\cdots,\eta_m;\,(\xi_1,s'_1),\cdots,(\xi_n,s'_n)} ) 
\big\|_{1,\infty}   \cr
}$$
and for $\check\et_1,\cdots,\check\et_m\in\check\cB$ and $s_1,\cdots,s_n\in \Si$
$$
\eqalign{
&\smsum_{\de\in \bbbn_0\times\bbbn_0^2} \sfrac{1}{\de!}
\max_{\rD\, {\rm dd-operator} \atop{\rm with\ } \de(\rD) =\de} \TN\rD f_\Si\,({\sst\check\eta_1,\cdots,\check\eta_m;(\xi_1,s_1),\cdots,(\xi_n,s_n)})
\TN_{1,\infty}\ t^\de\cr
&\hskip.5in\le  \abcst^n\, \cb_{j-1}
\sum_{s'_1,\cdots,s'_{n} \in \Si' \atop \tilde s_i' \cap \tilde s_i \ne \emptyset}
\smsum_{\de\in \bbbn_0\times\bbbn_0^2} \sfrac{1}{\de!}
\max_{\rD\, {\rm dd-operator} \atop{\rm with\ } \de(\rD) =\de} \TN\rD f\,({\sst\check\eta_1,\cdots,\check\eta_m;(\xi_1,s'_1),\cdots,(\xi_n,s'_n)})
\TN_{1,\infty}\ t^\de\cr
}$$

\Item ii)
If  $f$ is antisymmetric in its $(\xi,s)$ arguments
$$\eqalign{
&\v f_\Si \tv_{p,\Si} 
\le  \abcst^n\, \cb_{j-1}\,\v f \tv_{p,\Si'}
\sup_{\check\eta_1,\cdots \check\eta_m \in
\check\cB_m} 
\max_{s_1,\cdots,s_{p-m}\in\Si\atop s'_{p-m+1},\cdots,s'_n\in\Si'} 
\#{\rm Cons}({\sst \check\eta_1,\cdots \check\eta_m;
\, s_1,\cdots,s_{p-m}\,;\,s'_{p-m+1},\cdots,s'_n})
}$$
where ${\rm Cons}({\sst \check\eta_1,\cdots \check\eta_m;\, 
s_1,\cdots,s_{p-m}\,;\,s'_{p-m+1},\cdots,s'_n})$ 
denotes the set of all $(s_{p-m+1},\cdots,s_n)\in \Si^{m+n-p}$ such that
$\tilde s_i\cap \tilde s'_i \ne \emptyset \ {\rm for\ } i=p-m+1,\cdots,n$ 
and the configuration
$(\check\eta_1,\cdots,\check\eta_m;\,s_1,\cdots,s_n)$ 
is consistent with conservation of momentum in the sense of Definition
\defOSadmissablesectors.

\Item iii)
If $m=0$, $\om \ge \fl'$ and $\varphi$ is antisymmetric, then
$$\eqalign{
\V \varphi_\Si \V_{1,\Si,\om} 
&\le \abcst^n\, \cb_{j-1}\,\v \varphi \v_{1,\Si'} 
\ \ \max_{s_1\in\Si}
\max_{s'_2,\cdots,s'_n\in\Si'
\atop { {\rm dist}(s'_k,s'_\ell) \ge \om-2\fl' \ {\rm and} 
\atop { {\rm dist}(s'_k,a(s'_\ell)) \ge \om-2\al\fl'  
\atop {\rm for\ some\ } 2\le k\ne \ell \le n  }  }} 
\#{\rm Cons}(s_1\,;\,s'_2,\cdots,s'_n)
}$$
Here, $\al$ is the supremum of the derivative of the antipodal map $a$ on the Fermi curve $F$. 
If $m=1$, $\om \ge \fl'$ and $f$ is antisymmetric in its $(\xi,s)$ arguments, then
$$\eqalign{
\v f_\Si \tv_{1,\Si,\om} 
&\le \abcst^n\, \cb_{j-1}\,\v f \tv_{1,\Si'} 
\ \ \sup_{\check\eta\in\check\cB}
\max_{s'_1,\cdots,s'_n\in\Si'
\atop { {\rm dist}(s'_k,s'_\ell) \ge \om-2\fl' \ {\rm and} 
\atop { {\rm dist}(s'_k,a(s'_\ell)) \ge \om-2\al\fl'  
\atop {\rm for\ some\ } 1\le k\ne \ell \le n  }  }} 
\#{\rm Cons}(\check\eta;\,s'_1,\cdots,s'_n)
}$$
}

\prf i)
$$
\varphi_\Si({\sst\eta_1,\cdots,\eta_m;\,(\xi_1,s_1),\cdots,(\xi_n,s_n)} )
=\sum_{s'_1,\cdots,s'_{n} \in \Si' \atop \tilde s_i' \cap \tilde s_i \ne \emptyset} 
\int {\sst d\xi'_1\cdots d\xi'_{n}}\,
\varphi({\sst\eta_1,\cdots,\eta_m;\,(\xi'_1,s'_1),\cdots,(\xi'_n,s'_n)}
\smprod_{\ell=1}^n\hat\chi_{s_\ell}(\xi'_\ell,\xi_\ell)
$$
Hence, by Lemma \lemOSelloneinfty\ of [FKTo1] and Lemma \lemOSsectpartunit\ 
of [FKTo3], 
$$
\big\| \varphi_\Si({\sst\eta_1,\cdots,\eta_m;\,(\xi_1,s_1),\cdots,(\xi_n,s_n)})
\big\|_{1,\infty} 
 \le \abcst^n\, \cb_{j-1}^n\,
\sum_{s'_1,\cdots,s'_{n} \in \Si' \atop \tilde s_i' \cap \tilde s_i \ne \emptyset}
\big\| \varphi({\sst\eta_1,\cdots,\eta_m;\,(\xi_1,s'_1),\cdots,(\xi_n,s'_n)} ) 
\big\|_{1,\infty} 
$$
The proof of the second inequality is similar.

\Item ii) By part (i) and Remark \remOSadmissablesectors.i
$$\eqalign{
&\v f_\Si \tv_{p,\Si}
\le \sum_{\de\in \bbbn_0\times\bbbn_0^2} 
\sup_{s_1,\cdots,s_{p-m} \in \Si\atop 
         \check\eta_1,\cdots,\check\eta_m\in\check\cB_m} 
\ \sum_{s_{p-m+1},\cdots,s_n \in \Si}\cr
&\hskip1.5in
\max_{\rD\, {\rm dd-operator} \atop{\rm with\ } \de(\rD) =\de} \TN\rD f_\Si({\sst \check\eta_1,\cdots,\check\eta_m;\,(\ \cdot\ ,s_1),
\cdots,(\ \cdot\ ,s_n)}) 
\TN_{1,\infty} \sfrac{t^\de}{\de!}\cr
%%%
&\le  \abcst^n\, \cb_{j-1}\sum_{\de\in \bbbn_0\times\bbbn_0^2}
\sup_{s_1,\cdots,s_{p-m} \in \Si\atop 
         \check\eta_1,\cdots,\check\eta_m\in\check\cB_m} 
\ \sum_{s_{p-m+1},\cdots,s_n \in \Si}\ \ 
\sum_{s'_1,\cdots,s'_{n} \in \Si' \atop \tilde s_i' \cap \tilde s_i \ne \emptyset}\cr
&\hskip1.5in 
\max_{\rD\, {\rm dd-operator} \atop{\rm with\ } \de(\rD) =\de} \TN\rD f\,({\sst\check\eta_1,\cdots,\check\eta_m;\,(\ \cdot\ ,s'_1),
\cdots,(\ \cdot\ ,s'_n)})
\TN_{1,\infty}\ \sfrac{t^\de}{\de!}\cr
%%%
&=  \abcst^n\, \cb_{j-1}\sum_{\de\in \bbbn_0\times\bbbn_0^2}
\sup_{s_1,\cdots,s_{p-m} \in \Si\atop 
         \check\eta_1,\cdots,\check\eta_m\in\check\cB_m} 
\ \sum_{s'_1,\cdots,s'_{p-m} \in \Si' \atop 
{\tilde s_i' \cap \tilde s_i \ne \emptyset\atop i=1,\cdots,p-m}}
\ \sum_{s'_{p-m+1},\cdots,s'_{n} \in \Si'}
\ \sum_{s_{p-m+1},\cdots,s_n \in \Si\atop 
{\tilde s_i' \cap \tilde s_i \ne \emptyset\atop i=p-m+1,\cdots,n}}\cr
&\hskip1.5in 
\max_{\rD\, {\rm dd-operator} \atop{\rm with\ } \de(\rD) =\de} \TN\rD f\,({\sst\check\eta_1,\cdots,\check\eta_m;\,(\ \cdot\ ,s'_1),
\cdots,(\ \cdot\ ,s'_n)})
\TN_{1,\infty}\ \sfrac{t^\de}{\de!}\cr
%%%
&\le  \abcst^n\, \cb_{j-1}\hskip-5pt\sum_{\de\in \bbbn_0\times\bbbn_0^2} 
\sup_{s_1,\cdots,s_{p-m} \in \Si\atop 
         \check\eta_1,\cdots,\check\eta_m\in\check\cB_m} 
\ \sum_{s'_1,\cdots,s'_{p-m} \in \Si' \atop 
{\tilde s_i' \cap \tilde s_i \ne \emptyset\atop i=1,\cdots,p-m}}
\ \hskip-30pt\sum_{\hskip30pt s'_{p-m+1},\cdots,s'_{n} \in \Si'}\hskip-37pt
\#{\rm Cons}({\sst
\check\eta_1,\cdots,\check\eta_m;\,s_1,\cdots,s_{p-m}\,;\,s'_{p-m+1},\cdots,s'_n})
\cr
&\hskip1.5in
\max_{\rD\, {\rm dd-operator} \atop{\rm with\ } \de(\rD) =\de} \TN\rD f\,({\sst\check\eta_1,\cdots,\check\eta_m;\,(\ \cdot\ ,s'_1),
\cdots,(\ \cdot\ ,s'_n)})
\TN_{1,\infty}\ \sfrac{t^\de}{\de!}\cr
%%%
&\le  \abcst^n\, \cb_{j-1}\,\v f \tv_{p,\Si'}
\sup_{\check\eta_1,\cdots \check\eta_m \in
\check\cB_m} 
\max_{s_1,\cdots,s_{p-m}\in\Si\atop s'_{p-m+1},\cdots,s'_n\in\Si'} 
\#{\rm Cons}({\sst \check\eta_1,\cdots \check\eta_m;
\, s_1,\cdots,s_{p-m}\,;\,s'_{p-m+1},\cdots,s'_n})\cr
}$$
since, for $i =1,\cdots,p-m$, there are at most three sectors 
$s'_i \in \Si'$ with $\tilde s'_i \cap \tilde s_i \ne \emptyset $.
\Item iii)
If  $s_k,s_\ell\in \Si$ with 
${\rm dist}(s_k,s_\ell) \ge \om $,  ${\rm dist}(s_k,a(s_\ell)) \ge \om $ 
and $s_k',s_\ell' \in \Si'$ with
$ \tilde s_k \cap \tilde s'_k\ne \emptyset$, $\tilde s_\ell \cap \tilde s'_\ell
\ne \emptyset$ 
then ${\rm dist}(s'_k,s'_\ell) \ge \om-2\fl' $ and  
${\rm dist}(s'_k,a(s'_\ell)) \ge \om-2\al\fl' $. Using this observation, the proof
of (iii) is analogous to the proof of (ii).
\endproof

\lemma{\STM\lemOSseccountI}{
Let $j>i\ge 2$, $\sfrac{1}{M^{j-3/2}}\le\fl\le\sfrac{1}{M^{(j-1)/2}}$ and 
$\sfrac{1}{M^{i-3/2}}\le\fl'\le\sfrac{1}{M^{(i-1)/2}}$ with $\fl<\sfrac{1}{4}\fl'$. Let $\Si$ and $\Si'$ be 
sectorizations of length  $\fl$ at scale $j$ and length $\fl'$ 
at scale $i$, respectively.
\Item i)
There is a constant $\abcst$ independent of $M$ such that for every $s'\in\Si'$
$$
\#\{ s \in \Si\,\big|\,\tilde s \cap \tilde s' \ne \emptyset\,\}
\le \abcst \sfrac{\fl'}{\fl}
$$
\Item ii) 
Let $m\ge 0$, $p\ge m$, $n\ge p-m+1$, $\check \eta_1,\cdots,\check\eta_m \in\cB$,
$s_1,\cdots,s_{p-m}\in\Si$ and 
$s'_{p-m+1},\cdots,s'_n\in \Si'$. Then
$$
\#{\rm Cons}({\sst \check \eta_1,\cdots,\check\eta_m;
\,s_1,\cdots,s_{p-m}\,;\,s'_{p-m+1},\cdots,s'_n}) 
\le \abcst^n \big( \sfrac{\fl'}{\fl} \big)^{n+m-p-1}
$$
\Item iii) 
Let $\om'\ge 4\fl'$, and let $s_1 \in\Si$ and $s'_2,\cdots,s'_n\in \Si'$ such that 
${\rm dist}(s'_k,s'_\ell) \ge \om'$ and $ {\rm dist}(s'_k,a(s'_\ell)) \ge \om'$  
for some $ 2\le k\ne \ell \le n $. Then
$$
\#{\rm Cons}(s_1\,;\,s'_2,\cdots,s'_n) 
\le \abcst^n \big( \sfrac{\fl'}{\fl} \big)^{n-3}
\big(1+\sfrac{1}{M^{j-1}\fl\om'}\big)
$$
\Item iv) 
Let $\om'\ge 4\fl'$, $\check\eta=(q_o,\q,\si,a)\in\check B$ and
$s'_1,\cdots,s'_n\in\Si'$ such that 
${\rm dist}(s'_k,s'_\ell) \ge \om'$ and $ {\rm dist}(s'_k,a(s'_\ell)) \ge \om'$  
for some $ 1\le k\ne \ell \le n $. Then
$$
\#{\rm Cons}(\check\eta\,;\,s'_1,\cdots,s'_n) 
\le \abcst^n \big( \sfrac{\fl'}{\fl} \big)^{n-2}
\big(1+\sfrac{1}{M^{j-1}\fl\om'}\big)
$$
}

\prf
i) is trivial
\Item ii)
By part (i), there are at most $\abcst^n \big( \sfrac{\fl'}{\fl} \big)^{n+m-p-1}\,$
$(n+m-p-1)$--tuples $(s_{p-m+1},\cdots,s_{n-1})$ of sectors in $\Si$ such that
$\tilde s_i \cap \tilde s'_i \ne \emptyset$ for $i=p-m+1,\cdots,n-1$. Given such an
$(n+m-p-1)$--tuple $(s_{p-m+1},\cdots,s_{n-1})$ and a particle number
preserving sequence $(a_1,\cdots,a_n)$ of creation--annihilation indices, the set
$$
\{ \,-(-1)^{a_n} \big( \check\eta_1+\cdots+\check\eta_m+
(-1)^{a_1}\k_1+\cdots+(-1)^{a_{n-1}}\k_{n-1}\big)
\
\big| 
\ \k_i\in \tilde s_i\ {\rm for}\ i=1,\cdots,n-1\,\}
$$
has diameter at most $\abcst\,(n-1)\,\fl\ $ and therefore meets at most
$\abcst\,(n-1)$ extended sectors of $\Si$. This shows that there are at most
$\abcst^n$ sectors $s_n \in \Si$ such that $(s_1,\cdots, s_n)$ is consistent with 
conservation of momentum.
\Item iii)
Let $(a_1,\cdots,a_n)$ be a particle number
preserving sequence  of creation--annihilation indices.
For $i=1,\cdots,n-1$ let 
$I_i=\{\,\k\in F\,\big|\,{\rm dist}(\k,s'_{i+1}) \le\fl\,\}$.
We apply the first inequality of Proposition
\proSecXI\ with $\de=\fl'+2\fl$, $\La = \sfrac{\sqrt{2}}{M^{j-1}}$, 
$\Ga$ the set of centers of the intervals $s\cap F,\ s\in \Si$ 
and $\p$ the center of $s_1\cap F$. It follows that 
$$\eqalign{
\#{\rm Cons}(s_1\,;\,s'_2,\cdots,s'_n) 
&\le \abcst^n \big( \sfrac{\fl'}{\fl} \big)^{n-3}
\big(1+\sfrac{1}{M^{j-1}\fl\om'}\big)
}$$
\Item iv) is similar to (iii), using the second inequality of Proposition
\proSecXI\ instead.
\endproof

\proof{ of Proposition \propOSresectorI}
i) 
As $m\ne 0$, by Lemma \lemOSprepresectorI.i and Lemma \lemOSseccountI.i
$$\eqalign{
\V \varphi_\Si \V_{1,\Si} 
&= \smsum_{s_1,\cdots,s_n \in \Si}\, 
\|\varphi_\Si({\sst\eta_1,\cdots,\eta_m;\,(\xi_1,s_1),\cdots,(\xi_n,s_n)})\|_{1,\infty}
\cr
&\le  \abcst^n\, \cb_{j-1} \sum_{s_1,\cdots,s_n \in \Si
\atop{ s'_1,\cdots,s'_{n} \in \Si' \atop{ \tilde s_i' \cap \tilde s_i \ne
\emptyset}}} 
\big\| \varphi({\sst\eta_1,\cdots,\eta_m;\,(\xi_1,s'_1),\cdots,(\xi_n,s'_n)} ) 
\big\|_{1,\infty}  \cr 
&\le  \abcst^n\, \cb_{j-1}\,\big(\sfrac{\fl'}{\fl} \big)^n
 \smsum_{s'_1,\cdots,s'_n \in\Si'} 
\big\|\varphi({\sst\eta_1,\cdots,\eta_m;\,(\xi_1,s'_1),\cdots,(\xi_n,s'_n)} ) 
\big\|_{1,\infty}  \cr 
&=  \abcst^n\, \cb_{j-1}\,\big(\sfrac{\fl'}{\fl} \big)^n
 \v \varphi \v_{1,\Si'} 
}$$

\Item ii)
By Lemma \lemOSprepresectorI.ii and Lemma \lemOSseccountI.ii
$$\eqalign{
\V f_\Si \tV_{p,\Si} 
&\le \abcst^n\, \cb_{j-1}\,\big(\sfrac{\fl'}{\fl} \big)^{n+m-p-1}
\v f \tv_{p,\Si'}
}\EQN\eqnOSssumI$$
Now assume that $\ \fl \ge \sfrac{1}{M^{2/3(j-1)}}$, $\fl'\le\sfrac{1}{16}$
 and $n\ge 3$. Observe that $\V f_\Si \tV_{1,\Si}$ vanishes for $m\ge 2$,
so it suffices to consider $m=0,1$.
Set $\om = \al\sqrt{\fl}$. By Lemma \lemSecXI
$$
\V f_\Si \tV_{1,\Si} \le 
\V f_\Si\tV_{1,\Si,\om} 
+\abcst\, n^2 \, \sfrac{\omega^2}{\fl^2}\,\V f_\Si \tV_{3,\Si}
$$
By (\eqnOSssumI),
$$\eqalign{
n^2\,\sfrac{\omega^2}{\fl^2}\,\V f_\Si \tV_{3,\Si}
\ &\le\ \abcst^n\,
\cb_{j-1}\,\big(\sfrac{\fl'}{\fl}\big)^{n+m-4}\,\sfrac{\omega^2}{\fl^2}\,
\v f \tv_{3,\Si'} 
\ \le\ \abcst^n\,
\cb_{j-1}\,\big(\sfrac{\fl'}{\fl}\big)^{n+m-4}\,\sfrac{1}{\fl}\,
\v f \tv_{3,\Si'} \cr
\ &= \abcst^n\,
\cb_{j-1}\,\big(\sfrac{\fl'}{\fl}\big)^{n+m-3}\,\sfrac{1}{\fl'}\,
\v f \tv_{3,\Si'} 
}$$
If $m=0$, by Lemma \lemOSprepresectorI.iii and Lemma \lemOSseccountI.iii,
with $\om'=\om-2\al\fl'$,
$$\eqalign{
\V f_\Si \tV_{1,\Si,\om} 
&\le \abcst^n\, \cb_{j-1}\,\big( \sfrac{\fl'}{\fl} \big)^{n+m-3}
\big(1+\sfrac{1}{M^{j-1}\fl(\om-2\al\fl')}\big)\!
\v f \tv_{1,\Si'} \cr
&\le \abcst^n\, \cb_{j-1}\,\big( \sfrac{\fl'}{\fl} \big)^{n+m-3}
\v f \tv_{1,\Si'} 
}$$
since 
$\,M^{j-1}\fl(\om-2\al\fl') 
\ge M^{j-1}\fl\big(\al\sqrt{\fl}-\sfrac{\al}{3}\sqrt{\fl}\big)
= \sfrac{2}{3}\al M^{j-1}\fl^{3/2} \ge \sfrac{2}{3}\al$.
Similarly one sees, using Lemma \lemOSseccountI.iv, that also in the case $m=1$
$$
\V f_\Si \tV_{1,\Si,\om} 
\le \abcst^n\, \cb_{j-1}\,\big( \sfrac{\fl'}{\fl} \big)^{n+m-3}
\v f \tv_{1,\Si'} 
$$

\Item iii) Write
$$
f_{\Si'}({\sst\check\eta_1,\cdots,\check\eta_m;\,(\xi_1,s'_1),\cdots,(\xi_n,s'_n)} )
=\smsum_{s_1,\cdots,s_{n} \in \Si} g({\sst\check\eta_1,\cdots,\check\eta_m;\,(\xi_1,s'_1),\cdots,(\xi_n,s'_n)
;s_1,\cdots,s_{n}} )
$$
with
$$
g({\sst\check\eta_1,\cdots,\check\eta_m;\,(\xi_1,s'_1),\cdots,(\xi_n,s'_n)
;s_1,\cdots,s_{n}} )
= \int {\sst d\xi'_1\cdots d\xi'_{n}}\,
f({\sst\check\eta_1,\cdots,\check\eta_m;\,(\xi'_1,s_1),\cdots,(\xi'_n,s_n)} )
\smprod_{\ell=1}^n\hat\chi_{s'_\ell}(\xi'_\ell,\xi_\ell) 
$$
Then
$$\eqalign{
\v f_{\Si'}\tv_{p,\Si'} 
&= \sum_{\de\in \bbbn_0\times\bbbn_0^2}\sup_{{1\le i_1<\cdots<i_{p-m}\le n \atop s'_{i_1},\cdots,s'_{i_{p-m}}\in\Si'}
\atop \check\eta_1,\cdots,\check\eta_m \in \check \cB} 
\smsum_{s'_i \in \Si' \ {\rm for}\atop i\ne i_1,\cdots i_{p-m}}\cr
&\hskip1.5in 
\max_{\rD\, {\rm dd-operator} \atop{\rm with\ } \de(\rD) =\de} \TN\rD f_\Si({\sst\check\eta_1,\cdots,\check\eta_m;(\xi_1,s'_1),\cdots,(\xi_n,s'_n)})
\TN_{1,\infty}\ \sfrac{t^\de}{\de!}\cr
&\le \sum_{\de\in \bbbn_0\times\bbbn_0^2}\sup_{{1\le i_1<\cdots<i_{p-m}\le n \atop s'_{i_1},\cdots,s'_{i_{p-m}}\in\Si'}
\atop \check\eta_1,\cdots,\check\eta_m \in \check \cB} 
\smsum_{s'_i \in \Si' \ {\rm for}\atop i\ne i_1,\cdots i_{p-m}}
\smsum_{s_1,\cdots,s_{n} \in \Si\atop s_i\cap s'_i\ne\emptyset
{\rm \ for\ }1\le i\le n}\cr
&\hskip1.5in
\max_{\rD\, {\rm dd-operator} \atop{\rm with\ } \de(\rD) =\de} \TN\rD g({\sst\check\eta_1,\cdots,\check\eta_m;\,(\xi_1,s'_1),\cdots,(\xi_n,s'_n)
;s_1,\cdots,s_{n}} )
\TN_{1,\infty}\ \sfrac{t^\de}{\de!}\cr
}$$
For each fixed $\check\et_1,\cdots,\check\et_m,s'_1,\cdots,s'_n,s_1,\cdots,s_n$,
$$\eqalign{
&\smsum_{\de\in \bbbn_0\times\bbbn_0^2} \sfrac{1}{\de!}
\max_{\rD\, {\rm dd-operator} \atop{\rm with\ } \de(\rD) =\de} \TN\rD g({\sst\check\eta_1,\cdots,\check\eta_m;\,(\xi_1,s'_1),\cdots,(\xi_n,s'_n)
;s_1,\cdots,s_{n}} )
\TN_{1,\infty}\ t^\de\cr
&\hskip1in
\le\smsum_{\de\in \bbbn_0\times\bbbn_0^2} \sfrac{1}{\de!}
\max_{\rD\, {\rm dd-operator} \atop{\rm with\ } \de(\rD) =\de} \TN\rD f({\sst\check\eta_1,\cdots,\check\eta_m;\,(\xi'_1,s_1),\cdots,(\xi'_n,s_n)} )
\TN_{1,\infty}\ t^\de\prod_{\ell=1}^n \big\|\hat\chi_{s'_\ell}\big\|_{1,\infty}
}$$
as in Lemma \lemOSelloneinfty\ of [FKTo1]. 
Hence, by Lemma \lemOSsectpartunit\ of [FKTo3] and Example \exOSappMonoidI\ 
of [FKTo1],
$$\eqalign{
&\smsum_{\de\in \bbbn_0\times\bbbn_0^2} \sfrac{1}{\de!}
\max_{\rD\, {\rm dd-operator} \atop{\rm with\ } \de(\rD) =\de} \TN\rD g({\sst\check\eta_1,\cdots,\check\eta_m;\,(\xi_1,s'_1),\cdots,(\xi_n,s'_n)
;s_1,\cdots,s_{n}} )
\TN_{1,\infty}\ t^\de\cr
&\hskip1in
\le\smsum_{\de\in \bbbn_0\times\bbbn_0^2} \sfrac{1}{\de!}
\max_{\rD\, {\rm dd-operator} \atop{\rm with\ } \de(\rD) =\de} \TN\rD f({\sst\check\eta_1,\cdots,\check\eta_m;\,(\xi'_1,s_1),\cdots,(\xi'_n,s_n)} )
\TN_{1,\infty}\ t^\de\prod_{i=1}^n \abcst\,\cb_{i-1}\cr
&\hskip1in
\le\abcst^n\ \cb_{i-1}\smsum_{\de\in \bbbn_0\times\bbbn_0^2} \sfrac{1}{\de!}
\max_{\rD\, {\rm dd-operator} \atop{\rm with\ } \de(\rD) =\de} \TN\rD f({\sst\check\eta_1,\cdots,\check\eta_m;\,(\xi'_1,s_1),\cdots,(\xi'_n,s_n)} )
\TN_{1,\infty}\ t^\de\cr
}$$
uniformly in $s'_1,\cdots, s'_n,s_1,\cdots,s_n,
\check \et_1,\cdots,\check\et_m$. So 
$$\eqalign{
\v f_{\Si'}\tv_{p,\Si'} 
&\le \abcst^n\ \cb_{i-1}\sum_{\de\in \bbbn_0\times\bbbn_0^2} 
\sup_{{1\le i_1<\cdots<i_{p-m}\le n \atop s'_{i_1},\cdots,s'_{i_{p-m}}\in\Si'}
\atop \check\eta_1,\cdots,\check\eta_m \in \check \cB} 
\sum_{s'_i \in \Si' \ {\rm for}\atop i\ne i_1,\cdots i_{p-m}}
\sum_{s_1,\cdots,s_{n} \in \Si\atop s_i\cap s'_i\ne\emptyset
{\rm \ for\ }1\le i\le n}\cr
&\hskip1.5in
\max_{\rD\, {\rm dd-operator} \atop{\rm with\ } \de(\rD) =\de} \TN\rD f({\sst\check\eta_1,\cdots,\check\eta_m;\,(\xi'_1,s_1),\cdots,(\xi'_n,s_n)} )
\TN_{1,\infty}\ \sfrac{t^\de}{\de!}\cr
&\le \abcst^n\ \cb_{i-1}\sum_{\de\in \bbbn_0\times\bbbn_0^2}
\sup_{{1\le i_1<\cdots<i_{p-m}\le n \atop s'_{i_1},\cdots,s'_{i_{p-m}}\in\Si'}
\atop \check\eta_1,\cdots,\check\eta_m \in \check \cB} 
\sum_{{s_i \in \Si \atop s_i\cap s'_i\ne\emptyset}
                \atop{\rm for\ } i= i_1,\cdots i_{p-m}}
\sum_{s_i \in \Si \ {\rm for}\atop i\ne i_1,\cdots i_{p-m}}
\sum_{{s'_i \in \Si' \atop s_i\cap s'_i\ne\emptyset}
                \atop{\rm for\ } i\ne i_1,\cdots i_{p-m}}\cr
&\hskip1.5in
\max_{\rD\, {\rm dd-operator} \atop{\rm with\ } \de(\rD) =\de} \TN\rD f({\sst\check\eta_1,\cdots,\check\eta_m;\,(\xi'_1,s_1),\cdots,(\xi'_n,s_n)} )
\TN_{1,\infty}\ \sfrac{t^\de}{\de!}\cr
&\le \abcst^n\ \cb_{i-1}\sum_{\de\in \bbbn_0\times\bbbn_0^2}
\sup_{{1\le i_1<\cdots<i_{p-m}\le n \atop s'_{i_1},\cdots,s'_{i_{p-m}}\in\Si'}
\atop \check\eta_1,\cdots,\check\eta_m \in \check \cB} 
\sum_{{s_i \in \Si \atop s_i\cap s'_i\ne\emptyset}
                \atop{\rm for\ } i= i_1,\cdots i_{p-m}}
\sum_{s_i \in \Si \ {\rm for}\atop i\ne i_1,\cdots i_{p-m}}\cr
&\hskip1.5in
\max_{\rD\, {\rm dd-operator} \atop{\rm with\ } \de(\rD) =\de} \TN\rD f({\sst\check\eta_1,\cdots,\check\eta_m;\,(\xi'_1,s_1),\cdots,(\xi'_n,s_n)} )
\TN_{1,\infty}\ \sfrac{t^\de}{\de!}\cr
}$$
since the set $\set{s'_i \in \Si', i\ne i_1,\cdots i_{p-m}}
{s_i\cap s'_i\ne\emptyset{\rm\ for\ }i\ne i_1,\cdots i_{p-m}}$
contains at most $3^n$ terms. Finally, applying
$$
\sup_{s'_{i_1},\cdots,s'_{i_{p-m}}\in\Si'}
\sum_{{s_i \in \Si \atop s_i\cap s'_i\ne\emptyset}
                \atop{\rm for\ } i= i_1,\cdots i_{p-m}}
h(s_1,\cdots,s_{i_{p-m}})
\le\big(\abcst\sfrac{\fl'}{\fl}\big)^{p-m}
\sup_{s_{i_1},\cdots,s_{i_{p-m}}\in\Si}
h(s_1,\cdots,s_{i_{p-m}})
$$
yields
$$\eqalign{
\v f_{\Si'}\tv_{p,\Si'} 
&\le \abcst^n\ \big(\sfrac{\fl'}{\fl}\big)^{p-m}\ \cb_{i-1}
\sum_{\de\in \bbbn_0\times\bbbn_0^2}
\sup_{{1\le i_1<\cdots<i_{p-m}\le n \atop s_{i_1},\cdots,s_{i_{p-m}}\in\Si}
\atop \check\eta_1,\cdots,\check\eta_m \in \check \cB} 
\sum_{s_i \in \Si \ {\rm for}\atop i\ne i_1,\cdots i_{p-m}}\cr
&\hskip1.5in
\max_{\rD\, {\rm dd-operator} \atop{\rm with\ } \de(\rD) =\de} \TN\rD f({\sst\check\eta_1,\cdots,\check\eta_m;\,(\xi'_1,s_1),\cdots,(\xi'_n,s_n)} )
\TN_{1,\infty}\ \sfrac{t^\de}{\de!}\cr
&\le \abcst^n\ \big(\sfrac{\fl'}{\fl}\big)^{p-m}\ \cb_{i-1}\ 
\v f\tv_{p,\Si}
}$$
\endproof

\vfill\eject
%=====================================================================
%======== SECTOR COUNTING for PP LADDERS =============================
%=====================================================================

\chap{Sector Counting for Particle--Particle Ladders}\PG\pgOSXXII

In this Section we prove that, when the Fermi surface $F$ is strongly asymmetric
in the sense of Definition  \defModI, particle--particle ladders obey bounds
that are stronger than those given by standard power
counting (see Appendix \APappNaiveladder). These bounds are based on 
the following sector counting result.

\proposition{\STM\propOSLadA}{ 
Assume that the Fermi surface $F$ is strongly asymmetric.
There is a constant $\abcst$ independent of $M$ such that for all
sectorizations of scale $j\ge 2$ and length 
$\fl \ge\sfrac{1}{M^{j-1}}$ and all $s'_1,s'_2\in \Si$
and all $k_1,k_2\in\bbbr\times\bbbr^2$
$$\deqalign{
&\sharp \set{(s_1,s_2)\in\Si\times\Si}
{(\tilde s_1+\tilde s_2)\cap(\tilde s'_1+\tilde s'_2)\ne\emptyset}
&\le\abcst\,\sfrac{\fl^{\raise2pt\hbox{$\scriptscriptstyle{1/n_0}$}}}{\fl}\cr
&\sharp \set{(s_1,s_2)\in\Si\times\Si}
{(\tilde s_1+\tilde s_2)\cap(k_1+\tilde s'_1)\ne\emptyset}
&\le\abcst\,\sfrac{\fl^{\raise2pt\hbox{$\scriptscriptstyle{1/n_0}$}}}{\fl}\cr
&\sharp \set{(s_1,s_2)\in\Si\times\Si}
{k_1+k_2\in\tilde s_1+\tilde s_2}
&\le\abcst\,\sfrac{\fl^{\raise2pt\hbox{$\scriptscriptstyle{1/n_0}$}}}{\fl}\cr
}$$
}
\noindent
This proof of this Proposition, which is given after Lemma \:\propLadA, is 
based on the following four lemmata.

\lemma{\STM\lemLadAI}{ Assume that  $F$ is strongly asymmetric.
There exists a constant $\abcst$ such
that for all $\veps>0$ and all disks $D$ in $\bbbr^2$ of radius $\veps$
$$
{\rm length}\ \set{\k\in F}{\k+a(\k)\in D}\le\abcst\, \veps^{1/(n_0-1)}
$$ 
where $n_0$ is the constant of Definition \defModI\ and $a(\k)$ is the
antipode of $\k$.
}
\prf
Since $F$ is compact, it suffices to show that each point $\p\in F$ has
a neighbourhood $U$ in $F$ for which there exists $1\le n\le n_0-1$
and a constant $\abcst$ such that, for all $\veps>0$ and all disks $D$
in $\bbbr^2$ of radius $\veps$,
$$
{\rm length}\ \set{\k\in U}{\k+a(\k)\in D}\le\abcst\, \veps^{1/n}
$$
Fix $\p\in F$. Without loss of generality, we may assume that the oriented
unit tangent vector to $F$ at $\p$ is $(1,0)$ and that the unit inward
pointing normal vector to $F$ at $\p$ is $(0,1)$. 
Let $\varphi(t)=\varphi_\p(t)$, $\bar\varphi(t)=\varphi_{a(\p)}(t)$,
where $\varphi_\p$ is the parameterizing map of Definition \defModI. Precisely,
$t\mapsto\k(t)=\p+(t,\varphi(t))$ is a parameterization of $F$ near $\p$ and
$t\mapsto\bar\k(t)=a(\p)-(t,\bar\varphi(t))$ is a parameterization of $F$ near
 $a(\p)$.

By strict convexity, the slopes $\dot\varphi(t)$ and $\dot{\bar\varphi}(t)$ for
the Fermi curve at $\k(t)$ and $\bar\k(t)$, respectively, are strictly
increasing with $t$. Hence there is a strictly increasing function $\bar
t(t)$ such that
$$
\dot{\bar\varphi}\big(\bar t(t)\big)=\dot\varphi(t)
$$
and hence
$$
\bar\k\big(\bar t(t)\big)=a\big(\k(t)\big)
$$
so that 
$$
\k(t)+a\big(\k(t)\big) 
= \p+a(\p)+\big(t-\bar t(t),\varphi(t)-\bar\varphi(\bar t(t))\big)
$$
Since  $\ddot{\bar\varphi}(\bar t)$ does not vanish, $\bar t(t)$ is $C^{n_0-1}$.

By construction, $\varphi(0)=\bar\varphi(0)=\dot\varphi(0)=\dot{\bar\varphi}(0)
= 0$. Since $F$ is strongly asymmetric, there is a minimal $1\le n\le n_0-1$
such that $\bar\varphi^{(n+1)}(0)\ne \varphi^{(n+1)}(0)$. We may assume,
without loss of generality, that 
$$
\big|\bar\varphi^{(n+1)}(0)\big|<\big| \varphi^{(n+1)}(0)\big|
\EQN\eqnLadAI$$
Denote by $\de$ and $\bar\de$ the degrees of the zeroes of $\dot\varphi(t)$
and $\dot{\bar\varphi}(t)$, respectively, at $t=0$. By convexity, $\de$
and $\bar\de$ are odd. By (\eqnLadAI) and the minimality of $n$, $\de\le\bar\de$
and $\de\le n$.

We consider the cases $\de=\bar\de$ and $\de<\bar\de$ separately. First,
suppose that $1\le\de=\bar\de\le n$. Apply the elementary 
Lemma \:\lemLadAII.a, below, to $\dot\varphi$ and $\dot{\bar\varphi}$.
This gives the representations 
$$
\dot\varphi(t)=t^\de\nu(t)\qquad 
\dot{\bar\varphi}(\bar t)=\bar t^\de\bar\nu(\bar t)
$$
with
$$
\nu(0),\bar\nu(0)\ne 0,\qquad
\nu^{(i)}(0)=\bar\nu^{(i)}(0)\hbox{ for }0\le i<n-\de,\qquad
\nu^{(n-\de)}(0)\ne\bar\nu^{(n-\de)}(0)
$$
The $\de^{\rm th}$ roots $\psi(t)=\big(\dot\varphi(t)\big)^{1/\de}$
and $\bar\psi(\bar t)=\big(\dot{\bar\varphi}(\bar t)\big)^{1/\de}$ both have
zeroes of order precisely one at $0$ and obey
$$\eqalign{
\psi(0)=\bar\psi(0)= 0,\qquad
&\dot\psi(0),\dot{\bar\psi}(0)\ne 0,\qquad 
\psi^{(i)}(0)=\bar\psi^{(i)}(0)\hbox{ for }1\le i\le n-\de,\cr
&\psi^{(n-\de+1)}(0)\ne\bar\psi^{(n-\de+1)}(0)\cr
}$$
Since $\bar t(t)$ obeys
$$
\bar\psi\big(\bar t(t)\big)=\psi(t)
$$  
we conclude that $\bar t(t)$ is $C^{n-\de+1}$ and obeys
$$
\bar t^{(i)}(0)=\left.\cases{
1& if $i=1$\cr
0& if $1<i\le n-\de$\cr
\tilde b\ne 0 & if $i=n-\de+1$\cr
}\right\}\hbox{ if }\de<n\qquad\hbox{ and }\qquad
\dot{\bar t}(0)=\tilde b\ne 1\hbox{ if }\de=n
$$
Consequently, there is a neighbourhood $U'$ of $0$ and a $b>0$ such that
for all $t\in U'$
$$
\Big|\sfrac{d^{n-\de+1}\hfill}{dt^{n-\de+1}}\big(t-\bar t(t)\big)\Big|\ge b
\EQN{\eqnLadAII}$$

Now suppose that $\de<\bar\de$. Again denoting 
$\psi(t)=\big(\dot\varphi(t)\big)^{1/\de}$ and
$\bar\psi(\bar t)=\big(\dot{\bar\varphi}(\bar t)\big)^{1/\de}$, we have that
$\psi(t)$ and $\bar\psi(\bar t)$ are both $C^1$ (for $\bar t$ near zero,
$\bar\psi(\bar t)\approx\abcst\, \bar t^{\bar\de/\de}$ and $\bar\de/\de>1$)
and obey
$$
\psi(0)=\bar\psi(0)\qquad \dot\psi(0)>\dot{\bar\psi}(0)=0
$$
As $\bar\psi\big(\bar t(t)\big)=\psi(t)$ and hence
$$
\dot{\bar t}(t)=\sfrac{\dot\psi(t)}{\dot{\bar\psi}(\bar t(t))}
$$
there is a $b>0$ and a neighbourhood $U'$ of $0$ such that for all 
$t\in U'\setminus\{0\}$
$$
\dot{\bar t}(t)\ge 1+b\quad\Longrightarrow\quad 
\Big|\sfrac{d\hfill}{dt}\big(t-\bar t(t)\big)\Big|\ge b
\EQN{\eqnLadAIII}$$

Set $U=\set{\p+\big(t,\varphi(t)\big)}{t\in U'}$.
If $D$ is a disk of radius $\veps$, then its projection to the $x$--axis
is an interval $\cJ$ of length $2\veps$ and 
$$
{\rm length}\,\set{\k\in U}{\k+a(\k)\in D}
\le\abcst\, {\rm length}\,\set{t\in U'}{x_0+t-\bar t(t)\in\cJ}
$$
where $x_0$ is the $x$--component of $\p+a(\p)$.
Therefore, by (\eqnLadAII) and (\eqnLadAIII), this Lemma follows from 
Lemma \:\lemLadAIII\  below.
\endproof

\lemma{\STM\lemLadAII}{ Let $U$ be a neighbourhood of $0$.
\Item a)
Let $f\in C^n(U)$ have a zero of order at least $\de\le n$ at $0$. Then
there exists a function $g\in C^{n-\de}(U)\cap C^n(U\setminus\{0\})$ such
that 
$$
f(t)= t^\de g(t)
$$
and 
$$
\lim_{t\rightarrow 0\atop t\ne 0} t^j g^{(n-\de+j)}(t)=0
\qquad\hbox{for all }1\le j\le \de
$$
\Item b)
Let $f\in C(U)\cap C^1(U\setminus\{0\}).$ If $\lim\limits_{t\rightarrow
0\atop t>0}f'(t)$ and $\lim\limits_{t\rightarrow
0\atop t<0}f'(t)$ both exist and are equal, then $f\in C^1(U)$.

}

\lemma{\STM\lemLadAIII}{
Let $b$ be a strictly positive real number and $n$
be a strictly positive integer. Let $\cI\subset\bbbr$ be an interval (not
necessarily compact) and $f$ a $C^n$ function on $\cI$ obeying
$$
|f^{(n)}(x)|\ge b\qquad{\rm for\ all\ }x\in \cI
$$ 
Then for all $\veps>0$ and
all intervals $\cJ$ of length $2\veps$,
$$
{\rm length}\,\set{x\in \cI}{f(x)\in \cJ}\le  
2^{n+1}\big(\sfrac{\veps}{b}\big)^{1/n}
$$
}
\prf
Set $\al =\big(\sfrac{\veps}{b}\big)^{1/n}$ and $g(x)=f(x)-y_0$, where $y_0$
is the midpoint of $\cJ$. We must show
$$
|g^{(n)}(x)|\ge\sfrac{\veps}{\al^n}\quad{\rm for\ all\ }x\in \cI
\quad\Longrightarrow\quad
{\rm length}\set{x\in \cI}{|g(x)|\le\veps}\le 2^{n+1}\al
$$
Define $c_n$ inductively by $c_1=2$ and $c_n=2+2c_{n-1}$. Because
$d_n=2^{-n}c_n$ obeys $d_1=1$ and $d_n=2^{-n+1}+d_{n-1}$ we have 
$d_n\le 2$ and hence $c_n\le 2^{n+1}$. We shall prove
$$
|g^{(n)}(x)|\ge\sfrac{\veps}{\al^n}\quad{\rm for\ all\ }x\in \cI
\quad\Longrightarrow\quad
{\rm length}\set{x\in \cI}{|g(x)|\le\veps}\le c_n\,\al
$$
by induction on $n$.

Suppose that $n=1$ and let $x$ and $y$ be any two points in 
$\set{x\in \cI}{|g(x)|\le\veps}$. Then
$$
|x-y|=\sfrac{|x-y|}{|g(x)-g(y)|}|g(x)-g(y)|=\sfrac{|g(x)-g(y)|}{|g'(\ze)|}
\le \sfrac{2\veps}{|g'(\ze)|}
$$
for some $\ze\in \cI$. As $|g'(\ze)|\ge\sfrac{\veps}{\al}$ we have $|x-y|\le2\al$.
Thus $\set{x\in \cI}{|g(x)|\le\veps}$ is contained in an interval of length
at most $2\al$ as desired.

Now suppose that $|g^{(n)}(x)|\ge\sfrac{\veps}{\al^n}$ on $\cI$ and that
the induction hypothesis is satisfied for $n-1$. As in the last paragraph, the
set $\set{x\in \cI}{|g^{(n-1)}(x)|\le\sfrac{\veps}{\al^{n-1}}}$ is contained
in a subinterval $\cI_0$ of $\cI$ of length at most $2\al$. Then 
$\cI\setminus\cI_0$ is the union of at most two other intervals $\cI_+,\cI_-$ 
on which $|g^{(n-1)}(x)|\ge\sfrac{\veps}{\al^{n-1}}$. By the inductive hypothesis
$$\eqalign{
{\rm length}\set{x\in \cI}{|g(x)|\le\veps}
&\le {\rm length}(\cI_0)+\sum_{i=\pm}{\rm length}\set{x\in \cI_i}{|g(x)|\le\veps}\cr
&\le 2\al+2c_{n-1}\al=c_n\al\cr
}$$
\endproof

\proposition{\STM\propLadA}{ Assume that  $F$ is strongly asymmetric.
Let $\Ga$ be an $\veps$--separated set in $F$
and $R$ a square of side length $8\veps$ in $\bbbr^2$. Then
$$
\#\set{(\ga_1,\ga_2)\in\Ga\times\Ga}{\ga_1+\ga_2\in
R}\le\abcst\,\sfrac{\veps^{1/n_0}}{\veps}
$$
with $\abcst$ depending only on the geometry of $F$. Here $n_0$ is the
constant of Definition \defModI.
}
\prf Let $\om_1=\veps^{1-{1\over n_0}}$ and
$$\eqalign{
X_1&=\set{(\ga_1,\ga_2)\in\Ga\times\Ga}{\ga_1+\ga_2\in R,\ 
\min\big\{d(\ga_1,\ga_2),d\big(a(\ga_1),\ga_2\big)\big\}\ge\om_1}\cr
X_2&=\set{(\ga_1,\ga_2)\in\Ga\times\Ga}{\ga_1+\ga_2\in R,\ 
d(\ga_1,\ga_2)\le\om_1}\cr
X_3&=\set{(\ga_1,\ga_2)\in\Ga\times\Ga}{\ga_1+\ga_2\in R,\ 
d\big(a(\ga_1),\ga_2\big)\le\om_1}\cr
}$$
By Lemma \lemSecVIII, part iv, with an arbitrary point $\p$ and $\om_2$
large enough,
$$
\# X_1\le\sfrac{\abcst}{\om_1}=\abcst\,\sfrac{\veps^{1/n_0}}{\veps}
$$
Next observe that, for any given $\ga_1\in\Ga$, the length of 
$\set{\ga_1+\k}{\k\in F}\cap R$ is bounded by $\abcst\,\veps$, so that 
$$
\#\set{\ga_2\in\Ga}{\ga_1+\ga_2\in R}\le \abcst
\EQN\eqnLadAIV$$
If, for some $\ga_1\in\Ga$, there exists $\ga_2\in\Ga$ such that 
$(\ga_1,\ga_2)\in X_2$, then $2\ga_1=\ga_1+\ga_2+(\ga_1-\ga_2)$
lies in the disk $D$ of radius 
$8\veps+\om_1$ centered at the center of $R$. Since 
$$
{\rm length}\,\set{\k\in F}{2\k\in D}\le\abcst\,\om_1
$$
there are at most $\abcst\,\sfrac{\om_1}{\veps}$
choices of $\ga_1\in \Ga$ with $2\ga_1\in D$. By (\eqnLadAIV) this implies
that
$$
\# X_2\le \abcst\,\sfrac{\om_1}{\veps}=\abcst\, \veps^{-1/n_0}
\le\abcst\,\sfrac{\veps^{1/n_0}}{\veps}
$$
since $n_0\ge 2$.
If, for some $\ga_1\in\Ga$, there exists $\ga_2\in\Ga$ such that 
$(\ga_1,\ga_2)\in X_3$, then $\ga_1+a(\ga_1)\in D$. By Lemma \lemLadAI
$$
{\rm length}\,\set{\k\in F}{\k+a(\k)\in D}\le\abcst\,\om_1^{1\over n_0-1}
$$
Consequently
$$
\# X_3\le \abcst\,\sfrac{\om_1^{1\over n_0-1}}{\veps}
=\abcst\, \sfrac{\veps^{1/n_0}}{\veps}
$$
\endproof

\proof{ of Proposition \propOSLadA}
For each sector $s\in\Si_j$ let $\ga_s$ be the center of  $s\cap F$. Then 
$\Ga = \{\ga_s\, \big|\,s\in \Si_j\}$ is a $\sfrac{3}{4}\fl_j$ separated set.
Clearly $\tilde s'_1+\tilde s'_2$ is contained in the disk of radius 
$\abcst'\,\fl_j$ around $\ga_{s'_1} + \ga_{s'_2}$. Therefore
$(\tilde s_1+\tilde s_2)\cap(\tilde s'_1+\tilde s'_2)\ne\emptyset$ only 
if $\ga_{s_1} + \ga_{s_2}$
is contained in the disk of radius $2\abcst'\, \fl_j$ around 
$\ga_{s'_1} + \ga_{s'_2}$. 
So the first part of the Proposition follows directly from 
Proposition \propLadA, applied $\big(\sfrac{4\abcst'}{8}\times\sfrac{4}{3}\big)^2$ times. 
The other two parts are similar.
\endproof

\definition{\STM\defOSLadA}{ 
\Item i) The creation/annihilation index of $z\in\check\cB\dunion(\cB\times\Si)$
is
$$
b(z)=\cases{b & if $z=(k,\si,b)\in\check\cB$\cr
            b & if $z=(x,\si,b,s)\in\cB\times\Si$\cr}
$$
\Item ii)
Let $f\in\check\cF_{4;\Si}$. We say that $f$ is of particle--particle type if 
$$
f({\sst z_1,z_2,z_3,z_4})=0
\quad {\rm unless\ } b(z_1) = b(z_2) = 0,\  b(z_3) = b(z_2) = 1
$$
}

\lemma{\STM\lemOSchannelthree}{
Let $f\in\check\cF_{4;\Si}$ be of particle--particle type. Then,
$$
\v f\tv_{ch,\Si}\le \abcst\,\sfrac{\fl^{\raise2pt\hbox{$\scriptscriptstyle{1/n_0}$}}}{\fl}
\v f\tv_{3,\Si}
$$
with the channel norm $\v \ \cdot\ \tv_{ch,\Si}$ of 
Definition \defOSchannelnorm\ of [FKTo3].
}
\prf It suffices to consider $f\in\check\cF_r(4-r,\Si)$ with $r\le 2$.
As in the proof of Lemma \lemchannelnorm\ of [FKTo3], set
$$
F({\sst\check\eta_1,\cdots,\check\eta_r;s_1,\cdots,s_{4-r}}) 
= 
\smsum_{\de\in \bbbn_0\times\bbbn_0^2} \sfrac{1}{\de!}
\max_{\rD\, {\rm dd-operator} \atop{\rm with\ } \de(\rD) =\de} \TN\rD f\,({\sst\check\eta_1,\cdots,\check\eta_r;(\xi_1,s_1),\cdots,(\xi_{4-r},s_{4-r})})
\TN_{1,\infty}\ t^\de
$$
Then, by Proposition \propOSLadA,
$$\eqalign{
\v f\tv_{ch,\Si} 
&= \sup_{\check\eta_1,\cdots,\check\eta_r \in \check \cB
                              \atop s_{1},\cdots,s_{2-r}\in\Si} 
   \smsum_{s_{3-r},s_{4-r} \in \Si} \
   F({\sst\check\eta_1,\cdots,\check\eta_r;s_1,\cdots,s_{4-r}})\cr 
&\le\abcst\,\sfrac{\fl^{\raise2pt\hbox{$\scriptscriptstyle{1/n_0}$}}}{\fl}
   \sup_{\check\eta_1,\cdots,\check\eta_r \in \check \cB
                              \atop s_{1},\cdots,s_{3-r}\in\Si} 
   \smsum_{s_{4-r} \in \Si} \
   F({\sst\check\eta_1,\cdots,\check\eta_r;s_1,\cdots,s_{4-r}})\cr 
&\le  \abcst\,\sfrac{\fl^{\raise2pt\hbox{$\scriptscriptstyle{1/n_0}$}}}{\fl}
   \sup_{{1\le i_1<\cdots<i_{3-r}\le 4-r \atop s_{i_1},\cdots,s_{i_{3-r}}\in\Si}
\atop \check\eta_1,\cdots,\check\eta_r \in \check \cB} 
\smsum_{s_i \in \Si \ {\rm for}\atop i\ne i_1,\cdots i_{3-r}} \
   F({\sst\check\eta_1,\cdots,\check\eta_r;s_1,\cdots,s_{4-r}})
=\abcst\,\sfrac{\fl^{\raise2pt\hbox{$\scriptscriptstyle{1/n_0}$}}}{\fl}
\v f\tv_{3,\Si}\cr 
}$$

\endproof

\theorem{\STM\theoremOSLadA}{
Let $\Si$ be a sectorization of scale $j\ge 2$ and length
$\sfrac{1}{M^{j-3/2}}\le\fl\le\sfrac{1}{M^{(j-1)/2}}$. 
Let $u({\sst(\xi,s)},{\sst(\xi',s')}),v({\sst(\xi,s)},{\sst(\xi',s')})
\in\cF_0(2;\Si)$ be antisymmetric, spin independent, particle number 
conserving functions whose Fourier transforms  obey
 $|\check u(k)|,|\check v(k)| \le \half |\imath k_0-e(k)|$.
Furthermore, let $X\in\fN_3$ and assume that
$\ \v  u \v_{1,\Si} \le \half X\ $ and
$M^jX_\0\le\min\{\tau_1,\tau_2\}$, where $\tau_1$ and $\tau_2$ are
the constants of Proposition \propOSrealpropbound\ and
Lemma \lemOSdiffpropbound\ of [FKTo3], respectively. Set
$$
C(k) = \sfrac{\nu^{(j)}(k)}{\imath k_0 -e(\k) -\check u(k)}
\qquad,\qquad 
D(k) = \sfrac{\nu^{(\ge j+1)}(k)}{\imath k_0 -e(\k) -\check v(k)}
$$
and let $C(\xi,\xi'),\,D(\xi,\xi')$ be the Fourier transforms of $C(k)$, $D(k)$ as
in Definition \defOSftcov\ of [FKTo2]. Furthermore, let $f \in \check\cF_{4;\Si}$
be of particle--particle type. If the Fermi curve $F$ is strongly asymmetric in the sense of Definition \defModI, then for all $\ell \ge 1$
$$
\v L_\ell(f;C,D)\tv_{3,\Si}
 \le \big( \const \,\fl^{\raise2pt\hbox{$\scriptscriptstyle{1/n_0}$}}\,\fe_j(X)\big)^\ell\   
 \v f\tv_{3,\Si}^{\kern5pt\ell+1}
$$
where $\fe_j(X)=\sfrac{\cb_j}{1-M^jX}$
}
\prf
By Proposition \propOSNaiveLadder, with $X$ replaced by $M^jX$, and Lemma \lemOSchannelthree\ 
$$\eqalign{
\V L_\ell(f;C,D)\tV_{3,\Si} 
&\le  \Big( \const \sfrac{\fl\,\cb_j}{1-M^jX}\Big)^\ell\
\v f\tv_{ch,\Si}^{\kern5pt\ell}\ \v f\tv_{3,\Si}\cr
&\le\big( \const \,\fl^{\raise2pt\hbox{$\scriptscriptstyle{1/n_0}$}}\,\fe_j(X)\big)^\ell\   
 \v f\tv_{3,\Si}^{\kern5pt\ell+1}
}$$
\endproof

\vfill\eject
%=====================================================================
%======================= SPATIAL SECTORS =============================
%=====================================================================

\appendix{\APappSpatialSectors}{Sectors for  $\textfont1=\bigteni\scriptfont0=\bigsevenrm k_0$ Independent Functions }
\PG\pgOSE
In [FKTf1--f3] we shall implement a renormalization algorithm that uses
counterterms for the dispersion relation $e(\k)$ that are independent of 
$k_0$. In this appendix we adjust the discussion of sectorized norms in
\S\CHsectors\ and the discussion of resectorization, following Definition 
\defOSresector, to deal with such functions.

\definition{\STM\defOSzeroext}{Let $f(\x,\x')$ be a translation invariant function 
on $\bbbr^2\times\bbbr^2$ we define its extension $f_{\rm ext}(\xi,\xi')$
by
$$
f_{\rm ext}\big((x_0,\x,\si,a),(x_0',\x',\si',a')\big)
=\de_{\si,\si'}\de(x_0-x_0')\cases{
f(\x,\x')& if $a=1,\ a'=0$ \cr
-f(\x',\x)& if $a=0,\ a'=1$ \cr
0& otherwise \cr
}
$$
and its Fourier transform as
$$
\check f(\k) = \int d^2\x\  e^{-\imath(\k_1\x_1+\k_2\x_2)}\,f(\x,\0)
$$

}
\remark{\STM\remOSzeroext}{ If $\check f_{\rm ext}(k)$ is the Fourier 
transform of $f_{\rm ext}$ as in Definition \defOSfourtrans.i of [FKTo2], then
$$
\check f_{\rm ext}\big((k_0,\k)\big)=\check f(\k)
$$

}

\definition{\STM\defOSzerosectorext}{
Let $\Si$ be a sectorization at scale $j$ and $K\big((\x,s),(\x',s')\big)$ a translation invariant function on $\big(\bbbr^2\times\Si\big)^2$.
\Item i) We define its extension
$\ 
K_{\rm ext}\big((\xi,s),(\xi',s')\big)
\ $
on $(\cB\times\Si)^2$ by
$$
K_{\rm ext}\big((x_0,\x,\si,a,s),(x_0',\x',\si',a',s')\big)
=\de_{\si,\si'}\de(x_0-x_0')\cases{
K({\sst(\x,s),(\x',s')})& if $a=1,\ a'=0$ \cr
-K({\sst(\x',s'),(\x,s)})& if $a=0,\ a'=1$ \cr
0& otherwise \cr
}
$$

\Item ii) The function $K\big((\x,s),(\x',s')\big)$ is said to be sectorized if
its Fourier transform
$$
\int d^2\x\,d^2\x'\  e^{-\imath(\k_1\x_1+\k_2\x_2)}
e^{\imath(\k'_1\x'_1+\k'_2\x'_2)}\,K\big((\x,s),(\x',s')\big)
$$
vanishes unless $(0,\k)\in \tilde s$ and $(0,\k')\in \tilde s'$ where
$\tilde s$ and $\tilde s'$ are the extensions of $s$ and $s'$ of
Definition \defOSsectors.ii of [FKTo3].

\Item iii)  We define $\check K(\k)$ by
$$
\check K(\k)=
\sum_{s,s'\in\Si}\int d^2\x\  e^{-\imath(\k_1\x_1+\k_2\x_2)}
\,K\big((\x,s),(\0,s')\big)
$$

\Item iv) We set
$$
\| K\|_{1,\Si}=\V K_{\rm ext} \V_{1,\Si}
$$ 

}

\remark{\STM\remOSzerosectorext}{
\Item i)
If $\check K_{\rm ext}(k)$ is the Fourier transform of $K_{\rm ext}$ as in Definition \defOSsectrepr.iv, then 
$$
\check K_{\rm ext}\big((k_0,\k)\big)=\check K(\k)
$$
\Item ii) If $K$ is sectorized, then $K\big((\x,s),(\x',s')\big)$
and $K_{\rm ext}\big((x_0,\x,\si,a,s),(x_0',\x',\si',a',s')\big)$
vanish unless $s\cap s'\ne\emptyset$.

\Item iii) 
Suppose that $K$ is sectorized and write 
$\| K\|_{1,\Si}
=\sum_{\de\in\bbbn\times\bbbn^2}\sfrac{1}{\de!}\ka_\de t^\de$. Then $\ka_\de$
vanishes unless $\de_0=0$ and otherwise is given by
$$\eqalign{
\ka_{0,\bde}=\max\Big\{
\max_{s'\in\Si}\ \smsum_{s\in\Si}\int\! d^2\x\ 
\big| \x^{\bde}\,K\big((\x,s),(\0,s')\big)\big|\ ,\ 
\max_{s\in\Si}\ \smsum_{s'\in\Si}\int\! d^2\x\ 
\big| \x^{\bde}\,K\big((\x,s),(\0,s')\big)\big|
\Big\}
}$$ 
and obeys
$$
\max_{s,s'\in\Si}\ \int d^2\x\ 
\big|\x^{\bde}\,K\big((\x,s),(\0,s')\big)\big|
\le\ka_{0,\bde}
\le 3\max_{s,s'\in\Si}\ \int d^2\x\ 
\big| \x^{\bde}\,K\big((\x,s),(\0,s')\big)\big|
$$
}

\lemma{\STM\lemOSsectorext}{
Let $\Si$ be a sectorization of scale 
$j\ge 2$ and length $\sfrac{1}{M^{j-3/2}}\le\fl\le\sfrac{1}{M^{(j-1)/2}}$ and
 $K\big((\x,s),(\x',s')\big)$ be a sectorized, translation invariant function 
on $\big(\bbbr^2\times\Si\big)^2$. Let $\mu(t)$ be a $C_0^\infty$ function on $\bbbr$ and set, for each $\La>0$ 
$$\eqalign{
\mu_\La(k)&=\mu\big(\La^2[k_0^2+e(\k)^2]\big)\cr
(K_{\rm ext}*\hat\mu_\La)\big((\xi,s),(\xi',s')\big)
&=\int_\cB d\ze\ K_{\rm ext}\big((\xi,s),(\ze,s')\big)\hat\mu_\La(\ze,\xi')\cr
(\hat\mu_\La*K_{\rm ext})\big((\xi,s),(\xi',s')\big)
&=\int_\cB d\ze\ K_{\rm ext}\big((\ze,s),(\xi',s')\big)\hat\mu_\La(\ze,\xi)\cr
}$$
where $\hat\mu_\La$ was defined in Definition \defOSfourtransII\ of
[FKTo2]. Denote
$j(\La)=\min\set{i\in\bbbn}{M^i\ge\La}$. Then, there is a constant $\abcst$, depending on $\mu$, but not on $M$, $j$ or $\La$, such that
$$
\V K_{\rm ext}*\hat\mu_\La\V_{1,\Si}\ ,\ 
\V \hat\mu_\La*K_{\rm ext}\V_{1,\Si}
\le\abcst\,\cb_{j(\La)}\ \| K\|_{1,\Si}
$$

}
\noindent
This lemma is an immediate consequence of Lemma \lemOSumu\ of [FKTo3].
\remark{\STM\remOSsectorconv}{ In the notation of Lemma \lemOSsectorext,
$$
(K_{\rm ext}*\hat\mu_\La){\check{\,}}(k)=
(\hat\mu_\La*K_{\rm ext}){\check{\,}}(k)=
\check K(\k)\,\mu_\La(k)
$$

}

As in Definition \defOSresector, we define a resectorization for functions on $\big(\bbbr^2\times\Si\big)^2$. For a function $\chi(k)$ on
$\bbbr\times\bbbr^2$, set, as in Lemma \lemOSmorepartunity\ of [FKTo3],
$$
\chi^o(\x) 
 =\ \int e^{\imath\k\cdot\x}\,\chi(0,\k)\,\sfrac{d^2\k}{(2\pi)^2} 
\ =\ \int dx_0\, \hat\chi\big((x_0,\x,\uparrow,0),(0,\0,\uparrow,0) \big) 
$$
and let 
$$
\hat\chi^o(\xi,\xi')=\de_{\si,\si'}\de_{a,a'}\de(x_0-x'_0)
\chi^o\big((-1)^a(\x-\x'))
$$
Then 
$$\eqalign{
\hat\chi^o\big((x_0,\x,\si,a),(x'_0,\x',\si',a')\big) 
&= \de(x_0-x_0') \int dt\ \hat\chi\big((t,\x,\si,a),(0,\x',\si',a')\big) \cr
&=\hat\chi\big((x_0,\x,\si,a),(x'_0,\x',\si',a')\big) 
}$$

\definition{\STM\defOSindresector}{ Let $j,i\ge 2$. Let $\Si$ and $\Si'$ be  
sectorizations of scale $j$ and $i$,
respectively. If $i\ne j$, define, for each function $K$ on 
$\big( \bbbr^2\times\Si' \big)^2$,
$$
K_\Si\big( (\x,s_1),(\y,s_2) \big)
=\smsum_{s'_1,s'_2 \in \Si} \int  d\x'\, d\y'\,
\chi^o_{s_1}(\x-\x') \, K\big( (\x',s'_1),(\y',s'_2) \big)\,
 \chi^o_{s_2}(\y'-\y) 
$$
where $\chi_s,\ s\in\Si$ is the partition of unity of Lemma 
 \lemOSsectpartunit\ and (\eqnOSpartunit) of [FKTo3]. 
For $i= j$ and $\Si'=\Si$, define $K_{\Si}=K$.
}

\remark{\STM\defOSindresectorI}{
\Item i)
If $K$ is translation invariant, then
$\ \check K_\Si(\k,s_1,s_2)  =\smsum_{s'_1,s'_2 \in \Si}
\check K(\k,s_1',s_2')\, \chi_{s_1}(0,\k)\, \chi_{s_2}(0,\k) \,$.
\Item ii)
The resectorization $K_\Si$ is sectorized.
}

\remark{\STM\defOSindresectorII}{
Let $K({\sst(\x,s),(\x',s')})$ be a translation invariant sectorized function on $\big(\bbbr^2\times\Si'\big)^2$. Then
$$
\big( K_\Si\big)_{\rm ext}({\sst (\xi,s_1),(\eta,s_2)})
=\sum_{s'_1,s'_2 \in \Si \atop {s'_1\cap s_1 \ne \emptyset
        \atop {s'_2\cap s_2 \ne \emptyset}} } 
\int\hskip -2pt {\sst d\xi'\, d\eta'}\,K_{\rm ext}({\sst (\xi',s'_1),(\eta',s'_2)})\,
\hat\chi^o_{s_1}({\sst\xi',\xi}) \, \chi^o_{s_2}({\sst \eta',\eta})
$$
} 
\prf
Let $\xi=(x_0,\x,\si,a),\, \eta=(y_0,\y,\tau,b ) \in\cB$. We consider the case
$a=1, b=0$, the other cases are similar.
Fix any $s_1',s_2' \in \Si'$. If $s'_1\cap s_1 = \emptyset$ or 
$s'_2\cap s_2 = \emptyset$.
$$
\int\hskip -2pt {\sst d\xi'\, d\eta'}\,K_{\rm ext}({\sst (\xi',s'_1),(\eta',s'_2)})\,
\hat\chi^o_{s_1}({\sst\xi',\xi}) \, \hat\chi^o_{s_2}({\sst \eta',\eta})
=0 
= \hskip -4pt  \int \hskip -2pt {\sst d\x'\, d\y'}\,
\chi^o_{s_1}({\sst \x-\x'}) \, K( {\sst(\x',s'_1),(\y',s'_2) })\,
 \chi^o_{s_2}({\sst \y'-\y})  
$$
Otherwise
$$\eqalign{
\int  {\sst d\xi'\, d\eta'}&\,K_{\rm ext}({\sst (\xi',s'_1),(\eta',s'_2)})\,
\hat\chi^o_{s_1}({\sst\xi',\xi}) \, \hat\chi^o_{s_2}({\sst \eta',\eta}) \cr
& = \de_{\si,\tau} \int {\sst dx_0'\,dy_0'\,d\x'\,d\y'}\
\de({\sst x_0'-y_0'})\,\de({\sst x_0'-x_0})\,\de({\sst y_0'-y_0})\,
\chi^o_{s_1}({\sst \x-\x'}) \, K( {\sst(\x',s'_1),(\y',s'_2) })\,
 \chi^o_{s_2}({\sst \y'-\y})  \cr
& = \de_{\si,\tau}\,\de({\sst x_0-y_0}) \int {\sst d\x'\,d\y'}\
\chi^o_{s_1}({\sst \x-\x'}) \, K( {\sst(\x',s'_1),(\y',s'_2) })\,
 \chi^o_{s_2}({\sst \y'-\y})  
}$$
\endproof

\proposition{\STM\propOSindresectorI}{ Let $j>i\ge 2$, $\sfrac{1}{M^{j-3/2}}\le\fl\le\sfrac{1}{M^{(j-1)/2}}$ and 
$\sfrac{1}{M^{i-3/2}}\le\fl'\le\sfrac{1}{M^{(i-1)/2}}$ with $4\fl<\fl'$. 
Let $\Si$ and $\Si'$ be sectorizations of length  $\fl$ at scale $j$ and 
length $\fl'$ at scale $i$, respectively.
\Item i) 
Let $K\big((\x,s),(\x',s')\big)$ a translation invariant sectorized function on $\big(\bbbr^2\times\Si'\big)^2$. Then
$$
\big\| K_\Si \big\|_{1,\Si} \le \abcst\,\cb_{j-1}\,\big\| K \big\|_{1,\Si'}
$$
\Item ii) 
Let $K\big((\x,s),(\x',s')\big)$ a translation invariant sectorized function on $\big(\bbbr^2\times\Si\big)^2$. Then
$$
\big\| K_{\Si'} \big\|_{1,\Si'}
 \le \abcst\,\big[\sfrac{\fl'}{\fl}\big]\,\cb_{i-1}\,\big\| K \big\|_{1,\Si}
$$

}

\prf i)
By Definition \defOSzerosectorext.iv, Remark \defOSindresectorII, 
Lemma \lemOSelloneinfty\ of [FKTo1],
Lemma \lemOSmorepartunity\  and (\eqnOSprodcontrbound) of [FKTo3],
$$\eqalign{
\| K_\Si\|_{1,\Si}
&=\V \big(K_\Si\big)_{\rm ext} \V_{1,\Si}\cr
&\le \abcst\, \max_{s_1,s_2 \in \Si} 
\sum_{s'_1,s'_2 \in \Si \atop {s'_1\cap s_1 \ne \emptyset
        \atop {s'_2\cap s_2 \ne \emptyset}} }
\| K_{\rm ext}({\sst (\,\cdot\,,s'_1),(\,\cdot\,,s'_2)})\|_{1,\infty}
\|\hat\chi^o_{s_1}\|_{1,\infty} \, \| \chi^o_{s_2}\|_{1,\infty} \cr
&\le \abcst\, \max_{s_1,s_2 \in \Si}  \|\chi^o_{s_1}\|_{L_1}\, \|\chi^o_{s_2}\|_{L_1}\,
 \max_{s'_1,s'_2 \in \Si'} 
\| K_{\rm ext}({\sst (\,\cdot\,,s'_1),(\,\cdot\,,s'_2)})\|_{1,\infty} \cr
&\le \abcst\,\cb_{j-1}^2\,\V K_{\rm ext} \V_{1,\Si'} \cr
&\le \abcst\,\cb_{j-1}\,\big\| K \big\|_{1,\Si'} \cr
 }$$
since, for any $s\in \Si$, there are at most three sectors $s'\in \Si'$
with $s'\cap s \ne \emptyset$. 
\Item ii)
Similarly
$$\eqalign{
\| K_{\Si'}\|_{1,{\Si'}}
&=\V \big(K_{\Si'}\big)_{\rm ext} \V_{1,\Si'}\cr
&\le \abcst\, \max_{s'_1,s'_2 \in \Si'} 
\sum_{s_1,s_2 \in \Si \atop {s_1\cap s'_1 \ne \emptyset
        \atop {s_2\cap s'_2 \ne \emptyset}} }
\| K_{\rm ext}({\sst (\,\cdot\,,s_1),(\,\cdot\,,s_2)})\|_{1,\infty}
\|\hat\chi^o_{s'_1}\|_{1,\infty} \, \| \chi^o_{s'_2}\|_{1,\infty} \cr
&\le \abcst\,\max_{s'_1,s'_2 \in \Si'}  \|\chi^o_{s'_1}\|_{L_1}\, \|\chi^o_{s'_2}\|_{L_1}
\sum_{s_1 \in \Si \atop {s_1\cap s'_1 \ne \emptyset} }
\sum_{s_2 \in \Si}
\| K_{\rm ext}({\sst (\,\cdot\,,s_1),(\,\cdot\,,s_2)})\|_{1,\infty} \cr
&\le \abcst\,\cb_{i-1}^2\,\big[\sfrac{\fl'}{\fl}\big]\,
\max_{s_1 \in \Si} \smsum_{s_2 \in \Si}
\| K_{\rm ext}({\sst (\,\cdot\,,s_1),(\,\cdot\,,s_2)})\|_{1,\infty} \cr
&\le \abcst\,\cb_{i-1}\,\big[\sfrac{\fl'}{\fl}\big]\,\big\| K \big\|_{1,\Si} \cr
 }$$
since, for any $s_1'\in \Si'$, there are at most 
$\abcst\big[\sfrac{\fl'}{\fl}\big]$ sectors $s_1\in \Si$ with 
$s_1\cap s'_1 \ne \emptyset$. 
\endproof

\vfill\eject
%=====================================================================
%========================== REFERENCES ===============================
%=====================================================================

\titlea{ References}\PG\pgOSIVref

\item{[FKTa]} J. Feldman, H. Kn\"orrer, E. Trubowitz, 
{\bf Asymmetric Fermi Surfaces for Magnetic Schr\"odinger Operators}, 
      Communications in Partial Differential Equations {\bf 25} (2000),
      319-336.
\smallskip%
\item{[FKTf1]} J. Feldman, H. Kn\"orrer, E. Trubowitz, 
{\bf A Two Dimensional Fermi Liquid, Part 1: Overview}, preprint.
\smallskip%
\item{[FKTf2]} J. Feldman, H. Kn\"orrer, E. Trubowitz, 
{\bf A Two Dimensional Fermi Liquid, Part 2: Convergence}, preprint.
\smallskip%
\item{[FKTf3]} J. Feldman, H. Kn\"orrer, E. Trubowitz, 
{\bf A Two Dimensional Fermi Liquid, Part 3: The Fermi Surface}, preprint.
\smallskip%
\item{[FKTl]} J. Feldman, H. Kn\"orrer, E. Trubowitz, 
{\bf  Particle--Hole Ladders}, preprint.
\smallskip%
\item{[FKTo1]} J. Feldman, H. Kn\"orrer, E. Trubowitz, 
{\bf Single Scale Analysis of Many Fermion Systems, Part 1: Insulators}, preprint.
\smallskip%
\item{[FKTo2]} J. Feldman, H. Kn\"orrer, E. Trubowitz, 
{\bf Single Scale Analysis of Many Fermion Systems, Part 2: The First Scale}, preprint.
\smallskip%
\item{[FKTo3]} J. Feldman, H. Kn\"orrer, E. Trubowitz, 
{\bf Single Scale Analysis of Many Fermion Systems, Part 3: Sectorized Norms}, preprint.
\smallskip%
\item{[FKTr1]} J. Feldman, H. Kn\"orrer, E. Trubowitz, 
{\bf Convergence of Perturbation Expansions in Fermionic Models, Part 1: Nonperturbative Bounds}, preprint.
\smallskip%
\item{[FKTr2]} J. Feldman, H. Kn\"orrer, E. Trubowitz, 
{\bf Convergence of Perturbation Expansions in Fermionic Models, Part 2: Overlapping Loops}, preprint.
\smallskip%
\item{[FMRT]} J.Feldman, J. Magnen, V. Rivasseau, E. Trubowitz, 
{\bf Two Dimensional Many Fermion Systems as Vector Models},
Europhysics Letters, {\bf 24} (1993) 521-526.
\smallskip%
\item{[G]} M. Gromov, {\it Asymptotic Invariants of Infinite Groups}.
\smallskip%

\vfill\eject
%=====================================================================
%========================== NOTATION   ===============================
%=====================================================================

\vsize = 9.7truein
\hoffset=-0.2in
\voffset=-0.2in
\titlea{Notation }\PG\pgOSIVnot
\null\vskip-0.9in
\titleb{Norms}
\centerline{
\vbox{\offinterlineskip
\hrule
\halign{\vrule#&
         \strut\hskip0.05in\hfil#\hfil&
         \hskip0.05in\vrule#\hskip0.05in&
          #\hfil\hfil&
         \hskip0.05in\vrule#\hskip0.05in&
          #\hfil\hfil&
           \hskip0.05in\vrule#\cr
height2pt&\omit&&\omit&&\omit&\cr
&Norm&&Characteristics&&Reference&\cr
height2pt&\omit&&\omit&&\omit&\cr
\noalign{\hrule}
height2pt&\omit&&\omit&&\omit&\cr
&$\tn\ \cdot\ \tn_{1,\infty}$&&no derivatives, external positions, acts on functions&&Example \exOSSymmNorm&\cr
height4pt&\omit&&\omit&&\omit&\cr
&$\|\ \cdot\ \|_{1,\infty}$&&derivatives, external positions, acts on functions&&Example \exOSSymmNorm&\cr
height4pt&\omit&&\omit&&\omit&\cr
&$\|\ \cdot\ \cnorm_\infty$&&derivatives, external momenta, acts on functions
&&Definition \defOSderivmom&\cr
height4pt&\omit&&\omit&&\omit&\cr
&$\tn\ \cdot\ \tn_{\infty}$&&no derivatives, external positions, acts on functions&&Example \exOSelloneinftycontr&\cr
height4pt&\omit&&\omit&&\omit&\cr
&$\|\ \cdot\ \cnorm_1$&&derivatives, external momenta, acts on functions
&&Definition \defOSderivmom&\cr
height4pt&\omit&&\omit&&\omit&\cr
&$\|\ \cdot\ \cnorm_{\infty,B}$&&derivatives, external momenta, $B\subset\bbbr\times\bbbr^d$
&&Definition \defOSderivmom&\cr
height4pt&\omit&&\omit&&\omit&\cr
&$\|\ \cdot\ \cnorm_{1,B}$&&derivatives, external momenta, $B\subset\bbbr\times\bbbr^d$
&&Definition \defOSderivmom&\cr
height4pt&\omit&&\omit&&\omit&\cr
&$\|\ \cdot\ \|$&&$\rho_{m;n}\|\ \cdot\ \|_{1,\infty}$&&Lemma \lemOSscalednorm&\cr
height4pt&\omit&&\omit&&\omit&\cr
&$N(\cW;\cb,\ib,\al)$&&$\sfrac{1}{\ib^2}\,\cb\!\sum_{m,n\ge 0}\,
\al^{n}\,\ib^{n} \,\|\cW_{m,n}\|$&&Definition \defOSgrnorm&\cr
height4pt&\omit&&\omit&&\omit&\cr
& && &&Theorem  \thmOSinsulators&\cr
height4pt&\omit&&\omit&&\omit&\cr
&$N_0(\cW;\be;X,\vec\rho)$&&$\fe_0(X)\ \sum_{m+n\in 2\bbbn}\,
\be^{n}\rho_{m;n} \,\|\cW_{m,n}\|_{1,\infty}$
&&Theorem  \thmOSfirststep&\cr
height4pt&\omit&&\omit&&\omit&\cr
&$\|\ \cdot\ \|_{L^1}$&&derivatives, acts on functions on $\bbbr\times\bbbr^d$
&&before Lemma \lemOSprepintup&\cr
height4pt&\omit&&\omit&&\omit&\cr
&$\|\ \cdot\ \tnorm$&&derivatives, external momenta, acts on functions
&&Definition \defOSdiffdecaynorm&\cr
height4pt&\omit&&\omit&&\omit&\cr
&$N^\sim_0(\cW ;\be;X,\vec\rho)$&&$\fe_0(X)
\!\sum_{m+n\in 2\bbbn}\,\be^{m+n}\rho_{m;n} \,\| W^\sim_{m,n}\tnorm$
&&before Lemma \lemOSTZsourceterm&\cr
height4pt&\omit&&\omit&&\omit&\cr
&$\v \ \cdot\  \tv$&&like $\rho_{m;n}\|\ \cdot\ \tnorm$ but acts on $\tilde V^{\otimes n}$
&&Theorem  \thmOSTfirststep&\cr
height4pt&\omit&&\omit&&\omit&\cr
&$N^\sim(\cW ;\cb,\ib, \al)$&&$\sfrac{1}{\ib^2}\cb\,\smsum_{m,n} 
\al^{m+n}\, \ib^{m+n}\, \v W^\sim_{m,n} \tv$
&&Theorem  \thmOSTfirststep&\cr
height4pt&\omit&&\omit&&\omit&\cr
&$\v \ \cdot\ \v_{p,\Si}$&&derivatives, external positions, 
all but $p$ sectors summed
&&Definition \defOSsectnorm&\cr
height4pt&\omit&&\omit&&\omit&\cr
&$\|\ \cdot\ \|_{1,\Si}$&& like $\v \ \cdot\ \v_{1,\Si}$, but for functions
on $\big(\bbbr^2\times\Si\big)^2$
&&Definition \defOSzerosectorext&\cr
height4pt&\omit&&\omit&&\omit&\cr
&$\v \varphi \v_{\Si}$&&$\rho_{m;n}\cases{
\v \varphi \v_{1,\Si} + \sfrac{1}{\fl}\,\v \varphi \v_{3,\Si}
+ \sfrac{1}{\fl^2}\,\v \varphi \v_{5,\Si} 
    & if $m=0$ \cr
\sfrac{\fl}{M^{2j}}\,\v \varphi \v_{1,\Si} & if $m\ne0$} $
&&Definition \defOSscalednorms&\cr
height4pt&\omit&&\omit&&\omit&\cr
&$N_j(w;\al;\,X,\Si,\vec\rho)$&&$\sfrac{M^{2j}}{\fl}\,\fe_j(X) 
\smsum_{m,n\ge 0}\,
\al^{n}\,\big(\sfrac{\fl\,\IB}{M^j}\big)^{n/2} \,\v w_{m,n}\v_\Si$
&&Definition \defOSscalednorms&\cr
height4pt&\omit&&\omit&&\omit&\cr
&$\v \ \cdot\ \tv_{p,\Si}$&&derivatives, external momenta, all but $p$ sectors
summed
&&Definition \defOSsectdiffdecaynorm&\cr
height4pt&\omit&&\omit&&\omit&\cr
&$\v \ \cdot\ \tv_{p,\Si,\vec\rho}$&&weighted variant of $\v \ \cdot\ \tv_{p,\Si}$
&&Definition \defOSmomscalednorms.i&\cr
height4pt&\omit&&\omit&&\omit&\cr
&$\v f \tv_\Si$&&$\rho_{m;n}\cases{
\v f\tv_{1,\Si}+\sfrac{1}{\fl}\,\v f\tv_{3,\Si}
              +\sfrac{1}{\fl^2}\,\v f \tv_{5,\Si} 
    & if $m=0$ \cr
\smsum_{p=1}^6\sfrac{1}{\fl^{[(p-1)/2]}}\v f\tv_{p,\Si}
  & if $m\ne0$} $
&&Definition \defOSmomscalednorms.ii&\cr
height4pt&\omit&&\omit&&\omit&\cr
&$N_j^\sim(w;\al;\,X,\Si,\vec\rho)$&&$\sfrac{M^{2j}}{\fl}\,\fe_j(X) 
\smsum_{n\ge 0}\,
\al^{n}\,\big(\sfrac{\fl\,\IB}{M^j}\big)^{n/2} \,
\v f_n\tv_\Si$
&&Definition \defOSmomscalednorms.iii&\cr
height4pt&\omit&&\omit&&\omit&\cr
&$\v\ \cdot\ \tv_{ch,\Si}$&&channel variant of $\v\ \cdot\ \tv_{2,\Si}$
for ladders
&&Definition \defOSchannelnorm&\cr
height4pt&\omit&&\omit&&\omit&\cr
&$\v\ \cdot\ \v_{ch,\Si}$&&channel variant of $\v\ \cdot\ \v_{2,\Si}$
for ladders
&&Definition \defOSchannelnorm&\cr
height4pt&\omit&&\omit&&\omit&\cr
&$\v\ \cdot\ \v_{1,\Si,\om}$&&like $\v\ \cdot\ \v_{1,\Si}$
but excludes almost degenerate sectors
&&Lemma \lemSecXI&\cr
height4pt&\omit&&\omit&&\omit&\cr
&$\v\ \cdot\ \tv_{1,\Si,\om}$&&like $\v\ \cdot\ \tv_{1,\Si}$
but excludes almost degenerate sectors
&&Lemma \lemSecXI&\cr
height2pt&\omit&&\omit&&\omit&\cr
}\hrule}}

\vfil
\goodbreak
\titleb{Other Notation}
\null\vskip-0.3in
\centerline{
\vbox{\offinterlineskip
\hrule
\halign{\vrule#&
         \strut\hskip0.05in\hfil#\hfil&
         \hskip0.05in\vrule#\hskip0.05in&
          #\hfil\hfil&
         \hskip0.05in\vrule#\hskip0.05in&
          #\hfil\hfil&
           \hskip0.05in\vrule#\cr
height2pt&\omit&&\omit&&\omit&\cr
&Not'n&&Description&&Reference&\cr
height2pt&\omit&&\omit&&\omit&\cr
\noalign{\hrule}
height2pt&\omit&&\omit&&\omit&\cr
&$\Om_S(\cW)(\phi,\psi)$
&&$\log\sfrac{1}{Z} \int  e^{\cW(\phi,\psi+\ze)}\,d\mu_{S}(\ze)$
&&before (\eqnOSintroI)&\cr
height2pt&\omit&&\omit&&\omit&\cr
&$J$&&particle/hole swap operator&&(\eqnOSjdef)&\cr
height2pt&\omit&&\omit&&\omit&\cr
&$\tilde \Om_C(\cW)(\phi,\psi)$
&&$\log \sfrac{1}{Z}\int e^{\phi J\ze}\,e^{\cW(\phi,\psi +\ze)} d\mu_C(\ze)$
&&Definition \defOSrengrpmap&\cr
height2pt&\omit&&\omit&&\omit&\cr
&$r_0$&&number of $k_0$ derivatives tracked&&\S\CHintroII&\cr
height2pt&\omit&&\omit&&\omit&\cr
&$r$&&number of $\k$ derivatives tracked&&\S\CHintroII&\cr
height2pt&\omit&&\omit&&\omit&\cr
&$M$&&scale parameter, $M>1$&&before Definition \defOSscales&\cr
height2pt&\omit&&\omit&&\omit&\cr
&$\const$&&generic constant, independent of scale&& &\cr
height2pt&\omit&&\omit&&\omit&\cr
&$\abcst$&&generic constant, independent of scale and $M$&& &\cr
height2pt&\omit&&\omit&&\omit&\cr
&$\nu^{(j)}(k)$&&$j^{\rm th}$ scale function&&Definition \defOSscales&\cr
height2pt&\omit&&\omit&&\omit&\cr
&$\tilde\nu^{(j)}(k)$&&$j^{\rm th}$ extended scale function
&&Definition \defOSextendedshell.i&\cr
height2pt&\omit&&\omit&&\omit&\cr
&$\nu^{(\ge j)}(k)$&&$\varphi\big(M^{2j-1}(k_0^2+e(\k)^2)\big)$&&Definition \defOSscales&\cr
height2pt&\omit&&\omit&&\omit&\cr
&$\tilde\nu^{(\ge j)}(k)$
&&$\varphi\big(M^{2j-2}(k_0^2+e(\k)^2)\big)$
&&Definition \defOSextendedshell.ii&\cr
height2pt&\omit&&\omit&&\omit&\cr
&$\bar\nu^{(\ge j)}(k)$
&&$\varphi\big(M^{2j-3}(k_0^2+e(\k)^2)\big)$
&&Definition \defOSextendedshell.iii&\cr
height2pt&\omit&&\omit&&\omit&\cr
&$n_0$&&degree of asymmetry&&Definition \defModI&\cr
height2pt&\omit&&\omit&&\omit&\cr
&$\fl$&&length of sectors&&Definition \defOSsectors&\cr
height2pt&\omit&&\omit&&\omit&\cr
&$\Si$&&sectorization &&Definition \defOSsectors&\cr
height2pt&\omit&&\omit&&\omit&\cr
&$S(C)$&&$\sup_m\sup_{\xi_1,\cdots,\xi_m \in \cB}\
\Big(\ \Big| \int \psi(\xi_1)\cdots\psi(\xi_m)\,d\mu_C(\psi) \Big|\ \Big)^{1/m}$&&Definition \defIntBndsS&\cr
height2pt&\omit&&\omit&&\omit&\cr
&$\IB$&&$j$--independent constant&&Definitions \defOSscalednorms,\defOSmomscalednorms&\cr
height2pt&\omit&&\omit&&\omit&\cr
&$\cb_j$&& $
=\sum_{|\bde|\le r\atop |\de_0|\le r_0}  M^{j|\de|}\,t^\de
+\sum_{|\bde|> r\atop {\rm or\ }|\de_0|> r_0}\infty\, t^\de
\in\fN_{d+1}
$&&Definition \defOScbj&\cr
height2pt&\omit&&\omit&&\omit&\cr
&$\fe_j(X)$&& $= \sfrac{\cb_j}{1-M^j X}$&&Definition \defOSscalednorms.ii&\cr
height2pt&\omit&&\omit&&\omit&\cr
&$f_{\rm ext}$&& extends $f(\x,\x')$ to 
$f_{\rm ext}\big((x_0,\x,\si,a),(x_0',\x',\si',a')\big)$&&
Definition \defOSzeroext&\cr
height2pt&\omit&&\omit&&\omit&\cr
&$*$&& convolution&&before (\eqnOSexpandc)&\cr
height2pt&\omit&&\omit&&\omit&\cr
&$\circ$&& ladder convolution&&Definition \defOSbubbleprop.iv&\cr
height2pt&\omit&&\omit&&\omit&\cr
&$\bullet$&& ladder convolution&&Definitions \defOSsectbubbleprop,\defOSsectbubblepropII&\cr
height2pt&\omit&&\omit&&\omit&\cr
&$\check f$&&Fourier transform&&Definition \defOSfourtrans.i&\cr
height2pt&\omit&&\omit&&\omit&\cr
&$\check u$&&Fourier transform for sectorized $u$&&Definition \defOSsectrepr.iv&\cr
height2pt&\omit&&\omit&&\omit&\cr
&$f^\sim$&&partial Fourier transform&&Definition \defOSfourtrans.ii&\cr
height2pt&\omit&&\omit&&\omit&\cr
&$\hat\chi$&&Fourier transform&&Definition \defOSfourtransII&\cr
height2pt&\omit&&\omit&&\omit&\cr
&$\cB$&&$\bbbr \times \bbbr^d \times \{\uparrow, \downarrow\}\times\{0,1\}$ 
viewed as position space&&beginning of \S\CHnorms&\cr
height2pt&\omit&&\omit&&\omit&\cr
&$\check \cB$&&$\bbbr\times\bbbr^d\times\{\uparrow, \downarrow\}\times\{0,1\}$ 
viewed as momentum space&&beginning of \S \CHfourier&\cr
height2pt&\omit&&\omit&&\omit&\cr
&$\check \cB_m$&&$\set{(\check \eta_1,\cdots,\check \eta_m)\in \check \cB^m}
{\check \eta_1+\cdots+\check \eta_m=0}$&&before Definition \defOSamptransinv&\cr
height2pt&\omit&&\omit&&\omit&\cr
&$\fX_\Si$&&$\check \cB\dunion(\cB\times\Si)$&&Definition \defOSdisjointfield&\cr
height2pt&\omit&&\omit&&\omit&\cr
&$\cF_m(n)$&&functions on $\cB^m \times \cB^n$, antisymmetric in $\cB^m$
arguments&&Definition \defOSFmn&\cr
height2pt&\omit&&\omit&&\omit&\cr
&$\check\cF_m(n)$&&functions on $\check\cB^m \times \cB^n$, antisymmetric in $\check\cB^m$
arguments&&Definition \defOScheckcF&\cr
height2pt&\omit&&\omit&&\omit&\cr
&$\cF_m(n;\Si)$&&functions on $\cB^m \times  \big( \cB \times\Si \big)^n$,
internal momenta in sectors&&Definition \defOSsectrepr.ii&\cr
height2pt&\omit&&\omit&&\omit&\cr
&$\check\cF_m(n;\Si)$&&functions on $\check\cB^m \times  \big( \cB \times\Si \big)^n$,
internal momenta in sectors&&Definition \defOSsectcheckcF.i&\cr
height2pt&\omit&&\omit&&\omit&\cr
&$\check \cF_{n;\Si}$&&functions on $\fX_\Si^n$ that reorder to 
$\check \cF_{m}(n-m;\Si)$'s&&Definition \defOSsectcheckcF.iii&\cr
height4pt&\omit&&\omit&&\omit&\cr
}\hrule}}

\end

%% file: jfomacros.tex
%%%
%%%   jfmacros.tex
%%%   Version of 12/08/02
%%%   Joel Feldman  feldman@math.ubc.ca
%%%

\def\ifundefined#1{\expandafter\ifx\csname#1\endcsname\relax}
\ifundefined{ftmagnification}  \def\ftmagnification{1200} \fi
\ifundefined{spacingNumerator}  \def\spacingNumerator{5} \fi
\ifundefined{spacingDenominator}  \def\spacingDenominator{4} \fi

%=====================================================================
%======================= PRIMARY MACROS ==============================
%=====================================================================

%%%%%%% VARIOUS SETTINGS
\magnification\ftmagnification
\tolerance=10000
\hsize=17truecm\vsize=23truecm

\parindent=40pt
\mathsurround=0pt
     \multiply\baselineskip by \spacingNumerator
     \divide \baselineskip by \spacingDenominator 

%
%     HEADERS
%
\def\today{\ifcase\month\or January\or February\or March\or April\or
     May\or June\or July\or August\or September\or October\or November\or
     December\fi\space\number\day, \number\year}
%
%     STYLE SHORT FORMS
%
\def\dst{\displaystyle}
\def\sst{\scriptstyle}
\def\tst{\textstyle}
\def\ssst{\scriptscriptstyle}
%
%       EQUATIONS
%     
\def\frac#1#2{\dst {#1\over#2}}     % fractions in displaystyle
\def\sfrac#1#2{{\tst{#1\over#2}}}   % fractions in textstyle    

\def\deqalign#1{\vcenter{\openup1\jot \mathsurround=0pt \ialign{
                \strut\hfil$\displaystyle{##}$&&$\displaystyle{{}##}$\hfil
                \crcr
                #1\crcr}}}         %  double eqalign 

\def\meqalign#1{\vcenter{\openup1\jot \mathsurround=0pt \ialign{
                &\strut\hfil$\displaystyle{##}$&$\displaystyle{{}##}$\hfil&
                \quad$##$\crcr
                #1\crcr}}}         %  multiple eqalign

%
%      GREEK LETTERS
%
\def\al{\alpha}
\def\be{\beta}
\def\ga{\gamma}
\def\de{\delta}
\def\ep{\epsilon}
\def\ze{\zeta}
\def\et{\eta}

\def\ka{\kappa}
\def\la{\lambda}

\def\si{\sigma}

\def\om{\omega}
\def\Ga{\Gamma}

\def\Th{\Theta}
\def\La{\Lambda}
\def\Si{\Sigma}

\def\Om{\Omega}   
%
%        BOLD FACE, SCRIPT and ROMAN CHARACTERS
%
\def\pmb#1{\setbox0=\hbox{#1}       % generate bold face
     \kern-.025em\copy0\kern-\wd0
     \kern.05em\copy0\kern-\wd0
     \kern-.025em\box0}             %Knuth puts in \raise.0433em before box0 
\def\0{{\bf 0}}

\def\k{{\bf k}}

\def\q{{\bf q}}
\def\x{{\bf x}}
\def\y{{\bf y}}

\def\p{{\bf p}}

\def\cB{{\cal B}}

\def\cF{{\cal F}}

\def\cI{{\cal I}}
\def\cJ{{\cal J}}

%
%      FONT FAMILIES
%
\font\tenfrak                 = eufm10
\font\sevenfrak               = eufm7
\font\fivefrak                = eufb5
\newfam\frakfam
     \textfont\frakfam=\tenfrak
     \scriptfont\frakfam=\sevenfrak   
     \scriptscriptfont\frakfam=\fivefrak
\def\frak{\fam\frakfam\tenfrak}
\font \tensans                = cmss10
\font \fivesans               = cmss10 at 5pt
\font \sevensans              = cmss10 at 7pt
\newfam\sansfam
     \textfont\sansfam=\tensans
     \scriptfont\sansfam=\sevensans
     \scriptscriptfont\sansfam=\fivesans
\def\sans{\fam\sansfam\tensans}
%
%      Blackboard characters
%
\def\bbbr{{\rm I\!R}}  
\def\bbbn{{\rm I\!N}}

\def\bbbc{{\mathchoice {\setbox0=\hbox{$\displaystyle\rm C$}\hbox{\hbox %\bbbc
to0pt{\kern0.4\wd0\vrule height0.9\ht0\hss}\box0}}
{\setbox0=\hbox{$\textstyle\rm C$}\hbox{\hbox
to0pt{\kern0.4\wd0\vrule height0.9\ht0\hss}\box0}}
{\setbox0=\hbox{$\scriptstyle\rm C$}\hbox{\hbox
to0pt{\kern0.4\wd0\vrule height0.9\ht0\hss}\box0}}
{\setbox0=\hbox{$\scriptscriptstyle\rm C$}\hbox{\hbox
to0pt{\kern0.4\wd0\vrule height0.9\ht0\hss}\box0}}}}
\def\bbbq{{\mathchoice {\setbox0=\hbox{$\displaystyle\rm               %\bbbq
Q$}\hbox{\raise
0.15\ht0\hbox to0pt{\kern0.4\wd0\vrule height0.8\ht0\hss}\box0}}
{\setbox0=\hbox{$\textstyle\rm Q$}\hbox{\raise
0.15\ht0\hbox to0pt{\kern0.4\wd0\vrule height0.8\ht0\hss}\box0}}
{\setbox0=\hbox{$\scriptstyle\rm Q$}\hbox{\raise
0.15\ht0\hbox to0pt{\kern0.4\wd0\vrule height0.7\ht0\hss}\box0}}
{\setbox0=\hbox{$\scriptscriptstyle\rm Q$}\hbox{\raise
0.15\ht0\hbox to0pt{\kern0.4\wd0\vrule height0.7\ht0\hss}\box0}}}}
\def\bbbz{{\mathchoice {\hbox{$\sans\textstyle Z\kern-0.4em Z$}}       %\bbbz
{\hbox{$\sans\textstyle Z\kern-0.4em Z$}}
{\hbox{$\sans\scriptstyle Z\kern-0.3em Z$}}
{\hbox{$\sans\scriptscriptstyle Z\kern-0.2em Z$}}}}
%
%      SYMBOLS
%
\def\const{{\rm const}\,}

\def\half{\sfrac{1}{2}}

\def\optbar#1{\vbox{\ialign{##\crcr\hfil${\scriptscriptstyle(}\mkern -1mu
         \vrule height 1.2pt width 3pt depth -.8pt
         {\scriptscriptstyle)}$\hfil\crcr
          \noalign{\kern-1pt\nointerlineskip}$\hfil\displaystyle{#1}\hfil$\crcr}}}
\def\<{\left<}
\def\>{\right>}

\def\smprod{\mathop{\textstyle\prod}}
\def\smsum{\mathop{\textstyle\sum}}
\def\set#1#2{\big\{ \ #1\ \big|\ #2\ \big\}}
\def\eval#1{\big|\lower4pt\hbox{$\displaystyle\sst #1$}}
%
%       SECTION TITLES
%
\font \tafontt                = cmbx10 scaled\magstep2
\font \tbfontt                = cmbx10 scaled\magstep1
\def\titlea#1{\centerline{\tafontt #1 }\vskip.5truein}
\def\titleb#1{\removelastskip\vskip.3truein%
\noindent{\tbfontt #1 }\vskip.25truein}
\def\titlec#1{\removelastskip\vskip.15truein\noindent{\bf #1 }\vskip.1truein}

%
%          THEOREM etc.
%
\def\newenvironment#1#2#3#4{\long\def#1##1##2{%
\removelastskip\penalty-100\vskip\baselineskip%
\noindent{#3#2\if!##1!.\else\unskip\ \ignorespaces
##1\unskip\fi\ }{#4\ignorespaces##2\vskip\baselineskip}}}
\newenvironment\lemma{Lemma}{\bf}{\it}
\newenvironment\proposition{Proposition}{\bf}{\it}
\newenvironment\theorem{Theorem}{\bf}{\it}
\newenvironment\corollary{Corollary}{\bf}{\it}
\newenvironment\example{Example}{\bf}{\rm}
\newenvironment\problem{Problem}{\bf}{\rm}
\newenvironment\definition{Definition}{\bf}{\rm}
\newenvironment\remark{Remark}{\bf}{\rm}
\newenvironment\hypothesis{Hypothesis}{\bf}{\it}
\newenvironment\convention{Convention}{\bf}{\it}

\def\Item{\vskip.1in\noindent}

%
%         PROOF, QED  
%
\long\def\proof#1{\removelastskip\penalty-100\vskip\baselineskip\noindent{\bf
            Proof\if!#1!\else\ \ignorespaces#1\fi:\ }\ \ \ignorespaces}
\long\def\prf{\removelastskip\penalty-100\vskip\baselineskip\noindent{\bf
            Proof:\ }\ \ \ignorespaces}
\def\endproof{\hfill\vrule height .6em width .6em depth 0pt\goodbreak\vskip.25in }

%===========================================================================
%===================== Automatic numbering (Frozen) ========================
%===========================================================================
\ifundefined{warnForwardRef}  \def\warnForwardRef{n} \fi
\newcount\chapno
\newcount\sectno
\newcount\equano
\newcount\theono
\newcount\probno

\def\IgNoRe#1{}

\chapno=0
\sectno=0
\equano=0
\theono=0
\probno=0
\def\eqhead{}
\def\frefwarning{\if\warnForwardRef y\immediate\write16{   Forward reference on line \the\inputlineno}\fi}
\def\qqqrefwarning{\immediate\write16{   ??? reference on line \the\inputlineno}}

\def\chap#1{\equano=0\sectno=0\theono=0\probno=0\global\advance\chapno by 1%
\def\eqhead{\ifcase\chapno\or I\or II\or III\or IV\or V\or VI\or VII\or
VIII\or IX\or X\or XI\or XII\or XIII\or XIV\or XV\or XVI\or XVII\or XVIII\or
XIX\or XX\or XXI\or XXII\or XXIII\or XXIV\or XXV\or XXVI\or XXVII\or XXVIII\or XXIX\or XXX\or XXXI\or XXXII\or XXXIII\or XXXIV\or XXXV\or XXXVI\or XXXVII\or XXXVIII\or XXXIX\fi.}%
\titlea{\eqhead \hglue 5pt #1}%
}

\def\sect#1{\global\advance\sectno by 1%
\titleb{\eqhead\number\sectno  \hglue 5pt #1}%
}%

\def\appendix#1#2{\equano=0\sectno=0\theono=0\probno=0\def\eqhead{#1.}
\titlea{Appendix #1: #2}%
}

\def\:#1{\def\temp{\expandafter\IgNoRe\string#1}%
\expandafter\ifx\csname\temp\endcsname\relax%
\expandafter\gdef#1{\qqqrefwarning ???}\fi#1}

\def\Eqn{{\hbox{\global\advance\equano by 1}}%
\eqno ({\rm \eqhead\number\equano})}%

\def\Eqno{{\hbox{\global\advance\equano by 1}}%
 ({\rm \eqhead\number\equano})}%

\def\EQN#1{\Eqn\edef\Zwi{\eqhead\number\equano}%
\global\let #1=\Zwi
}

\def\EQNO#1{\Eqno\edef\Zwi{\eqhead\number\equano}%
\global\let #1=\Zwi
}

\def\STM#1{{\global\advance \theono by 1}% 
\eqhead\number\theono
\edef\Zwi{\eqhead\number\theono }
\global\let#1=\Zwi
}

\def\PRB#1{{\global\advance \probno by 1}% 
\eqhead\number\probno
\edef\Zwi{\eqhead\number\probno }
\global\let#1=\Zwi
}

\def\PG#1{\def\Zwi{\number\pageno }
\global\let#1=\Zwi
}

\def\Stm{{\global\advance \theono by 1}% 
\eqhead\number\theono
}

\def\Prb{{\global\advance \probno by 1}% 
\eqhead\number\probno
}

\def\EDEF#1#2{%   Export definition to .txs file
\def\tEmP{#1}\expandafter\gdef\tEmP{#2}
}

%%%%%%%%%%%%%%%%%%%%%%%%%%%%%%%%%%%%%%%%%%%%%%%%%%%%%%%%%%%%%%%%%%%%%%%%%%%%%%%
%%%%%%%%%%%%%   Macros for figure insertion
%%%%%%%%%%%%%%%%%%%%%%%%%%%%%%%%%%%%%%%%%%%%%%%%%%%%%%%%%%%%%%%%%%%%%%%%%%%%%%%
%%%%%%%
%%%%%%%  The two main figure insertion macros are
%%%%%%%
%                \figput{<filename w/o extension>}
%                \figplace{<filename w/o extension>}{<hor shift>}{<vert shift>}
%%%%%%%
%%%%%%%  The first just inserts the figure at the current location. The
%%%%%%%  second inserts the figure at the current location but then shifts 
%%%%%%%  horizontally by the second argument and vertically by the third.
%%%%%%%
%%%%%%%  Some typical TeX commands for inserting figures are
%%%%%%%      \centerline{\figput{<filename w/o extension>}}
%%%%%%%      \vadjust{\centerline{\figput{<filename w/o extension>}}}
%%%%%%%      \midinsert\centerline{\figput{<filename w/o extension>}}\endinsert
%%%%%%%      \topinsert\centerline{\figput{<filename w/o extension>}}\endinsert

%%%%%%%
%%%%%%%   TO SET A FIGURE DIRECTORY INSERT, FOR EXAMPLE,
%%%%%%%                 \def\figdir{figures/}
%%%%%%%   IN YOUR SOURCE FILE. REMEMBER THE TAILING /
%%%%%%%

%%%%%%%
%%%%%%%     SELECT (a) YOUR POSTSCRIPT FILE SUFFIX AND (b) YOUR SYSTEM  NOW!
%%%%%%%
\def\suffix{ps}
\newcount\system
%\global\system=1   % for textures 
%\global\system=2   % for msdos
\global\system=3   % for unix(dvips)
%\global\system=4   % for unix(dvips) scaled by a factor of 1.2
%\global\system=6   % for xdvik

\def\ifundefined#1{\expandafter\ifx\csname#1\endcsname\relax}
\ifundefined{figdir}\def\figdir{}\fi
%
% Now for the definitions and main macro for figure inclusion.
%
\newcount\firstline
\newdimen\pswidth  \newdimen\xleft
\newdimen\psheight \newdimen\ytop \newdimen\ybot
\newcount\justx \newcount\justy
\global\justx=0 \global\justy=0
\newdimen\vpos \newtoks\labeL 
\newread\labeLfile \newdimen\xcoord \newdimen\ycoord
\newif\ifdoit 
\newbox\labox
%  variables for use with xdvik
\newdimen\xdvikwid 
\newdimen\xdvikht
\newdimen\pspoints
\newdimen\rwi
\pspoints=1bp
\newcount\temp
\def\readdim#1{\global\read\labeLfile to \temp
\global #1=\temp pt}
%
% 
%    figcrop{<filename,w/o extension>} treats the first two labels as marking
%    the upper left and lower right corners of the figure. This is for
%    positioning purposes only. The figure may extend beyond the corners.
%    The corner markers are not printed.
%
%
\def\figcrop#1{\par%  #1=filename
\openin\labeLfile=\figdir#1.lbl                                              
\global\read\labeLfile to\firstline\message{#1}               
\global\read\labeLfile to\temp%read overall dimensions                                     
\readdim{\ybot}
\readdim{\xleft}%               read upper left point
\readdim{\ytop}
\global\read\labeLfile to\justx%ignore
\global\read\labeLfile to\justy%ignore
\global\read\labeLfile to\labeL%ignore
\readdim{\pswidth}%            read lower right point
\global\advance\pswidth by -\xleft
\readdim{\psheight}
\global\advance\ybot by -\psheight
\global\advance\psheight by -\ytop
\global\read\labeLfile to\justx%ignore
\global\read\labeLfile to\justy%ignore
\global\read\labeLfile to\labeL%ignore                                    
\vbox to\psheight{\vfill
%%%
%%% NOTE: next line may have to be changed for your DVIPS driver %%%
\ifnum\system=1
\fi %textures
\ifnum\system=2
\fi %msdos
\ifnum\system=3
  %%  \special{" grestore newpath gsave}
                                                 \fi         %%unix:dvips
\ifnum\system=4
\fi         %%unix:dvips,scaled
\ifnum\system=1
\hbox to \pswidth{\kern-\xleft\special{postscriptfile \figdir#1.\suffix }\hfil}\fi
                                                              %textures
\ifnum\system=2
\hbox to \pswidth{\kern-\xleft\special{ps: plotfile \figdir#1.\suffix }\hfil}\fi
                                                              %mdos 
\ifnum\system=3
\hbox to \pswidth{\kern-\xleft\includegraphics{\figdir#1.\suffix}\hfil}\fi
                                                             %unix:dvips 
\ifnum\system=4
\hbox to \pswidth{\kern-\xleft\includegraphics{\figdir#1.\suffix}\hfil}\fi
                                                             %unix:dvips,scaled
\ifnum\system=5
\hbox to \pswidth{\kern-\xleft\includegraphics{\figdir#1.\suffix}\hfil}\fi %orphee
\ifnum\system=6
   \xdvikwid=\pswidth
   \xdvikht=\psheight
   {\global\divide\xdvikwid by \pspoints}
   {\global\divide\xdvikht by \pspoints}
   \rwi=\xdvikwid
    {\global\multiply\rwi by 10}
\hbox to \pswidth{\kern-\xleft\includegraphics{\figdir#1.\suffix\space}\hfil}\fi                   %xdvik
%%%
\vskip -\baselineskip
\vskip -\ybot 
\vskip-\psheight %                                     
\hbox to\pswidth  {\hss}%                                            
\parindent=0pt\offinterlineskip                                       
\vpos=0 pt%                                                              
\loop\readdim{\xcoord}                                 
\ifdim \xcoord < -999pt \doitfalse\else\doittrue\fi                        
\ifdoit \advance \xcoord by -\xleft
\readdim{\ycoord}
\advance \ycoord by -\ytop                              
\global\read\labeLfile to\justx                                       
\global\read\labeLfile to\justy                                       
\global\read\labeLfile to\labeL
\global\setbox\labox=\hbox{\labeL\hskip-0.3em}%    
\advance\vpos by-\ycoord                                              
\vskip-\vpos \vpos=\ycoord                                         
\hbox to\pswidth{\hskip\xcoord %                                 
\hbox to 0pt{\ifnum\justx>0\hss\fi%                                   
\vbox to0pt{%                                                         
\ifnum\justy<2\vss\fi%                                                
\copy\labox\kern0pt%  
\ifnum\justy>0\vss\fi}%                                               
\ifnum\justx<2\hss\fi}%                                               
\hss}%                                                                
\repeat%                                                              
\advance\vpos by-\psheight%                                           
\vskip-\vpos %                                                     
}\closein\labeLfile}
%
%
%     \figplace{<filename w/o extension>}{<hor shift>}{<vert shift>}
%     moves to the right by <hor shift> and down by <vert shift>
%     and then applies \figcrop
% 
\def\figplace#1#2#3{
\openin\labeLfile=\figdir#1.lbl
\ifeof \labeLfile
       \immediate\write16{***Can't find \figdir#1.lbl; Skipping it.***}
\else  \closein\labeLfile
       \null\hskip#2\raise #3 \hbox{\figcrop{#1}}
\fi
}
%
%
%     \figput{<filename w/o extension>}
%     
%     just applies \figcrop
% 

%%%%%%%%%%%%%%%%%%%%%%%%%%%%%%%%%%%%%%%%%%%%%%%%%%%%%%%%%%%%%%%%%%%%%%%
%%%%%%%%    omacros
%%%%%%%%%%%%%%%%%%%%%%%%%%%%%%%%%%%%%%%%%%%%%%%%%%%%%%%%%%%%%%%%%%%%%%%

    \def\squiggle{\raise2pt\hbox{${\scriptstyle\sim}$}}
    \def\stoday{\number\day\space\ifcase\month\or Jan\or Feb\or 
                      Mar\or Apr\or May\or Jun\or Jul\or Aug\or Sep\or 
                      Oct\or Nov\or Dec\fi, \number\year}

    \def\veps{{\varepsilon}}
    
    \def\abcst{{\sst const}}

    \def\tanV{\vec {\rm t}}
    \def\normV{\vec {\rm n}}
    \def\abcst{{\sst const}}
    \def\cb{{\frak c}}
    \def\ib{{\rm b}}
    \def\IB{{\rm\sst B}}

    \def\ord{{\rm Ord}\,}
    
    \def\dunion{\cup\kern-0.7em\cdot\kern0.45em}

    \def\cD{{\cal D}}
    \def\rD{{\rm D}}

    \def\cW{{\cal W}}

    \def\fe{{\frak e}}
    \def\fl{{\frak l}}

    \def\fN{{\frak N}}
    \def\fX{{\frak X}}

    \def\bde{{\mathchoice{\pmb{$\de$}}{\pmb{$\de$}}
                              {\pmb{$\sst\de$}}{\pmb{$\ssst\de$}}}}

    \def\lW{\mathopen{{\tst{\hbox{.}\atop\raise 2.5pt\hbox{.}}}}}
    \def\rW{\mathclose{{\tst{{.}\atop\raise 2.5pt\hbox{.}}}}}
    \def\lww{\mathopen{{\tst{\raise 1pt\hbox{.}\atop\raise 1pt\hbox{.}}}}}
    \def\rww{\mathclose{{\tst{\raise 1pt\hbox{.}\atop\raise 1pt\hbox{.}}}}}

    \def\v{\pmb{$\vert$}}
    \def\V{\pmb{$\big\vert$}}

    \def\tn{|\kern-1pt|\kern-1pt|}
    \def\TN{\big|\kern-1.5pt\big|\kern-1.5pt\big|}
    \def\TTN{\Big|\kern-2pt\Big|\kern-2pt\Big|}

    \def\cnorm{\kern8pt\check{\kern-8pt\|}}
    \def\Cnorm{\kern8pt\check{\kern-8pt\big\|}}
    \def\CNorm{\kern8pt\check{\kern-8pt\Big\|}}

    \def\tnorm{\kern8pt\tilde{\kern-8pt\|}}
    \def\Tnorm{\kern8pt\tilde{\kern-8pt\big\|}}
    \def\TNorm{\kern8pt\tilde{\kern-8pt\Big\|}}

    \def\tv{\kern8pt\tilde{\kern-8pt\pmb{$\vert$}}}
    \def\tV{\kern8pt\tilde{\kern-8pt\pmb{$\big\vert$}}}
    \def\tVV{\kern8pt\tilde{\kern-8pt\pmb{$\Big\vert$}}}

    \def\jbar{{\mathchoice
                   {{\smash{\lower1ex\hbox{$\mathchar'26$}}\mkern-9mu j}}
                   {{\smash{\lower1ex\hbox{$\mathchar'26$}}\mkern-9mu j}}
                   {{\smash{\lower1.2ex\hbox{$\mathchar'26$}}\mkern-10.2mu j}}
                   {{\smash{\lower1.2ex\hbox{$\mathchar'26$}}\mkern-10.2mu j}}}}

\def\Eqnb{{\hbox{\global\advance\equano by 1}}%
\eqno ({\rm \eqhead\number\equano}}%

\def\EQNB#1{\Eqnb\edef\Zwi{\eqhead\number\equano}%
\global\let #1=\Zwi
}

   \font\sixrm=cmr6   \font\eightrm=cmr8  
   \font\sixi=cmmi6   \font\eighti=cmmi8  
  \font\sixsy=cmsy6  \font\eightsy=cmsy8 
  \font\sixbf=cmbx6  \font\eightbf=cmbx8 
                     \font\eightit=cmti8 
                     \font\eightsl=cmsl8 
                     \font\eighttt=cmtt8 

\font\eightfrak=eufm7 at 8pt

\def\eightpoint{\def\rm{\fam0\eightrm}% switch to 8-point type
 \textfont0=\eightrm \scriptfont0=\sixrm \scriptscriptfont0=\fiverm
 \textfont1=\eighti \scriptfont1=\sixi \scriptscriptfont1=\fivei
 \textfont2=\eightsy \scriptfont2=\sixsy \scriptscriptfont2=\fivesy
 \textfont3=\tenex \scriptfont3=\tenex \scriptscriptfont3=\tenex
 \textfont\itfam=\eightit \def\it{\fam\itfam\eightit}%
 \textfont\slfam=\eightsl \def\sl{\fam\slfam\eightsl}%
 \textfont\ttfam=\eighttt \def\tt{\fam\ttfam\eighttt}%
 \textfont\frakfam=\eightfrak \def\frak{\fam\frakfam\tenfrak}%
 \textfont\bffam=\eightbf \scriptfont\bffam=\sixbf
 \scriptscriptfont\bffam=\fivebf \def\bf{\fam\bffam\eightbf}%
 \normalbaselineskip=9pt
 \setbox\strutbox=\hbox{\vrule height7pt depth2pt width0pt}%
 \let\sc=\sixrm \let\big=\eightbig \normalbaselines\rm}
\catcode`@=11
\def\footnote#1{\edef\@sf{\spacefactor\the\spacefactor}#1\@sf
     \insert\footins\bgroup\eightpoint
     \interlinepenalty100 \let\par=\endgraf
     \leftskip=0pt \rightskip=0pt
     \splittopskip=10pt plus 1pt minus 1pt \floatingpenalty=20000
     \smallskip\item{#1}\bgroup\strut\aftergroup\@foot\let\next}
\skip\footins=12pt plus 2pt minus 4pt
\dimen\footins=30pc
\catcode`@=12

%%%%%%%%%%%%%%%%%%%%%%%%%%%%%%%%%%%%%%%%%%%%%%%%%%%%%%%%%%%%%%%%%%%%%%%
%%%%%%%%    allr.txs - automatic numbering data from FTKr1,2
%%%%%%%%%%%%%%%%%%%%%%%%%%%%%%%%%%%%%%%%%%%%%%%%%%%%%%%%%%%%%%%%%%%%%%%

  \IgNoRe{PG}
  \IgNoRe{STM Assertion }
  \IgNoRe{PG}
  \IgNoRe{PG}
  \IgNoRe{STM Assertion }
  \IgNoRe{PG}
  \IgNoRe{STM Assertion }
  \IgNoRe{STM Assertion }
  \IgNoRe{EQN}
  \IgNoRe{STM Assertion }
  \IgNoRe{STM Assertion }
  \IgNoRe{PG}
  \IgNoRe{STM Assertion }
  \IgNoRe{STM Assertion }
  \IgNoRe{EQN}
  \IgNoRe{STM Assertion }
  \IgNoRe{STM Assertion }
  \IgNoRe{STM Assertion }
  \IgNoRe{STM Assertion }
  \IgNoRe{STM Assertion }
  \IgNoRe{STM Assertion }
  \IgNoRe{PG}
  \IgNoRe{STM Assertion }
  \IgNoRe{STM Assertion }
  \IgNoRe{STM Assertion }
  \IgNoRe{STM Assertion }
  \IgNoRe{STM Assertion }
  \IgNoRe{STM Assertion }
  \IgNoRe{STM Assertion }
  \IgNoRe{STM Assertion }
  \IgNoRe{STM Assertion }
  \IgNoRe{STM Assertion }
  \IgNoRe{STM Assertion }
  \IgNoRe{STM Assertion }
  \IgNoRe{PG}
  \IgNoRe{EQN}
  \IgNoRe{STM Assertion }
  \IgNoRe{STM Assertion }
  \IgNoRe{STM Assertion }
  \IgNoRe{PG}
  \IgNoRe{STM Assertion }
  \IgNoRe{STM Assertion }
  \IgNoRe{STM Assertion }
  \IgNoRe{STM Assertion }
  \IgNoRe{EQN}
  \IgNoRe{EQN}
  \IgNoRe{STM Assertion }
  \IgNoRe{PG}
  \IgNoRe{PG}
  \IgNoRe{STM Assertion }
  \IgNoRe{EQN}
  \IgNoRe{STM Assertion }
  \IgNoRe{STM Assertion }
  \IgNoRe{STM Assertion }
  \IgNoRe{EQN}
  \IgNoRe{STM Assertion }
  \IgNoRe{STM Assertion }
  \IgNoRe{STM Assertion }
  \IgNoRe{PG}
  \IgNoRe{STM Assertion }
  \IgNoRe{STM Assertion }
  \IgNoRe{PG}
  \IgNoRe{STM Assertion }
  \IgNoRe{STM Assertion }
  \IgNoRe{STM Assertion }
  \IgNoRe{PG}
  \IgNoRe{STM Assertion }
  \IgNoRe{STM Assertion }
  \IgNoRe{STM Assertion }
  \IgNoRe{STM Assertion }
  \IgNoRe{STM Assertion }
  \IgNoRe{STM Assertion }
  \IgNoRe{STM Assertion }
  \IgNoRe{STM Assertion }
  \IgNoRe{PG}
  \IgNoRe{STM Assertion }
  \IgNoRe{STM Assertion }
  \IgNoRe{STM Assertion }
  \IgNoRe{STM Assertion }
  \IgNoRe{STM Assertion }
  \IgNoRe{STM Assertion }
  \IgNoRe{STM Assertion }
  \IgNoRe{STM Assertion }
  \IgNoRe{PG}
  \IgNoRe{STM Assertion }
  \IgNoRe{STM Assertion }
  \IgNoRe{PG}
  \IgNoRe{PG}
  \IgNoRe{STM Assertion }
  \IgNoRe{STM Assertion }
  \IgNoRe{PG}
  \IgNoRe{PG}
  \IgNoRe{STM Assertion }
  \IgNoRe{STM Assertion }
  \IgNoRe{EQN}
  \IgNoRe{STM Assertion }
  \IgNoRe{PG}
  \IgNoRe{STM Assertion }
  \IgNoRe{STM Assertion }
  \IgNoRe{STM Assertion }
  \IgNoRe{PG}
  \IgNoRe{STM Assertion }
  \IgNoRe{STM Assertion }
  \IgNoRe{STM Assertion }
  \IgNoRe{EQN}
  \IgNoRe{STM Assertion }
  \IgNoRe{PG}
  \IgNoRe{EQN}
  \IgNoRe{STM Assertion }
  \IgNoRe{STM Assertion }
  \IgNoRe{STM Assertion }
  \IgNoRe{STM Assertion }
  \IgNoRe{EQN}
  \IgNoRe{EQN}
  \IgNoRe{PG}
  \IgNoRe{PG}
  \IgNoRe{STM Assertion }
  \IgNoRe{STM Assertion }
  \IgNoRe{STM Assertion }
  \IgNoRe{STM Assertion }
  \IgNoRe{STM Assertion }
  \IgNoRe{STM Assertion }
  \IgNoRe{STM Assertion }
  \IgNoRe{STM Assertion }
  \IgNoRe{PG}
  \IgNoRe{STM Assertion }
  \IgNoRe{STM Assertion }
  \IgNoRe{STM Assertion }
  \IgNoRe{STM Assertion }
  \IgNoRe{STM Assertion }
  \IgNoRe{STM Assertion }
  \IgNoRe{STM Assertion }
  \IgNoRe{STM Assertion }
  \IgNoRe{STM Assertion }
  \IgNoRe{STM Assertion }
  \IgNoRe{PG}
  \IgNoRe{STM Assertion }
  \IgNoRe{STM Assertion }
  \IgNoRe{STM Assertion }
  \IgNoRe{PG}
  \IgNoRe{STM Assertion }
  \IgNoRe{STM Assertion }
  \IgNoRe{STM Assertion }
  \IgNoRe{STM Assertion }
  \IgNoRe{PG}
  \IgNoRe{STM Assertion }
  \IgNoRe{STM Assertion }
  \IgNoRe{PG}
  \IgNoRe{STM Assertion }
  \IgNoRe{PG}
  \IgNoRe{STM Assertion }
  \IgNoRe{PG}
  \IgNoRe{STM Assertion }
  \IgNoRe{STM Assertion }
  \IgNoRe{STM Assertion }
  \IgNoRe{STM Assertion }
  \IgNoRe{PG}
  \IgNoRe{PG}
  \IgNoRe{STM Assertion }
  \IgNoRe{STM Assertion }
  \IgNoRe{EQN}
  \IgNoRe{EQN}
  \IgNoRe{STM Assertion }
  \IgNoRe{STM Assertion }
  \IgNoRe{STM Assertion }
  \IgNoRe{STM Assertion }
  \IgNoRe{PG}
  \IgNoRe{STM Assertion }
  \IgNoRe{STM Assertion }
  \IgNoRe{STM Assertion }
  \IgNoRe{STM Assertion }
  \IgNoRe{STM Assertion }
  \IgNoRe{STM Assertion }
  \IgNoRe{STM Assertion }
  \IgNoRe{PG}
  \IgNoRe{STM Assertion }
  \IgNoRe{EQN}
  \IgNoRe{EQN}
  \IgNoRe{PG}
  \IgNoRe{STM Assertion }
  \IgNoRe{EQN}
  \IgNoRe{STM Assertion }
  \IgNoRe{STM Assertion }
  \IgNoRe{STM Assertion }
  \IgNoRe{PG}
  \IgNoRe{STM Assertion }
  \IgNoRe{EQN}
  \IgNoRe{STM Assertion }
  \IgNoRe{PG}
  \IgNoRe{PG}

%%%%%%%%%%%%%%%%%%%%%%%%%%%%%%%%%%%%%%%%%%%%%%%%%%%%%%%%%%%%%%%%%%%%%%%
%%%%%%%%    fl-all.txs - automatic numbering data from FTKf1,2,3
%%%%%%%%%%%%%%%%%%%%%%%%%%%%%%%%%%%%%%%%%%%%%%%%%%%%%%%%%%%%%%%%%%%%%%%

  \IgNoRe{STM Assertion }
  \IgNoRe{PG}
  \IgNoRe{EQN}
  \IgNoRe{EQN}
  \IgNoRe{EQN}
  \IgNoRe{EQN}
  \IgNoRe{EQN}
  \IgNoRe{STM Assertion }
  \IgNoRe{STM Assertion }
  \IgNoRe{STM Assertion }
  \IgNoRe{STM Assertion }
  \IgNoRe{STM Assertion }
  \IgNoRe{STM Assertion }
  \IgNoRe{STM Assertion }
  \IgNoRe{STM Assertion }
  \IgNoRe{STM Assertion }
 \def\remModII{\frefwarning I.11} \IgNoRe{STM Assertion }
  \IgNoRe{STM Assertion }
  \IgNoRe{PG}
  \IgNoRe{PG}
  \IgNoRe{PG}
  \IgNoRe{PG}
  \IgNoRe{EQN}
  \IgNoRe{EQN}
  \IgNoRe{EQN}
  \IgNoRe{EQN}
  \IgNoRe{PG}
  \IgNoRe{PG}
  \IgNoRe{PG}
  \IgNoRe{EQN}
  \IgNoRe{PG}
  \IgNoRe{PG}
  \IgNoRe{EQN}
  \IgNoRe{EQN}
  \IgNoRe{EQN}
  \IgNoRe{EQN}
  \IgNoRe{EQN}
  \IgNoRe{EQN}
  \IgNoRe{EQN}
  \IgNoRe{EQN}
  \IgNoRe{EQN}
  \IgNoRe{EQN}
  \IgNoRe{PG}
  \IgNoRe{PG}
  \IgNoRe{EQN}
  \IgNoRe{EQN}
  \IgNoRe{EQN}
  \IgNoRe{STM Assertion }
  \IgNoRe{PG}
  \IgNoRe{EQN}
  \IgNoRe{EQN}
  \IgNoRe{EQN}
  \IgNoRe{EQN}
  \IgNoRe{STM Assertion }
  \IgNoRe{STM Assertion }
  \IgNoRe{STM Assertion }
  \IgNoRe{STM Assertion }
  \IgNoRe{STM Assertion }
  \IgNoRe{STM Assertion }
  \IgNoRe{STM Assertion }
  \IgNoRe{STM Assertion }
  \IgNoRe{STM Assertion }
  \IgNoRe{EQN}
  \IgNoRe{EQN}
  \IgNoRe{EQN}
  \IgNoRe{EQN}
  \IgNoRe{STM Assertion }
  \IgNoRe{EQN}
  \IgNoRe{STM Assertion }
  \IgNoRe{EQN}
  \IgNoRe{STM Assertion }
  \IgNoRe{PG}
  \IgNoRe{EQN}
  \IgNoRe{STM Assertion }
  \IgNoRe{STM Assertion }
  \IgNoRe{EQN}
  \IgNoRe{PG}
  \IgNoRe{PG}
  \IgNoRe{STM Assertion }
  \IgNoRe{STM Assertion }
  \IgNoRe{PG}
  \IgNoRe{STM Assertion }
  \IgNoRe{STM Assertion }
  \IgNoRe{STM Assertion }
  \IgNoRe{STM Assertion }
  \IgNoRe{STM Assertion }
  \IgNoRe{STM Assertion }
  \IgNoRe{STM Assertion }
  \IgNoRe{STM Assertion }
  \IgNoRe{PG}
  \IgNoRe{STM Assertion }
  \IgNoRe{STM Assertion }
  \IgNoRe{STM Assertion }
  \IgNoRe{STM Assertion }
  \IgNoRe{EQN}
  \IgNoRe{STM Assertion }
  \IgNoRe{STM Assertion }
  \IgNoRe{STM Assertion }
  \IgNoRe{EQN}
  \IgNoRe{STM Assertion }
  \IgNoRe{STM Assertion }
  \IgNoRe{STM Assertion }
  \IgNoRe{STM Assertion }
  \IgNoRe{PG}
  \IgNoRe{STM Assertion }
  \IgNoRe{STM Assertion }
  \IgNoRe{STM Assertion }
  \IgNoRe{STM Assertion }
  \IgNoRe{STM Assertion }
  \IgNoRe{STM Assertion }
  \IgNoRe{STM Assertion }
  \IgNoRe{STM Assertion }
  \IgNoRe{STM Assertion }
  \IgNoRe{EQN}
  \IgNoRe{EQN}
  \IgNoRe{EQN}
  \IgNoRe{STM Assertion }
  \IgNoRe{PG}
  \IgNoRe{STM Assertion }
  \IgNoRe{STM Assertion }
  \IgNoRe{STM Assertion }
  \IgNoRe{EQN}
  \IgNoRe{EQN}
  \IgNoRe{EQN}
  \IgNoRe{STM Assertion }
  \IgNoRe{STM Assertion }
  \IgNoRe{EQN}
  \IgNoRe{EQN}
  \IgNoRe{EQN}
  \IgNoRe{STM Assertion }
  \IgNoRe{PG}
  \IgNoRe{PG}
  \IgNoRe{STM Assertion }
  \IgNoRe{STM Assertion }
  \IgNoRe{PG}
  \IgNoRe{STM Assertion }
  \IgNoRe{STM Assertion }
  \IgNoRe{EQN}
  \IgNoRe{EQN}
  \IgNoRe{EQN}
  \IgNoRe{EQN}
  \IgNoRe{EQN}
  \IgNoRe{EQN}
  \IgNoRe{EQN}
  \IgNoRe{EQN}
  \IgNoRe{EQN}
  \IgNoRe{EQN}
  \IgNoRe{EQN}
  \IgNoRe{EQN}
  \IgNoRe{EQN}
  \IgNoRe{EQN}
  \IgNoRe{STM Assertion }
  \IgNoRe{EQN}
  \IgNoRe{PG}
  \IgNoRe{EQN}
  \IgNoRe{STM Assertion }
  \IgNoRe{EQN}
  \IgNoRe{STM Assertion }
  \IgNoRe{STM Assertion }
  \IgNoRe{EQN}
  \IgNoRe{EQN}
  \IgNoRe{EQN}
  \IgNoRe{STM Assertion }
  \IgNoRe{EQN}
  \IgNoRe{EQN}
  \IgNoRe{EQN}
  \IgNoRe{EQN}
  \IgNoRe{EQN}
  \IgNoRe{EQN}
  \IgNoRe{EQN}
  \IgNoRe{EQN}
  \IgNoRe{EQN}
  \IgNoRe{EQN}
  \IgNoRe{EQN}
  \IgNoRe{EQN}
  \IgNoRe{EQN}
  \IgNoRe{EQN}
  \IgNoRe{EQN}
  \IgNoRe{EQN}
  \IgNoRe{EQN}
  \IgNoRe{EQN}
  \IgNoRe{EQN}
  \IgNoRe{EQN}
  \IgNoRe{EQN}
  \IgNoRe{STM Assertion }
  \IgNoRe{PG}
  \IgNoRe{PG}
  \IgNoRe{STM Assertion }
  \IgNoRe{PG}
  \IgNoRe{EQN}
  \IgNoRe{EQN}
  \IgNoRe{EQN}
  \IgNoRe{EQN}
  \IgNoRe{STM Assertion }
  \IgNoRe{EQN}
  \IgNoRe{PG}
  \IgNoRe{EQN}
  \IgNoRe{EQN}
  \IgNoRe{EQN}
  \IgNoRe{EQN}
  \IgNoRe{EQN}
  \IgNoRe{EQN}
  \IgNoRe{EQN}
  \IgNoRe{EQN}
  \IgNoRe{EQN}
  \IgNoRe{EQN}
  \IgNoRe{EQN}
  \IgNoRe{EQN}
  \IgNoRe{STM Assertion }
  \IgNoRe{STM Assertion }
  \IgNoRe{EQN}
  \IgNoRe{EQN}
  \IgNoRe{PG}
  \IgNoRe{PG}
  \IgNoRe{STM Assertion }
  \IgNoRe{EQN}
  \IgNoRe{STM Assertion }
  \IgNoRe{PG}
  \IgNoRe{EQN}
  \IgNoRe{EQN}
  \IgNoRe{EQN}
  \IgNoRe{STM Assertion }
  \IgNoRe{STM Assertion }
  \IgNoRe{EQN}
  \IgNoRe{EQN}
  \IgNoRe{EQN}
  \IgNoRe{EQN}
  \IgNoRe{EQN}
  \IgNoRe{STM Assertion }
  \IgNoRe{EQN}
  \IgNoRe{EQN}
  \IgNoRe{EQN}
  \IgNoRe{EQN}
  \IgNoRe{STM Assertion }
  \IgNoRe{STM Assertion }
  \IgNoRe{EQN}
  \IgNoRe{STM Assertion }
  \IgNoRe{STM Assertion }
  \IgNoRe{STM Assertion }
  \IgNoRe{STM Assertion }
  \IgNoRe{PG}
  \IgNoRe{STM Assertion }
  \IgNoRe{STM Assertion }
  \IgNoRe{STM Assertion }
  \IgNoRe{STM Assertion }
  \IgNoRe{STM Assertion }
  \IgNoRe{STM Assertion }
  \IgNoRe{STM Assertion }
  \IgNoRe{STM Assertion }
  \IgNoRe{STM Assertion }
  \IgNoRe{STM Assertion }
  \IgNoRe{STM Assertion }
  \IgNoRe{STM Assertion }
  \IgNoRe{STM Assertion }
  \IgNoRe{STM Assertion }
  \IgNoRe{STM Assertion }
  \IgNoRe{PG}
  \IgNoRe{STM Assertion }
  \IgNoRe{STM Assertion }
  \IgNoRe{STM Assertion }
  \IgNoRe{STM Assertion }
  \IgNoRe{STM Assertion }
  \IgNoRe{STM Assertion }
  \IgNoRe{STM Assertion }
  \IgNoRe{STM Assertion }
  \IgNoRe{STM Assertion }
  \IgNoRe{STM Assertion }
  \IgNoRe{STM Assertion }
  \IgNoRe{STM Assertion }
  \IgNoRe{EQN}
  \IgNoRe{STM Assertion }
  \IgNoRe{STM Assertion }
  \IgNoRe{STM Assertion }
  \IgNoRe{STM Assertion }
  \IgNoRe{STM Assertion }
  \IgNoRe{STM Assertion }
  \IgNoRe{EQN}
  \IgNoRe{STM Assertion }
  \IgNoRe{PG}
  \IgNoRe{PG}
  \IgNoRe{STM Assertion }
  \IgNoRe{STM Assertion }
  \IgNoRe{STM Assertion }
  \IgNoRe{EQN}
  \IgNoRe{STM Assertion }
  \IgNoRe{PG}
  \IgNoRe{EQN}
  \IgNoRe{STM Assertion }
  \IgNoRe{STM Assertion }
  \IgNoRe{EQN}
  \IgNoRe{EQN}
  \IgNoRe{EQN}
  \IgNoRe{EQN}
  \IgNoRe{EQN}
  \IgNoRe{EQN}
  \IgNoRe{EQN}
  \IgNoRe{EQN}
  \IgNoRe{EQN}
  \IgNoRe{EQN}
  \IgNoRe{EQN}
  \IgNoRe{EQN}
  \IgNoRe{EQN}
  \IgNoRe{EQN}
  \IgNoRe{EQN}
  \IgNoRe{EQN}
  \IgNoRe{EQN}
  \IgNoRe{EQN}
  \IgNoRe{EQN}
  \IgNoRe{EQN}
  \IgNoRe{STM Assertion }
  \IgNoRe{STM Assertion }
  \IgNoRe{PG}
  \IgNoRe{EQN}
  \IgNoRe{EQN}
  \IgNoRe{EQN}
  \IgNoRe{EQN}
  \IgNoRe{EQN}
  \IgNoRe{EQN}
  \IgNoRe{EQN}
  \IgNoRe{EQN}
  \IgNoRe{EQN}
  \IgNoRe{EQN}
  \IgNoRe{EQN}
  \IgNoRe{EQN}
  \IgNoRe{EQN}
  \IgNoRe{EQN}
  \IgNoRe{EQN}
  \IgNoRe{EQN}
  \IgNoRe{EQN}
  \IgNoRe{EQN}
  \IgNoRe{EQN}
  \IgNoRe{EQN}
  \IgNoRe{EQN}
  \IgNoRe{STM Assertion }
  \IgNoRe{PG}
  \IgNoRe{STM Assertion }
  \IgNoRe{STM Assertion }
  \IgNoRe{EQN}
  \IgNoRe{EQN}
  \IgNoRe{PG}
  \IgNoRe{EQN}
  \IgNoRe{EQN}
  \IgNoRe{EQN}
  \IgNoRe{EQN}
  \IgNoRe{EQN}
  \IgNoRe{EQN}
  \IgNoRe{EQN}
  \IgNoRe{EQN}
  \IgNoRe{STM Assertion }
  \IgNoRe{STM Assertion }
  \IgNoRe{STM Assertion }
  \IgNoRe{STM Assertion }
  \IgNoRe{STM Assertion }
  \IgNoRe{PG}
  \IgNoRe{PG}

%%%%%%%%%%%%%%%%%%%%%%%%%%%%%%%%%%%%%%%%%%%%%%%%%%%%%%%%%%%%%%%%%%%%%%%
%%%%%%%%    fl-chap.tex - chapter numbering data from FTKf1,2,3
%%%%%%%%%%%%%%%%%%%%%%%%%%%%%%%%%%%%%%%%%%%%%%%%%%%%%%%%%%%%%%%%%%%%%%%

\newcount\CHAPNO
\newcount\APPNO
\CHAPNO=0
\APPNO=1
\def\advCHAPNO{\advance\CHAPNO by 1}
\def\advAPPNO{\advance\APPNO by 1}

\def\caproman#1{\ifcase#1\or I\or II\or III\or IV\or V\or VI\or VII\or
VIII\or IX\or X\or XI\or XII\or XIII\or XIV\or XV\or XVI\or XVII\or XVIII\or
XIX\or XX\or XXI\or XXII\or XXIII\or XXIV\or XXV\or XXVI\or XXVII\or XXVIII\or XXIX\or XXX\or XXXI\or XXXII\or XXXIII\or XXXIV\or XXXV\or XXXVI\or XXXVII\or XXXVIII\or XXXIX\fi}%

\def\capletter#1{\ifcase#1\or A\or B\or C\or D\or E\or F\or G\or
H\or I\or J\or K\or L\or M\or N\or O\or P\or Q\or R\or
S\or T\or U\or V\or W\or X\or Y\or Z\fi}%

\newcount\cHintroI \cHintroI=\CHAPNO \advCHAPNO 
                                       %I
\newcount\cHintroOverview  \cHintroOverview=\CHAPNO \advCHAPNO 
                              \edef\CHintroOverview{\caproman\CHAPNO}  %II
\newcount\cHrenmap \cHrenmap=\CHAPNO \advCHAPNO 
                                       %III

 \advAPPNO

\newcount\cHintroII \cHintroII=\CHAPNO \advCHAPNO 
                              \edef\CHintroII{\caproman\CHAPNO}
\newcount\cHfirstscale \cHfirstscale=\CHAPNO \advCHAPNO
                              
\newcount\cHnewsectors \cHnewsectors=\CHAPNO \advCHAPNO
                              
\newcount\cHphladders \cHphladders=\CHAPNO \advCHAPNO
                              
\newcount\cHfinitescale \cHfinitescale=\CHAPNO \advCHAPNO
                              
\newcount\cHstep \cHstep=\CHAPNO \advCHAPNO
                              
\newcount\cHrecurs \cHrecurs=\CHAPNO \advCHAPNO
                              
 \advAPPNO

\newcount\cHintroIII \cHintroIII=\CHAPNO \advCHAPNO
                              \edef\CHintroIII{\caproman\CHAPNO}
\newcount\cHtildefinitescale \cHtildefinitescale=\CHAPNO \advCHAPNO
                              
\newcount\cHtildenewsectors \cHtildenewsectors=\CHAPNO \advCHAPNO
                              
\newcount\cHtildephladders \cHtildephladders=\CHAPNO \advCHAPNO
                              
\newcount\cHtildestep  \cHtildestep=\CHAPNO \advCHAPNO

 \advAPPNO
 \advAPPNO

%%%%%%%%%%%%%%%%%%%%%%%%%%%%%%%%%%%%%%%%%%%%%%%%%%%%%%%%%%%%%%%%%%%%%%%
%%%%%%%%    os-all.txs - automatic numbering data from FTKo1,2,3,4
%%%%%%%%%%%%%%%%%%%%%%%%%%%%%%%%%%%%%%%%%%%%%%%%%%%%%%%%%%%%%%%%%%%%%%%

  \IgNoRe{PG}
 \def\eqnOSintroI{\frefwarning I.1} \IgNoRe{EQN}
  \IgNoRe{STM Assertion }
  \IgNoRe{PG}
  \IgNoRe{STM Assertion }
  \IgNoRe{STM Assertion }
  \IgNoRe{STM Assertion }
  \IgNoRe{STM Assertion }
 \def\exOSSymmNorm{\frefwarning II.6} \IgNoRe{STM Assertion }
 \def\lemOSelloneinfty{\frefwarning II.7} \IgNoRe{STM Assertion }
  \IgNoRe{EQN}
  \IgNoRe{STM Assertion }
 \def\defOSFmn{\frefwarning II.9} \IgNoRe{STM Assertion }
  \IgNoRe{STM Assertion }
  \IgNoRe{STM Assertion }
  \IgNoRe{STM Assertion }
  \IgNoRe{STM Assertion }
 \def\exOSelloneinftycontr{\frefwarning III.4} \IgNoRe{STM Assertion }
  \IgNoRe{STM Assertion }
  \IgNoRe{PG}
  \IgNoRe{STM Assertion }
  \IgNoRe{STM Assertion }
  \IgNoRe{STM Assertion }
 \def\defOSgrnorm{\frefwarning III.9} \IgNoRe{STM Assertion }
  \IgNoRe{STM Assertion }
  \IgNoRe{STM Assertion }
 \def\defIntBndsS{\frefwarning IV.1} \IgNoRe{STM Assertion }
  \IgNoRe{STM Assertion }
  \IgNoRe{EQN}
  \IgNoRe{PG}
  \IgNoRe{PG}
  \IgNoRe{STM Assertion }
  \IgNoRe{STM Assertion }
  \IgNoRe{STM Assertion }
 \def\defOSderivmom{\frefwarning IV.6} \IgNoRe{STM Assertion }
  \IgNoRe{STM Assertion }
  \IgNoRe{STM Assertion }
  \IgNoRe{PG}
  \IgNoRe{EQN}
  \IgNoRe{EQN}
  \IgNoRe{EQN}
  \IgNoRe{EQN}
  \IgNoRe{EQN}
  \IgNoRe{EQN}
  \IgNoRe{STM Assertion }
  \IgNoRe{STM Assertion }
  \IgNoRe{STM Assertion }
 \def\lemOSscalednorm{\frefwarning V.1} \IgNoRe{STM Assertion }
  \IgNoRe{PG}
 \def\thmOSinsulators{\frefwarning V.2} \IgNoRe{STM Assertion }
  \IgNoRe{EQN}
  \IgNoRe{STM Assertion }
  \IgNoRe{STM Assertion }
  \IgNoRe{STM Assertion }
 \def\exOSappMonoidI{\frefwarning A.3} \IgNoRe{STM Assertion }
  \IgNoRe{PG}
  \IgNoRe{STM Assertion }
  \IgNoRe{STM Assertion }
  \IgNoRe{STM Assertion }
  \IgNoRe{STM Assertion }
  \IgNoRe{PG}
 \def\eqnOSjdef{\frefwarning VI.1} \IgNoRe{EQN}
  \IgNoRe{EQN}
  \IgNoRe{PG}
 \def\defOSrengrpmap{\frefwarning VII.1} \IgNoRe{STM Assertion }
  \IgNoRe{STM Assertion }
  \IgNoRe{STM Assertion }
  \IgNoRe{EQN}
  \IgNoRe{PG}
  \IgNoRe{STM Assertion }
  \IgNoRe{STM Assertion }
  \IgNoRe{EQN}
  \IgNoRe{STM Assertion }
  \IgNoRe{STM Assertion }
  \IgNoRe{STM Assertion }
 \def\defOSscales{\frefwarning VIII.1} \IgNoRe{STM Assertion }
  \IgNoRe{PG}
  \IgNoRe{STM Assertion }
  \IgNoRe{STM Assertion }
 \def\defOSextendedshell{\frefwarning VIII.4} \IgNoRe{STM Assertion }
  \IgNoRe{STM Assertion }
 \def\thmOSfirststep{\frefwarning VIII.6} \IgNoRe{STM Assertion }
  \IgNoRe{STM Assertion }
  \IgNoRe{STM Assertion }
 \def\defOSfourtrans{\frefwarning IX.1} \IgNoRe{STM Assertion }
  \IgNoRe{PG}
  \IgNoRe{STM Assertion }
 \def\defOSftcov{\frefwarning IX.3} \IgNoRe{STM Assertion }
 \def\defOSfourtransII{\frefwarning IX.4} \IgNoRe{STM Assertion }
  \IgNoRe{STM Assertion }
 \def\lemOSprepintup{\frefwarning IX.6} \IgNoRe{STM Assertion }
  \IgNoRe{EQN}
 \def\defOSamptransinv{\frefwarning X.1} \IgNoRe{STM Assertion }
  \IgNoRe{STM Assertion }
  \IgNoRe{PG}
  \IgNoRe{STM Assertion }
 \def\defOSdiffdecaynorm{\frefwarning X.4} \IgNoRe{STM Assertion }
  \IgNoRe{STM Assertion }
  \IgNoRe{STM Assertion }
  \IgNoRe{STM Assertion }
 \def\defOScheckcF{\frefwarning X.8} \IgNoRe{STM Assertion }
  \IgNoRe{STM Assertion }
  \IgNoRe{STM Assertion }
 \def\lemOSTZsourceterm{\frefwarning X.11} \IgNoRe{STM Assertion }
  \IgNoRe{EQN}
 \def\thmOSTfirststep{\frefwarning X.12} \IgNoRe{STM Assertion }
 \def\defOSsymmetries{\frefwarning B.1} \IgNoRe{STM Assertion }
  \IgNoRe{PG}
  \IgNoRe{STM Assertion }
  \IgNoRe{STM Assertion }
  \IgNoRe{STM Assertion }
  \IgNoRe{STM Assertion }
  \IgNoRe{STM Assertion }
  \IgNoRe{STM Assertion }
  \IgNoRe{STM Assertion }
  \IgNoRe{PG}
  \IgNoRe{STM Assertion }
  \IgNoRe{PG}
  \IgNoRe{PG}
 \def\defOSsectors{\frefwarning XII.1} \IgNoRe{STM Assertion }
 \def\defOScbj{\frefwarning XII.2} \IgNoRe{STM Assertion }
  \IgNoRe{PG}
 \def\lemOSsectpartunit{\frefwarning XII.3} \IgNoRe{STM Assertion }
 \def\defOSsectrepr{\frefwarning XII.4} \IgNoRe{STM Assertion }
 \def\exOSsectrepr{\frefwarning XII.5} \IgNoRe{STM Assertion }
  \IgNoRe{STM Assertion }
  \IgNoRe{STM Assertion }
  \IgNoRe{STM Assertion }
 \def\defOSsectnorm{\frefwarning XII.9} \IgNoRe{STM Assertion }
 \def\exOScommsectnorms{\frefwarning XII.10} \IgNoRe{STM Assertion }
  \IgNoRe{STM Assertion }
  \IgNoRe{STM Assertion }
  \IgNoRe{STM Assertion }
  \IgNoRe{STM Assertion }
  \IgNoRe{STM Assertion }
  \IgNoRe{STM Assertion }
  \IgNoRe{STM Assertion }
  \IgNoRe{STM Assertion }
  \IgNoRe{EQN}
  \IgNoRe{STM Assertion }
  \IgNoRe{STM Assertion }
  \IgNoRe{PG}
  \IgNoRe{STM Assertion }
  \IgNoRe{EQN}
 \def\eqnOSpartunit{\frefwarning XIII.2} \IgNoRe{EQN}
 \def\lemOSmorepartunity{\frefwarning XIII.3} \IgNoRe{STM Assertion }
  \IgNoRe{EQN}
 \def\eqnOSprodcontrbound{\frefwarning XIII.4} \IgNoRe{EQN}
  \IgNoRe{STM Assertion }
  \IgNoRe{EQN}
 \def\propOSrealpropbound{\frefwarning XIII.5} \IgNoRe{STM Assertion }
 \def\eqnOSexpandc{\frefwarning XIII.6} \IgNoRe{EQN}
 \def\lemOSdiffpropbound{\frefwarning XIII.6} \IgNoRe{STM Assertion }
 \def\lemOSumu{\frefwarning XIII.7} \IgNoRe{STM Assertion }
  \IgNoRe{STM Assertion }
 \def\defOSbubbleprop{\frefwarning XIV.1} \IgNoRe{STM Assertion }
  \IgNoRe{PG}
  \IgNoRe{STM Assertion }
 \def\defOSsectbubbleprop{\frefwarning XIV.3} \IgNoRe{STM Assertion }
  \IgNoRe{STM Assertion }
  \IgNoRe{STM Assertion }
  \IgNoRe{EQN}
 \def\eqnOSrhomn{\frefwarning XV.1} \IgNoRe{EQN}
 \def\defOSscalednorms{\frefwarning XV.1} \IgNoRe{STM Assertion }
 \def\remOSscalednorms{\frefwarning XV.2} \IgNoRe{STM Assertion }
  \IgNoRe{PG}
 \def\thOSrengroupestimate{\frefwarning XV.3} \IgNoRe{STM Assertion }
  \IgNoRe{STM Assertion }
  \IgNoRe{STM Assertion }
  \IgNoRe{EQN}
  \IgNoRe{STM Assertion }
  \IgNoRe{EQN}
  \IgNoRe{STM Assertion }
  \IgNoRe{STM Assertion }
  \IgNoRe{EQN}
  \IgNoRe{STM Assertion }
  \IgNoRe{EQN}
  \IgNoRe{EQN}
  \IgNoRe{EQN}
  \IgNoRe{EQN}
  \IgNoRe{STM Assertion }
  \IgNoRe{STM Assertion }
 \def\defOSdisjointfield{\frefwarning XVI.1} \IgNoRe{STM Assertion }
 \def\defOSdisjointOrd{\frefwarning XVI.2} \IgNoRe{STM Assertion }
  \IgNoRe{PG}
 \def\remOSbigdisjointunion{\frefwarning XVI.3} \IgNoRe{STM Assertion }
 \def\defOSsectdiffdecaynorm{\frefwarning XVI.4} \IgNoRe{STM Assertion }
 \def\remOSsecdiffdecaynorm{\frefwarning XVI.5} \IgNoRe{STM Assertion }
  \IgNoRe{STM Assertion }
 \def\defOSsectcheckcF{\frefwarning XVI.7} \IgNoRe{STM Assertion }
  \IgNoRe{STM Assertion }
 \def\defOSsectbubblepropII{\frefwarning XVI.9} \IgNoRe{STM Assertion }
  \IgNoRe{STM Assertion }
  \IgNoRe{STM Assertion }
  \IgNoRe{STM Assertion }
  \IgNoRe{EQN}
 \def\eqnOStilderhomn{\frefwarning XVII.1} \IgNoRe{EQN}
 \def\defOSmomscalednorms{\frefwarning XVII.1} \IgNoRe{STM Assertion }
  \IgNoRe{PG}
  \IgNoRe{STM Assertion }
  \IgNoRe{STM Assertion }
  \IgNoRe{EQN}
  \IgNoRe{STM Assertion }
  \IgNoRe{STM Assertion }
  \IgNoRe{EQN}
  \IgNoRe{EQN}
  \IgNoRe{EQN}
  \IgNoRe{EQN}
  \IgNoRe{EQN}
  \IgNoRe{STM Assertion }
  \IgNoRe{STM Assertion }
  \IgNoRe{STM Assertion }
  \IgNoRe{EQN}
  \IgNoRe{EQN}
  \IgNoRe{EQN}
  \IgNoRe{EQN}
  \IgNoRe{EQN}
  \IgNoRe{EQN}
  \IgNoRe{STM Assertion }
 \def\defOSchannelnorm{\frefwarning D.1} \IgNoRe{STM Assertion }
 \def\lemchannelnorm{\frefwarning D.2} \IgNoRe{STM Assertion }
  \IgNoRe{PG}
  \IgNoRe{STM Assertion }
  \IgNoRe{STM Assertion }
  \IgNoRe{EQN}
  \IgNoRe{EQN}
  \IgNoRe{STM Assertion }
  \IgNoRe{EQN}
  \IgNoRe{EQN}
  \IgNoRe{STM Assertion }
  \IgNoRe{EQN}
  \IgNoRe{EQN}
 \def\propOSNaiveLadder{\frefwarning D.7} \IgNoRe{STM Assertion }
  \IgNoRe{STM Assertion }
  \IgNoRe{PG}
 \def\defOSantp{\frefwarning XVIII.1} \IgNoRe{STM Assertion }
 \def\exOSantp{\frefwarning XVIII.2} \IgNoRe{STM Assertion }
 \def\defModI{\frefwarning XVIII.3} \IgNoRe{STM Assertion }
 \def\pgOSXVIII{\frefwarning 1} \IgNoRe{PG}
 \def\remModII{\frefwarning XVIII.4} \IgNoRe{STM Assertion }
 \def\propOSthreetoonenorm{\frefwarning XIX.1} \IgNoRe{STM Assertion }
 \def\defOSresector{\frefwarning XIX.2} \IgNoRe{STM Assertion }
 \def\remOSresector{\frefwarning XIX.3} \IgNoRe{STM Assertion }
 \def\pgOSXIX{\frefwarning 3} \IgNoRe{PG}
 \def\propOSresectorI{\frefwarning XIX.4} \IgNoRe{STM Assertion }
 \def\remOStoresectorI{\frefwarning XIX.5} \IgNoRe{STM Assertion }
 \def\defOSresectorII{\frefwarning XIX.6} \IgNoRe{STM Assertion }
 \def\corOSresectorI{\frefwarning XIX.7} \IgNoRe{STM Assertion }
 \def\corOSirrelevantresect{\frefwarning XIX.8} \IgNoRe{STM Assertion }
 \def\corOStildeirrelevantresect{\frefwarning XIX.9} \IgNoRe{STM Assertion }
 \def\defOSvanishkzero{\frefwarning XIX.10} \IgNoRe{STM Assertion }
 \def\propOSIntUp{\frefwarning XIX.11} \IgNoRe{STM Assertion }
 \def\corOSIntUp{\frefwarning XIX.12} \IgNoRe{STM Assertion }
 \def\corOSresectorvanishkzero{\frefwarning XIX.13} \IgNoRe{STM Assertion }
 \def\defOScreateSectoriz{\frefwarning XIX.14} \IgNoRe{STM Assertion }
 \def\propOScreateSectoriz{\frefwarning XIX.15} \IgNoRe{STM Assertion }
 \def\defOSadmissablesectors{\frefwarning XX.1} \IgNoRe{STM Assertion }
 \def\remOSadmissablesectors{\frefwarning XX.2} \IgNoRe{STM Assertion }
 \def\pgOSXX{\frefwarning 13} \IgNoRe{PG}
 \def\defSecI{\frefwarning XX.3} \IgNoRe{STM Assertion }
 \def\pgOSXXa{\frefwarning 14} \IgNoRe{PG}
 \def\lemSecIII{\frefwarning XX.4} \IgNoRe{STM Assertion }
 \def\propSecVI{\frefwarning XX.5} \IgNoRe{STM Assertion }
 \def\pgOSXXb{\frefwarning 16} \IgNoRe{PG}
 \def\propOSSecV{\frefwarning XX.6} \IgNoRe{STM Assertion }
 \def\lemSecVII{\frefwarning XX.7} \IgNoRe{STM Assertion }
 \def\eqnSecII{\frefwarning XX.1} \IgNoRe{EQN}
 \def\pgOSXXc{\frefwarning 18} \IgNoRe{PG}
 \def\eqnSecIII{\frefwarning XX.2} \IgNoRe{EQN}
 \def\lemSecVIII{\frefwarning XX.8} \IgNoRe{STM Assertion }
 \def\lemSecIX{\frefwarning XX.9} \IgNoRe{STM Assertion }
 \def\pgOSXXd{\frefwarning 22} \IgNoRe{PG}
 \def\OSthreemomI{\frefwarning XX.3} \IgNoRe{EQN}
 \def\OSthreemomII{\frefwarning XX.4} \IgNoRe{EQN}
 \def\OSthreemomIII{\frefwarning XX.5} \IgNoRe{EQN}
 \def\OSthreemomIV{\frefwarning XX.6} \IgNoRe{EQN}
 \def\eqnOSSecsumI{\frefwarning XX.7} \IgNoRe{EQN}
 \def\propSecX{\frefwarning XX.10} \IgNoRe{STM Assertion }
 \def\proSecXI{\frefwarning XX.11} \IgNoRe{STM Assertion }
 \def\eqnSecI{\frefwarning XX.8} \IgNoRe{EQN}
 \def\lemOSonetothree{\frefwarning XXI.1} \IgNoRe{STM Assertion }
 \def\pgOSXXI{\frefwarning 28} \IgNoRe{PG}
 \def\pgOSXXIa{\frefwarning 28} \IgNoRe{PG}
 \def\pgOSXXIb{\frefwarning 28} \IgNoRe{PG}
 \def\lemSecXI{\frefwarning XXI.2} \IgNoRe{STM Assertion }
 \def\eqnOSauxnorms{\frefwarning XXI.1} \IgNoRe{EQN}
 \def\lemOSprepresectorI{\frefwarning XXI.3} \IgNoRe{STM Assertion }
 \def\pgOSXXIc{\frefwarning 30} \IgNoRe{PG}
 \def\lemOSseccountI{\frefwarning XXI.4} \IgNoRe{STM Assertion }
 \def\eqnOSssumI{\frefwarning XXI.2} \IgNoRe{EQN}
 \def\propOSLadA{\frefwarning XXII.1} \IgNoRe{STM Assertion }
 \def\lemLadAI{\frefwarning XXII.2} \IgNoRe{STM Assertion }
 \def\pgOSXXII{\frefwarning 37} \IgNoRe{PG}
 \def\eqnLadAI{\frefwarning XXII.1} \IgNoRe{EQN}
 \def\eqnLadAII{\frefwarning XXII.2} \IgNoRe{EQN}
 \def\eqnLadAIII{\frefwarning XXII.3} \IgNoRe{EQN}
 \def\lemLadAII{\frefwarning XXII.3} \IgNoRe{STM Assertion }
 \def\lemLadAIII{\frefwarning XXII.4} \IgNoRe{STM Assertion }
 \def\propLadA{\frefwarning XXII.5} \IgNoRe{STM Assertion }
 \def\eqnLadAIV{\frefwarning XXII.4} \IgNoRe{EQN}
 \def\defOSLadA{\frefwarning XXII.6} \IgNoRe{STM Assertion }
 \def\lemOSchannelthree{\frefwarning XXII.7} \IgNoRe{STM Assertion }
 \def\theoremOSLadA{\frefwarning XXII.8} \IgNoRe{STM Assertion }
 \def\defOSzeroext{\frefwarning E.1} \IgNoRe{STM Assertion }
 \def\remOSzeroext{\frefwarning E.2} \IgNoRe{STM Assertion }
 \def\defOSzerosectorext{\frefwarning E.3} \IgNoRe{STM Assertion }
 \def\pgOSE{\frefwarning 43} \IgNoRe{PG}
 \def\remOSzerosectorext{\frefwarning E.4} \IgNoRe{STM Assertion }
 \def\lemOSsectorext{\frefwarning E.5} \IgNoRe{STM Assertion }
 \def\remOSsectorconv{\frefwarning E.6} \IgNoRe{STM Assertion }
 \def\defOSindresector{\frefwarning E.7} \IgNoRe{STM Assertion }
 \def\defOSindresectorI{\frefwarning E.8} \IgNoRe{STM Assertion }
 \def\defOSindresectorII{\frefwarning E.9} \IgNoRe{STM Assertion }
 \def\propOSindresectorI{\frefwarning E.10} \IgNoRe{STM Assertion }
 \def\pgOSIVref{\frefwarning 48} \IgNoRe{PG}
  \IgNoRe{PG}
  \IgNoRe{PG}
  \IgNoRe{PG}
 \def\pgOSIVnot{\frefwarning 49} \IgNoRe{PG}

%%%%%%%%%%%%%%%%%%%%%%%%%%%%%%%%%%%%%%%%%%%%%%%%%%%%%%%%%%%%%%%%%%%%%%%
%%%%%%%%    os-chap.tex - chapter numbering data from FTKo1,2,3,4
%%%%%%%%%%%%%%%%%%%%%%%%%%%%%%%%%%%%%%%%%%%%%%%%%%%%%%%%%%%%%%%%%%%%%%%

\newcount\CHAPNO
\newcount\APPNO
\CHAPNO=0
\APPNO=1
\def\advCHAPNO{\advance\CHAPNO by 1}
\def\advAPPNO{\advance\APPNO by 1}

\def\caproman#1{\ifcase#1\or I\or II\or III\or IV\or V\or VI\or VII\or
VIII\or IX\or X\or XI\or XII\or XIII\or XIV\or XV\or XVI\or XVII\or XVIII\or
XIX\or XX\or XXI\or XXII\or XXIII\or XXIV\or XXV\or XXVI\or XXVII\or XXVIII\or XXIX\or XXX\or XXXI\or XXXII\or XXXIII\or XXXIV\or XXXV\or XXXVI\or XXXVII\or XXXVIII\or XXXIX\fi}%

\def\capletter#1{\ifcase#1\or A\or B\or C\or D\or E\or F\or G\or
H\or I\or J\or K\or L\or M\or N\or O\or P\or Q\or R\or
S\or T\or U\or V\or W\or X\or Y\or Z\fi}%

\newcount\cHintroI \cHintroI=\CHAPNO \advCHAPNO 
                              
\newcount\cHnorms  \cHnorms=\CHAPNO \advCHAPNO 
                              \edef\CHnorms{\caproman\CHAPNO}
\newcount\cHproprengrp \cHproprengrp=\CHAPNO \advCHAPNO 
                              
\newcount\cHcovbounds  \cHcovbounds=\CHAPNO \advCHAPNO 
                              
\newcount\cHinsulator \cHinsulator=\CHAPNO \advCHAPNO

 \advAPPNO

\newcount\cHintroII \cHintroII=\CHAPNO \advCHAPNO 
                              \edef\CHintroII{\caproman\CHAPNO}
\newcount\cHamputate \cHamputate=\CHAPNO \advCHAPNO
                              
\newcount\cHscales \cHscales=\CHAPNO \advCHAPNO
                              
\newcount\cHfourier \cHfourier=\CHAPNO \advCHAPNO
                              \edef\CHfourier{\caproman\CHAPNO}
\newcount\cHmomentum \cHmomentum=\CHAPNO \advCHAPNO

 \advAPPNO
 \advAPPNO

\newcount\cHintroIII \cHintroIII=\CHAPNO \advCHAPNO
                              \edef\CHintroIII{\caproman\CHAPNO}
\newcount\cHsectors \cHsectors=\CHAPNO \advCHAPNO
                              \edef\CHsectors{\caproman\CHAPNO}
\newcount\cHsecpropbounds \cHsecpropbounds=\CHAPNO \advCHAPNO
                              
\newcount\cHladdersNotn  \cHladdersNotn=\CHAPNO \advCHAPNO
                              
\newcount\cHestren  \cHestren=\CHAPNO \advCHAPNO
                              
\newcount\cHsecmomnorm \cHsecmomnorm=\CHAPNO \advCHAPNO
                              
\newcount\cHmomestren \cHmomestren=\CHAPNO \advCHAPNO

\edef\APappNaiveladder{\capletter\APPNO} \advAPPNO

\newcount\cHintroIV  \cHintroIV=\CHAPNO \advCHAPNO
                              
\newcount\cHcomparison   \cHcomparison=\CHAPNO \advCHAPNO
                              
\newcount\cHsumsmom  \cHsumsmom=\CHAPNO \advCHAPNO
                              \edef\CHsumsmom{\caproman\CHAPNO}
\newcount\cHsectorsmom   \cHsectorsmom=\CHAPNO \advCHAPNO
                              
\newcount\cHppladsect    \cHppladsect=\CHAPNO \advCHAPNO
                              \edef\CHppladsect{\caproman\CHAPNO}

\edef\APappSpatialSectors{\capletter\APPNO} \advAPPNO